\documentclass{aa}  

\usepackage{graphicx}
\usepackage{longtable}
\usepackage{txfonts}
\usepackage{amsmath}
\usepackage{pifont}
\usepackage{comment}
\usepackage{color}
\usepackage[dvipsnames]{xcolor}
\usepackage{array}
\usepackage{tabularx}
\usepackage{booktabs}
\usepackage{multirow}

\newcommand{\msun}{\mbox{M$_{\odot}$}}
\newcommand{\rsun}{\mbox{R$_{\odot}$}}
\newcommand{\lsun}{\mbox{L$_{\odot}$}}
\newcommand{\zsun}{\mbox{Z$_{\odot}$}}
\newcommand{\teff}{\mbox{$T_{\mathrm{eff}}$}}
\newcommand{\logg}{\mbox{$\log g$}}

\newcommand{\gammae}{\mbox{$\Gamma_{\mathrm{e}} $}}
\newcommand{\Mz}{\mbox{$M_\mathrm{ZAMS}$}}
\newcommand{\parsec}{\textsc{parsec}}
\newcommand{\sevn}{\textsc{sevn}}
\newcommand{\mrdw}{\mbox{$\dot{M}_{\mathrm{rdw}}$}}

\newcommand{\mzams}{\mbox{$M_\mathrm{ZAMS}$}}
\newcommand{\mgam}{$\dot{M}_{\mathrm{\Gamma_e}}$}
\newcommand{\mlm}{$\dot{M}_{\mathrm{L/M}}$}

\newcommand{\xs}{$X_{\rm H}^{\rm S}$}
\newcommand{\mprim}{$M_\mathrm{ZAMS, 1}$}

\newcommand{\vinko}{$\dot{M}_{\mathrm{Vink01}}$}
\newcommand{\vinkeleven}{$\dot{M}_{\mathrm{Vink11}}$}
\newcommand{\sabg}{$\dot{M}_{\mathrm{Sab, \Gamma_e}}$}
\newcommand{\sablm}{$\dot{M}_{\mathrm{Sab, L/M}}$}
\usepackage[colorlinks]{hyperref}

\definecolor{tiffany}{RGB}{79, 166, 158}

\begin{document}

    \title{Enhanced Mass Loss of Very Massive Stars:} 
    \subtitle{Impact on the Evolution, Binary Processes, and Remnant Mass Spectrum}

    \author{Kendall G. Shepherd\inst{1, 2}\fnmsep\thanks{corresponding author: kshepher@sissa.it}
        \and
        Guglielmo Costa\inst{3, 2}
         \and  
       Cristiano Ugolini\inst{1}
       \and
        Guglielmo Volpato\inst{4, 2}
        \and  
        Diego Bossini\inst{3, 2}
        \and \\
        Cecilia Sgalletta\inst{1, 5, 6}
        \and
        Francesco Addari\inst{1}
         \and  
        Alessandro Bressan\inst{1}
         \and  
        Leo Girardi\inst{2}
        \and
        Mario Spera\inst{1, 5, 6}
         }
%\affiliation{SISSA \\
% \\
%}
    \institute{
        SISSA, via Bonomea 365, I--34136 Trieste, Italy
    \and
        INAF - Osservatorio Astronomico di Padova, Vicolo dell'Osservatorio 5, Padova, Italy
    \and
        Dipartimento di Fisica e Astronomia, Universit\`a degli studi di Padova,
        Vicolo dell'Osservatorio 3, Padova, Italy
    \and
        Institut d'Astronomie et d'Astrophysique, Université Libre de Bruxelles,
        CP 226 - Boulevards du Triomphe - B 1050 Bruxelles, Belgium
    \and
        INAF - Osservatorio Astronomico di Roma, Via Frascati 33, I-00040, Monteporzio Catone, Italy
    \and
        INFN - Sezione di Trieste, I-34127 Trieste, Italy
    }

    \date{Received 9 May 2025 / Accepted --}

\hypersetup{
    %draft,
    %colorlinks=true,
    linkcolor=blue,
    citecolor=blue,
    filecolor=magenta,      
    urlcolor=blue
}

\abstract{
Very massive stars (VMS) play a fundamental role in astrophysics due to their stellar feedback through winds and supernovae, and their role as massive black hole (BH) progenitors, though their origin and evolution remain a significant challenge. Recent theoretical work and observations suggest that VMS approaching the Eddington limit may experience mass loss exceeding the standard wind predictions.
This study investigates how enhanced winds influence single and binary VMS evolution, their observable properties, and resulting BH populations. New stellar wind prescriptions, sensitive to the Eddington parameter (\gammae) and the luminosity-to-mass ratio, were implemented into the stellar evolution code \textsc{parsec} v2.0. These updated single-star tracks (100 - 600~\msun\ at LMC metallicity), were used to model the VMS population in the Tarantula Nebula, and integrated into the \sevn\ binary evolution code.
The \gammae\-enhanced single-star tracks match observed VMS properties in the Tarantula Nebula better than standard wind models. Explaining the most massive star, R136a1, through single-star evolution alone suggests a zero-age main sequence (ZAMS) mass limit of $\lesssim$ 400 \msun, regardless of the wind recipe used. However, binary stellar mergers offer a suitable origin for R136a1 and other VMS in the Tarantula Nebula, potentially lowering the upper ZAMS mass limit by $\sim$ 100 \msun.
Stronger winds lead to smaller pre-supernova and BH masses, altering the predicted BH mass spectrum. In binaries, enhanced winds inhibit main-sequence stellar mergers and limit BH production exceeding the pair-instability mass gap’s lower edge ($\sim 50\,\msun$). 
For binary BHs merging within a Hubble time, enhanced-wind models yield more primary BHs above 30 \msun, and enable secondary BHs between 30-40 solar masses, a range not found with standard stellar winds at LMC metallicity.
This study highlights the crucial role of stellar wind physics and binary interactions in VMS evolution and BH populations, offering predictions relevant for interpreting VMS observations and gravitational wave source origins.

}

\maketitle

\section{Introduction} \label{sec:intro}
Stellar winds are one of the most influential aspects that drive the evolution of Very Massive Stars (VMS), defined as stars with masses $\geq$ 100 \msun. 
Their strong winds chemically enrich the surrounding interstellar medium (ISM), inject energy into the environment, and ionize the surrounding matter. Stellar winds also play a key role in forming Wolf-Rayet stars (WR), peculiar stripped stars that display enhanced mass loss rates. Stellar winds, which are driven by radiation pressure and metallicity-dependent processes, regulate the mass loss over the star's lifetime. These winds directly influence the initial-final mass relation of stars, which links the mass at the zero-age main-sequence (ZAMS) with the final remnant mass.

%In recent years, significant efforts have been made to better study and constrain the relationship between initial and remnant masses of VMS, yet this crucial connection remains poorly defined. This because many physical processes are yet to be fully understood, while few observational constraints exist. VMS are much less numerous than their low mass counterparts, thus providing a challenge in understanding their nature.

A variety of mass-loss prescriptions for massive stars have been developed, such as those by \citet{CAK1975, deJager1988, Nugis2000, Vink2000, Vink2001, Grafener2008, Vink2011, Grafener2011}, and are widely used in stellar evolution models. However, accurately determining the mass-loss rates of VMS remains a major open question in astrophysics, particularly in regimes near the Eddington limit where theoretical uncertainties and observational constraints are both significant.
In recent years, considerable effort has been made to better constrain the relationship between VMS's initial and remnant masses. Driven by new observational data and advances in hydrodynamical modeling, updated mass-loss prescriptions are being introduced regularly \citep[e.g.,][]{Bestenlehner2020a, Sabhahit2022, Bjorklund2023, Gormaz2024}. As a result, it is essential to implement and test these theoretical prescriptions in stellar evolution models to evaluate their impact on the structure and fate of VMS.

The Tarantula Nebula (30 Doradus cluster) in the Large Magellanic Cloud (LMC) is a treasure trove of massive stars, as it hosts the most massive single stars ever observed, with several stars having $M \geq$ 100 \msun\ \citep{Crowther2010, Bestenlehner2020, Brands2022, shenar2023}. Observations show that the effective temperatures of VMS in the Tarantula Nebula fall within a relatively narrow range, 4.6 $\lesssim$ log (\teff [K]) $\lesssim$ 4.74 \citep{Crowther2010, Schneider2018, Brands2022, Bestenlehner2014}. A similar trend is observed in the Arches cluster, where VMS lie in the range 4.5 $\lesssim$ log (\teff) $\lesssim$ 4.6 \citep{Martins2008}. While one might attribute this tight temperature distribution to a recent burst of star formation, standard stellar evolution models predict that VMS should undergo significant envelope expansion toward the end of the main sequence. This would shift them to cooler effective temperatures than those currently observed in the Tarantula cluster. In such models, VMS are expected to evolve toward large radii and lower \teff\, potentially crossing the Humphreys-Davidson (HD) limit \citep{Humphreys1979}, in contradiction with observations. This discrepancy suggests that revised mass-loss prescriptions may be required to prevent such expansion and more accurately reproduce the observed properties of VMS.

A potential solution, proposed by \citet{Sabhahit2022}, involves enhancing stellar wind mass loss rates in such a way that balances envelope expansion, thereby allowing stars to maintain nearly constant effective temperatures during the main sequence (MS).
As the evolution of VMS when in close proximity to the Eddington limit can be uncertain, even small changes in mass loss strength can significantly affect their evolutionary properties. Thus, adjusting the wind prescription in this regime may reconcile theoretical models with the observations.

%%%%Many stars, especially the more massive ones, are born in binaries or higher multiple stellar systems. \cite{Moe_2017} showed that the multiplicity of stellar systems increases with stellar mass, with more than $90\% $ of massive stars that are expected to be in binaries (see \citealt{offner2022}).
However, winds are not the only source of uncertainty in VMS evolution. 
Binary interactions can be consequential in shaping VMS's evolution and final outcomes, which are frequently found in binary or multiple systems \citep{Mason2009, Sana2013}.
Around $\sim$ 70\% of massive stars exist in binary systems, which will interact and exchange mass within their lifetime \citep[][and references therein]{Smith2014}, and have a preference for massive companions \citep{Kobulnicky2007}. In this scenario, stellar winds play a key role. For instance, enhanced mass loss can prevent a star from expanding to the red supergiant (RSG) stage \citep{Josiek2024}, thus avoiding eventual mass transfer episodes. Products of binary interactions have also been proposed as an explanation for various stellar objects, such as luminous blue variable stars \citep{Gallagher, Justham2014, Smith2015} and long-lived blue supergiants (BSGs) \citep[e.g., for the progenitor of SN1987A,][]{Podsiadlowski1990, Menon2017}.
Moreover, binary interactions can alter the physical appearance of stars; Roche Lobe overflow (RLOF) can strip a massive star of nearly its entire hydrogen envelope, making it resemble a WR star \citep{Vanbeveren1980, Petrovic2005}.
In some cases, stars in binary systems may even collide and merge, forming a more massive star that continues stellar evolution and can appears similar to one born with a higher ZAMS mass \citep{PortegiesZwart2002, Freitag2006, Costa2022}.
These wind-driven and binary interaction processes not only shape the evolution and final masses of VMS, but also influence the formation of compact object binaries. In particular, massive binary systems can ultimately evolve into black hole binaries, which may merge within a Hubble time and produce gravitational wave signals detectable by current observatories \citep{Abbot2016}.

%The standard picture of single stellar evolution predicts a stars with initial mass $\sim$ 140 - 260 \msun\ (corresponding to final He core masses of $\sim$ 64 - 130 \msun) will end evolution as a pair instability supernova (PISN), leaving no remnant behind \citep{Heger2002, Woosley2017}. This result predicts a gap in the mass spectrum of black holes between $\approx$ 40 - 65 \msun\ and $\approx$ 120 \msun. This picture has been challenged by recent detections of BHs with masses within the so-called mass gap, such as GW190521 and GW190426, with masses ... .Many studies have been done to try to explain such masses, including variations in the $^{12}\mathrm{C(\alpha, \gamma)^{16}O}$ reaction rates \citep{Costa2021}, pulsation driven mass loss \citep{Volpato2023} with rotation \citep{Volpato2024}, and other uncertainties in massive star evolution \citep{Farmer2019,Woosley2021,Vink2021}.
%Moreover, many studies have shown that the edges if the PI mass gap depend on uncertainties in the evolution of massive stars \citep{Farmer2019, Vink2021, Woosley2021, Volpato2023, Volpato2024}. These studies show that the edges of the PI mass gap remain uncertain, corroborating the need for further investigations to improve our models and their predictive capabilities.

In this work, we introduce newly implemented wind prescriptions into the \parsec\ v2.0 stellar evolution code, designed to capture the enhanced mass-loss behavior of VMS near the Eddington limit, following \citet{Sabhahit2022}. The prescriptions, not previously included in the code, allow us to explore how stronger wind dependencies affect the evolution of VMS. As a testbed, we focus on the Tarantula Nebula in the LMC, whose rich population of massive stars, especially in R136, offers valuable observational constraints. We compute stellar evolution tracks for single stars across a wide mass range (100 - 600 \msun), at LMC metallicity, and for the first time, integrate these updated tracks into the binary population synthesis code \sevn. This allows us to systematically explore how enhanced winds influence both single and binary evolution of VMS, including their remnant masses and the formation pathways of merging BHs. In doing so, we aim to address key uncertainties in the mass-loss behavior of VMS, evaluate whether the models can reproduce the observed stellar properties in the Tarantula Nebula, and assess their implications for the population of gravitational-wave sources at LMC-like metallicity.

The paper is organized as follows. We describe the observational data, input physics of our stellar tracks, and binary population synthesis simulations setup in Sec. \ref{sect:methods}. In Sec. \ref{sect:results}, we present the stellar tracks with the new wind implementation. We discuss their evolution on the Hertzsprung-Russell diagram (HRD) in Sec. \ref{sect:HRD}, chemical evolution in Sec. \ref{sect:chemi}, and pre-supernova and BH masses in Sec. \ref{sect:fates}. In Sec. %\ref{sect:binary} and 
\ref{subsec:wind_accretion} we discuss the impacts on binary evolution. In particular, we discuss stellar mergers in Sec. \ref{subsec:stellar mergers}, and the resulting BH mass distribution in Sec. \ref{subsec:Remnants}. Then in Sec. \ref{sect:discussion}, we discuss the implications, applicability, and uncertainties of our models and results. Finally, our results are summarized in Sec. \ref{sect:concl}.

\section{Methods}\label{sect:methods}

\subsection{Data}\label{subsec:data}

The primary focus of this study is on VMS, and as such, we focus our analysis on R136, a concentration of stars in the NGC 2070 star cluster located in the Tarantula Nebula. Initial studies suggested that the central object in R136 is a single star with a mass $>$ 1\,000 \msun\ \citep{Cassinelli1981, Savage1983}. However, follow-up studies showed that the central region is instead comprised of distinct stellar sources \citep{Weigelt1985, Lattanzi1994}. Even as individual stars, the central region still hosts the most massive stars known to date, with current mass estimates of $\sim$ 150 - 300 \msun\ \citep{Crowther2010}. In particular, it hosts three stars (R136a1, R136a2, and R136a3) with masses above $\sim$ 150 \msun, cataloged as WNh type, i.e., rare WR of nitrogen type (WN) that show the presence of hydrogen lines in their spectra, and are likely still undergoing core-hydrogen burning \citep{Massey1998, dekoter1998, Smith2008}.

The LMC's metallicity is estimated to be between a quarter and half that of the Sun ([Fe/H] $\simeq$ -0.37 to [Fe/H] $\simeq$ -0.41) \citep{Choudhury2016, Choudhury2021} and stellar modelers frequently use $Z \simeq 0.5 \zsun$ (i.e., Z $\approx$ 0.006) when simulating stars in the Tarantula Nebula \citep{Bestenlehner2014, Schneider2018, Brands2022, Martinet2023, Josiek2024}.

We compare our models with data from \citet{Schneider2018}, which is composed of $>$ 400 stars in the 30 Doradus region, taken from the VLT-FLAMES Tarantula Survey \citep{Evans2011}. This data were obtained using the Hubble Space Telescope and the Multi Unit Spectroscopic Explorer (MUSE) on the Very Large Telescope (VLT) and contain a compilation of results from \citet{Bestenlehner2014, SabinSanjulian2014, McEvoy2015, SabinSanjulian2017, Ramirez2017}.
We also compare our models with data from \cite{Brands2022} of the central regions of the R136 star cluster taken with the Hubble Space Telescope/STIS.
%\citet{Brands2022} provided surface abundances from fits of optical and UV data, in addition to the initial and evolutionary mass they derived using the \textsc{bonnsai} tool.

\begin{figure}[t]
     \centering
    \includegraphics[width=\linewidth]{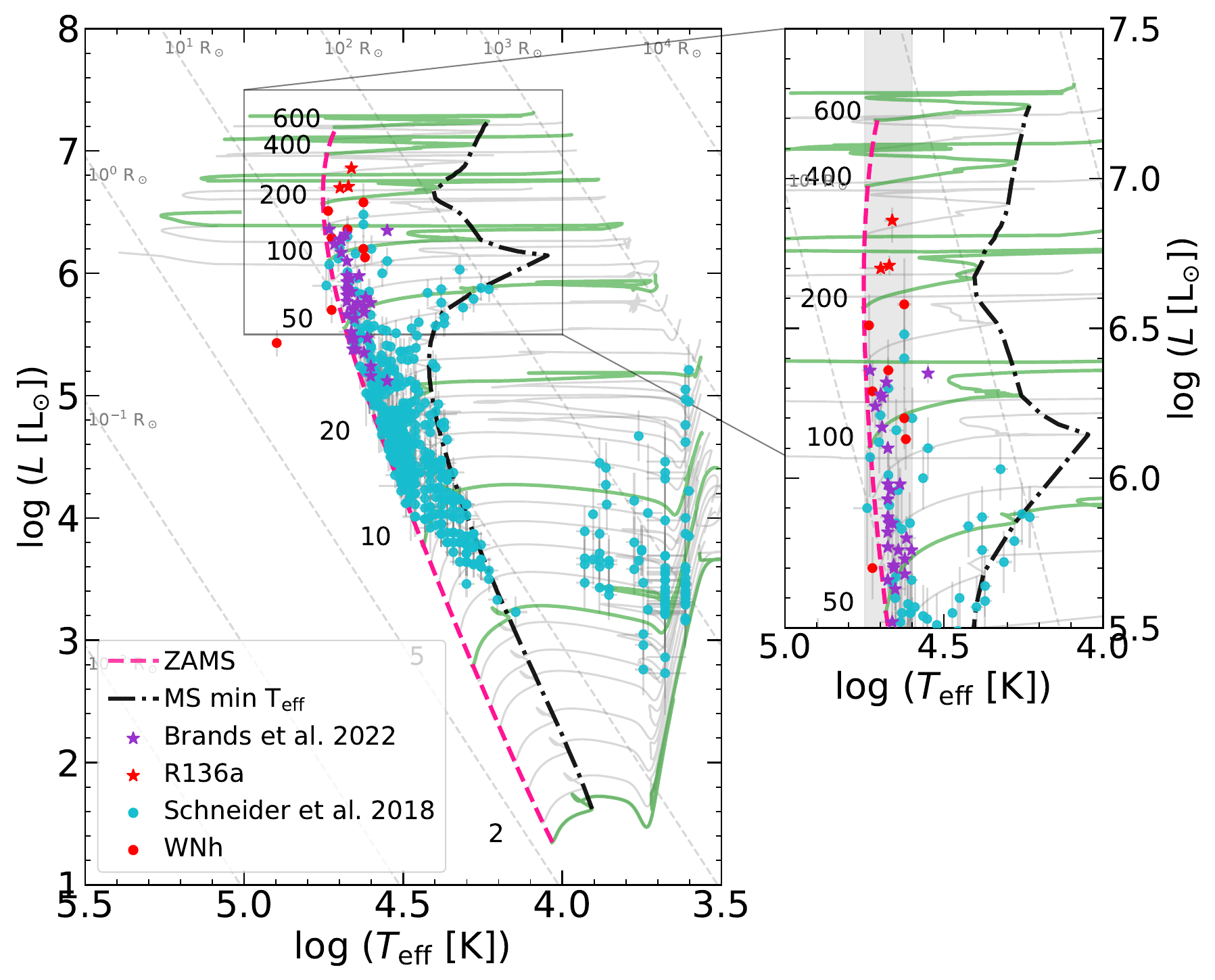}
    \caption{HRD adapted from \citet{CostaShepherd2025} showing stars in the LMC. The green and gray lines are tracks from \citet{CostaShepherd2025}, for tracks with Z = 0.006 and the standard mass loss rate. 
    Purple and red stars are from \citet{Brands2022} (red stars indicate R136a) and blue and red dots are from \citet{Schneider2018} (red dots are WNh stars). The dashed pink line and dash-dotted black line show the ZAMS and TAMS, respectively. The zoomed-in plot shows the most massive stars, with the gray-shaded region showing the width of the observed temperature range of VMS in the LMC. }
    \label{fig:HRD_LMC}
\end{figure}

Figure \ref{fig:HRD_LMC} shows the HRD of tracks using the standard mass loss recipe from \citet{CostaShepherd2025} and stellar data from \citet{Schneider2018} and \citet{Brands2022}.
The figure shows excellent agreement between the data and the evolutionary tracks for stars with ZAMS masses $\lesssim 80\msun$, as the ZAMS and terminal-age main sequence (TAMS) of the tracks effectively bracket the observed stars. More massive stars show a different picture: assuming that these stars are all part of the same starburst, the width between the ZAMS and TAMS in the stellar models is significantly broader than suggested by the data, and, at the TAMS, the models are cooler than the observed ones (4.75 $\leq$ log \teff\ [K] $\leq$ 4.6).
%If we assume these stars are all part of the same starburst, the models suggest that these massive stars should expand and be observed at lower \teff.% than the lower mass stars of the cluster.This is not the case, as the most massive stars are observed in a narrow effective temperature range (4.75 $\leq$ log \teff\ [K] $\leq$ 4.6). 
The discrepancy can be alleviated with very young ages of the cluster ($<$ 1 Myr), where stars populate only the hotter side of the MS, rapid rotation, or with increased mass loss rates, which can suppress the occurrence of envelope expansion. Such a young age seems unlikely, as stars would still retain their H envelope and would not appear as WNh yet. Additionally, different studies have suggested older stellar ages of 1 - 2.5 Myr \citep{Crowther2010, Bestenlehner2020}. The VMS observed are not rapidly rotating \citep{Bestenlehner2014, Bestenlehner2020} and VMS are not largely affected by moderate rotational mixing \citep{Yusof2013}.
In contrast, to explain this observed feature of VMS, we investigate an alternative wind prescription with increased mass loss.

%\KS{slow to moderate rotation rates for R136a, + not high v sin i values (Bestenlehner 2014, 2020.), For luminous VMS, mass loss is more responsible for surface He enrichment than rotational mixing (Bestenlehner 2014)}

%The discrepancy can be alleviated with a.) very young ages of the cluster ($<$ 1 Myr), b.) stellar mergers aggregating to form VMS in the narrow \teff\ range, or c.) increased mass loss rate which can suppress the occurrence of envelope expansion. Such a young age seems unlikely, and different studies have suggested ages of 1 - 2.5 Myr \citep{Crowther2010, Bestenlehner2020}. We investigate the latter option, that increased mass loss of VMS may explain this observed feature. We also investigate the possibility that stellar mergers may produce massive stars that occupy this \teff\ range. 

\subsection{Bayesian statistical analysis}\label{subsec:param}
We used an updated version of the \textsc{param}\footnote{\href{https://stev.oapd.inaf.it/cgi-bin/param}{https://stev.oapd.inaf.it/cgi-bin/param}} code \citep{daSilva2006, Rodrigues2014, Rodrigues2017}, which will be soon released (Bossini et al. in prep.), to fit the observed data with our stellar tracks (described in the following section). The code exploits Bayesian statistics, using 
the observed metallicity, effective temperature, and luminosities (and their uncertainties) to estimate to estimate stellar parameters of interest, like age, initial and current mass, and \logg, using theoretical isochrones derived from our stellar evolutionary tracks.
The output of \textsc{param} gives the best fitting value with its correlated error, i.e., the associated credible intervals that include 68\% of the distributions.

\subsection{Stellar code and input physics}\label{subsec:physics}
In this work, we present new stellar evolution tracks computed with the PAdova and TRieste Stellar Evolution Code \cite[\parsec,][]{Bressan2012} v2.0 with new mass loss implementations for VMS. %with the goal of describing VMS in the LMC, and determining their compact remnant masses with \sevn.
\parsec\ v2.0\footnote{\href{https://stev.oapd.inaf.it/PARSEC/}{https://stev.oapd.inaf.it/PARSEC/}} has been described in detail by \citet{Costa2021, Nguyen2022, CostaShepherd2025}, and \parsec\ v1.2 in previous works \citep{Chen2015, Fu2018}. Below, we briefly summarize the main input physics of the code

The stellar tracks are evolved from the pre-main sequence (PMS) to either the oxygen burning stage or until entering the electron-positron (e$^{+}$ - e$^{-}$) pair instability (PI) regime. We compute non-rotating\footnote{See App. \ref{app:rotating} for rotating tracks} tracks with initial masses 100 $\leq \mzams/\msun \leq$ 600 with a mass step of 10 \msun\ for tracks with 100 $\leq \mzams/\msun \leq$ 400, and a mass step of 50 \msun\ above. We focus on this mass range to investigate the effect of enhanced stellar winds of VMS near the Eddington limit, since stars with smaller masses are less affected by the differences in stellar winds. 

% , which use the standard wind prescription and are available on the \parsec\ web database\footnote{http://stev.oapd.inaf.it/PARSEC/}. 

Stellar tracks are computed with metallicity Z = 0.006 and Y = 0.259 in mass fraction (i.e., [M/H] = -0.404\footnote{Approximated value from [M/H] = log(($Z/X$)/0.0207) \citep{Bressan2012}}), to investigate the massive-star hosting environment of the LMC. We adopt the solar-scaled mixtures from \citet{Caffau2011} and take the solar metallicity to be Z$_{\odot}$ = 0.01524.

Energy transport via convection is described using the mixing length theory \citep{Bohm-Vitense1958}. We set the mixing length parameter $\alpha_{\mathrm{MLT}}$ = 1.74 \citep[as suggested by][]{Bressan2012}. We test stability against convection using the Schwarzschild criterion \citep{Schwarzschild1958}.

Core overshooting is included in our calculations, which allows for convective elements to penetrate into stable regions\footnote{Penetrative overshooting is described in \citet{Bressan1981}.}. The overshooting parameter is set to $\lambda_{\mathrm{ov}}$ = 0.5 in pressure scale height, $H_{\mathrm{P}}$, which indicates the maximum distance an overshooting bubble can travel across the unstable border. Therefore, in our framework the overshooting length is then $l_{\mathrm{ov}} \approx 0.5 \lambda_{\mathrm{ov}} H_{\mathrm{P}} = 0.25 H_{\mathrm{P}} $. Overshooting from the convective envelope is also taken into account, extending from the base of the convective envelope downward by $\Lambda_{\mathrm{env}} = 0.7 H_{\mathrm{P}}$.

The stellar type is tracked throughout the evolution of the star and is determined by its burning phases, surface abundances, and \teff. The definitions of the stellar types referenced in this work are listed in Table \ref{tab:phase_types}.

\begin{table}
  \centering
\caption{Definitions used for different stellar types.}
  \begin{tabular}{c|ccc}
  \toprule
    \multicolumn{1}{c|}{Type}&  \multicolumn{1}{c}{$\log \teff$ [K]} & \multicolumn{1}{c}{\xs} & \multicolumn{1}{c}{ } \\ 
    \midrule
    OB &  $\geq$ 4.0  &  > 0.3 & MS \\
    
    BSG & $\geq$ 4.0 & > 0.3 & Post-MS \\ 
    YSG & 3.7 - 4.0  & &  \\
    RSG & $\leq$ 3.7 & & \\ 
    \hline
    WR & $\geq$ 4.0 & $<$ 0.3 \\ 
    \hline
    WNh & $\geq$ 4.0 & 10$^{-5} - 0.3$ & \\ 
    WN & $\geq$ 4.0 & $\leq 10^{-5}$ & $X\mathrm{_N^{S}}$ $\geq$ $X\mathrm{_C^{S}}$\\ 
    WC &  4.0  - 5.25 & $\leq 10^{-5}$ & $X\mathrm{_C^{S}}$  >  $X\mathrm{_N^{S}}$ \\ 
    WO & $\geq$ 5.25 & $\leq 10^{-5}$ & $X\mathrm{_O^{S}}$  $\gtrsim$ $X\mathrm{_C^{S}}$  > $X\mathrm{_N^{S}}$  \\ 
    \bottomrule

  \end{tabular}
  
  \footnotesize{$X^{\rm S}$ indicates the percentage of H, C, N, and O at the star's surface. BSG, YSG, and RSG stand for blue, yellow, and red supergiant, respectively.
  }
  \label{tab:phase_types}
\end{table}

\subsection{Mass loss descriptions}\label{sect:winds}
\subsubsection{Standard winds}\label{subsec:standard winds}
This section outlines the mass loss prescriptions implemented in our models.
The standard mass loss rate for massive stars (\mzams $\geq$ 10 \msun) used in \parsec\ v2.0 is the same as that used in previous works \citep{Chen2015, Volpato2023, Volpato2024, CostaShepherd2025}. This recipe is based on the prescription formulated by \cite{Vink2001}, which provides the mass loss rate as a function of stellar parameters for massive O and B type stars for a range of metallicities and temperatures. For hot stars (\teff $\geq$ 10\,000 K), the main dependencies are:

%Here we describe the usual mass loss implementation for massive stars (\mzams $\geq$ 10 \msun) used in \parsec\ v2.0. The details are described and used in the models presented in \citet{CostaShepherd2025}, here we include an overview. We refer to this mass loss recipe as \mrdw, to keep consistency among other works \citep{Volpato2023, Volpato2024}, however we acknowledge that other mass loss mechanisms are at work in RSGs.

\begin{equation}
\begin{split}
    \dot{M}_{\mathrm{Vink2001}} \propto \frac{L^{2.194} Z^{0.85}} {M^{ 1.313} }
\end{split} 
\label{eq:vink_01_dep}
\end{equation}
Where $L, M$, and $Z$ are the star's current luminosity, mass, and metallicity. The full equation is given in App. \ref{app:massloss}, and also includes a dependence on \teff, the escape velocity $\nu_{\mathrm{esc}}$, and terminal velocity $\nu_{\infty}$.

To account for the increase in mass loss due to recombination or ionization of iron at different temperatures (i.e., 
the bistability jump), we adjust the mass loss rate as described in \citep{Vink2001}.
We also account for the increase in mass loss rates when the Eddington limit is approached \citep{Grafener2008, Vink2011}. The electron scattering Eddington parameter is calculated as:

\begin{equation}
    \Gamma_e = \frac{L \kappa_\mathrm{e}}{4 \pi \mathrm{c G} M} %= 10^{-4.813} (1+X) \Big(\frac{L_*}{\lsun} \Big) \Big( \frac{M_*}{\msun}\Big)^{-1}
    \label{eq:gamma_e}
\end{equation}
where $\kappa_{\mathrm{e}}$ is the opacity due to electron scattering and G is the gravitational constant. The mass loss rates follow different dependencies for high and low \gammae\ regimes. For the high \gammae\ regime, the dependencies are given as \citep{Vink2011}:

\begin{equation}
    \Dot{M}_{\mathrm{Vink11}}  \propto M^{0.78} \gammae^{4.77} \quad \text{for} \quad 0.7 < \gammae < 0.95
    \label{V11_highgamma}
\end{equation}

%However, \cite{Vink2011} does not provide the full dependencies for the mass loss rates. We take the maximum between the \cite{Vink2001} and \cite{Vink2011}.

For cool RSG stars (\teff $<$ 10\,000 K) we take the maximum between the \citet{Vink2001} rates and the rates provided by \cite{deJager1988}.
For WR stars, defined here as those with relative surface hydrogen abundance < 0.3, we apply the luminosity-dependent mass loss prescription from \cite{Sander2019}, \begin{equation}
\mathrm{Log} \frac{\dot{M}}{\msun\ \mathrm{yr}^{-1}} = -8.31 + 0.68 \cdot \mathrm{Log} \frac{L}{\lsun }
\label{eq:mloss_sander}
\end{equation}
which is adjusted to include a metallicity dependence based on \cite{Vink2015}. The combination of the above mass-loss recipes are referred to as \mrdw.

\subsubsection{New winds for VMS}
\cite{Sabhahit2022} proposed enhanced mass loss rates for VMS transitioning from O-type to WNh stars based on theoretical and observational evidence, marking a shift from optically thin to optically thick winds. The proposed rates use the so-called transition mass loss rate, which is a model-independent mass loss describing the transition from O to WNh winds.% and relies on the assumption that ($\Gamma$ -1)/$\Gamma$ is close to unity. 
The transition mass loss rate is given as 
\begin{equation}
\begin{split}
      \Dot{M}_{\text{trans }} = f \frac{L_{\text{trans}}}{\nu_{\infty}c}
     \label{eq:mtrans}
\end{split}
\end{equation}
where $f$ is a metallicity dependent correction factor, $f$ = 0.6 \citep{Vink2012}. One can use the average values of luminosities and terminal velocities of observed Of/WNh stars to calculate the $\Gamma_{\mathrm{e, trans}}$, which is the location of the upturn in mass loss rates. By computing the $\dot{M}_{\mathrm{trans}}$, a model-independent transition mass loss rate is obtained in which the winds switch from optically thin to optically thick when above the transition point. \citet{Sabhahit2022} empirically obtained the transition parameters of the Of/WNh stars in the Tarantula Nebula to be $L$/\lsun\ = 10$^{6.31}$, $\nu_{\infty}$ = 2550 km/s, log \teff\ = 4.64, and $X_s$ = 0.62. For the LMC, the location of the increase in mass loss is thus calculated to be $\Gamma_{\mathrm{e, trans}}$ = 0.42, and the transition mass loss rate is $\log \dot{M}_{\mathrm{trans}}$ = -5.0 \citep{Sabhahit2022}.
The main dependencies for the enhanced winds of VMS are given as:

\begin{equation}
\begin{split}
      \Dot{M}_{\text{Sab, } \Gamma_e} \propto M^{0.78}\Gamma_{\mathrm{e}}^{4.77} \propto \frac{L^{4.77} (1+X_{\rm H}^{\rm S})^{4.77}Z^{0.5}}{M^{3.99}}
     \label{eq:gammae_sab_dep}
\end{split}
\end{equation}

\begin{equation}
\begin{split}
      \Dot{M}_{\text{Sab, L/M}} \propto M^{0.78}(L/M)^{4.77} \propto \frac{L^{4.77} Z^{0.71}}{M^{3.99}}
     \label{eq:LM_sab_dep}
\end{split}
\end{equation}
Where \xs\ is the surface hydrogen abundance (in mass fraction), included only in the \sabg\ recipe. The \sablm\ recipe omitting a surface \xs\ dependence is an extreme case, used to show that the $L/M$ has a larger effect on the mass loss rates than the \xs, and the \xs\ dependence may not be strictly necessary, like the recipe of Eq. \ref{eq:vink_01_dep}. 
The full equations are given in Eqs \ref{eq:gammae_sab_rate} and \ref{eq:LM_sab_rate} in App. \ref{app:massloss}.
Our models adopt two variations, which we refer to as \mgam\ and \mlm:

\begin{equation}
    \dot{M}_{\mathrm{\Gamma_e}} =
      \begin{cases}
       \mathrm{max} (\dot{M}_{\mathrm{Vink2001}} ,  \dot{M}_{\mathrm{Sab, ~\Gamma_e}} ) & \mathrm{for}  \quad \teff > 10,000 ~\mathrm{K} \\
       \mathrm{max} (\dot{M}_{\mathrm{deJager}} ,   \dot{M}_{\mathrm{Sab, ~ \Gamma_e}} ) & \mathrm{for} \quad \teff < 10,000 ~\mathrm{K}
     
     \end{cases}
     \label{eq:mgam}
\end{equation}

\begin{equation}
    \dot{M}_{\mathrm{L/M}} =
      \begin{cases}
       \mathrm{max} (\dot{M}_{\mathrm{Vink2001}} ,  \dot{M}_{\mathrm{Sab, ~L/M}} ) & \mathrm{for} \quad \teff > 10,000 ~\mathrm{K} \\
       \mathrm{max} (\dot{M}_{\mathrm{deJager}} ,   \dot{M}_{\mathrm{Sab, ~L/M}} ) & \mathrm{for} \quad \teff < 10,000 ~\mathrm{K}
     
     \end{cases}
     \label{eq:mlm}
\end{equation}

Choosing the maximum of the \citet{Vink2001} and the \citet{Sabhahit2022} recipes automatically uses the 'low \gammae' recipe (\vinko) below the transition point, and switches to the 'high \gammae' recipe of \sabg\ and \sablm above it.
The full equations, additional discussions, and details on the winds are provided in App \ref{app:massloss}.

\subsubsection{Differences in wind prescriptions}\label{subsec:mdot_diffs}
\begin{figure}
     \centering
    \includegraphics[width=\linewidth]{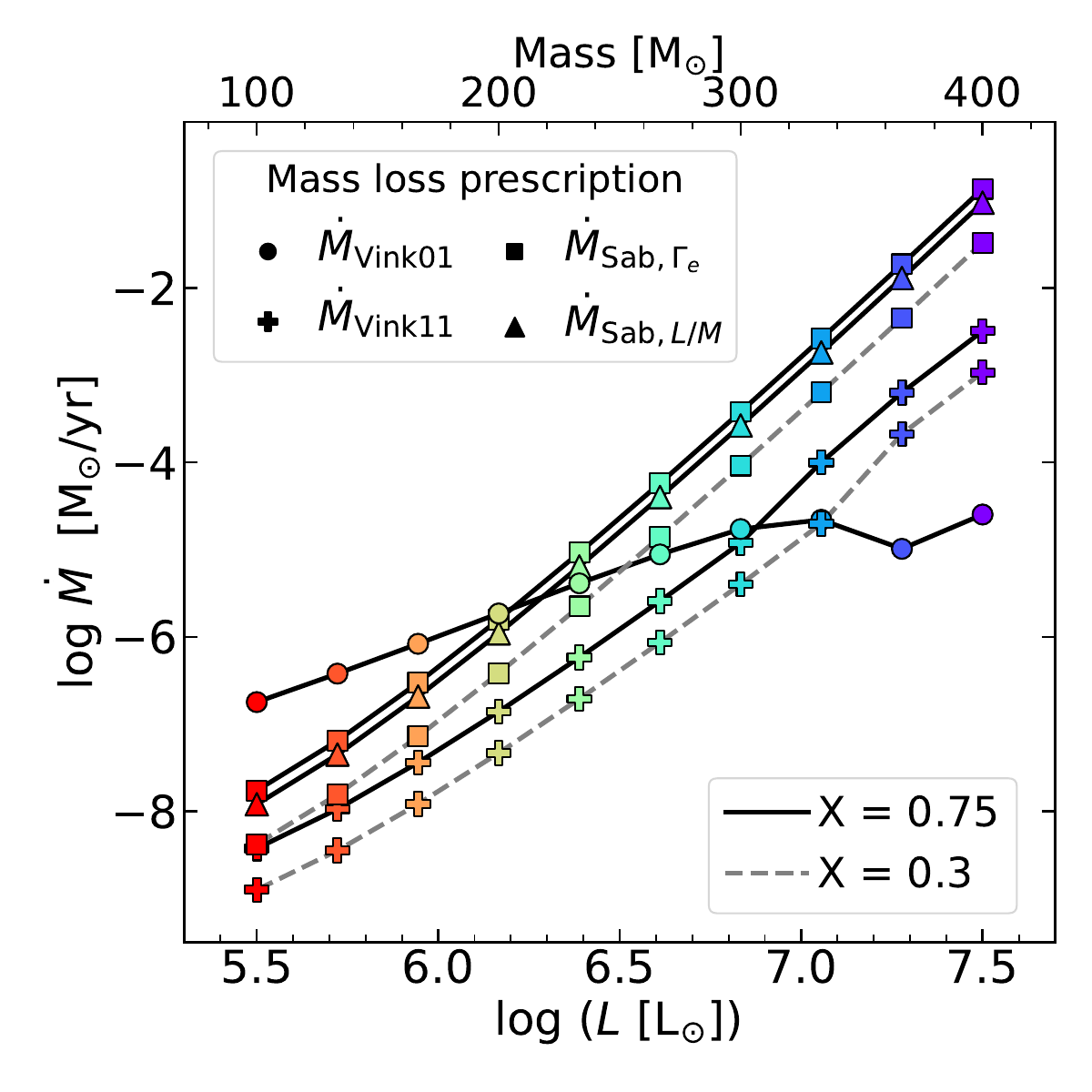}
    \caption{Mass loss rates predicted for different prescriptions as a function of the mass and luminosity for two \xs\ values, indicated in solid black and dashed gray. The circle and cross markers show predictions based on \citet{Vink2001} and \citet{Vink2011}, respectively, and the square and triangle refer to predictions from \citet{Sabhahit2022}, (i.e., equations \ref{eq:gammae_sab_rate} and \ref{eq:LM_sab_rate} ). Values are computed with a fixed log \teff = 4.8 K, Z = 0.006, and $\nu_{\infty}$ = 2550 km/s.  }
    \label{fig:wind_rx}
\end{figure}

To illustrate the main differences in the mass loss treatment, Fig. \ref{fig:wind_rx} shows the mass loss predictions for different recipes and stellar parameters\footnote{See Fig. 3 of \citet{Smith2014} for an overview including other mass loss prescriptions at solar Z.}. The values were computed for a fixed $Z$, \teff\, and $\nu_{\rm inf}$. All the $\dot{M}$ values increase with increasing luminosity (and therefore mass). Smaller surface H abundances (at the same metallicity) decrease the mass loss rates for the \vinkeleven\ and \sabg\ predictions, while they do not affect the \vinko\ or \sablm\ predictions, as they have no dependence on \xs. The \sabg\ and \sablm\ rates show a similar trend when \xs\ $ \approx$ 0.75, but begin to diverge with decreasing \xs.

As the luminosity increases, the \vinkeleven\ surpasses the \vinko\ around log $L$/L$_{\odot}$ $\sim$ 6.7 - 7.0, (depending on the \xs) due to the star approaching the Eddington limit. Similarly, the \sabg\ and \sablm\ rates are greater than the \vinko\ rates only when log $L$/L$_{\odot} \gtrsim$ 6.25, (i.e., entering the high \gammae\ regime). The \sabg\ and \sablm\ rates are always greater than the \vinkeleven\ rate at log luminosities $\geq$ 5.5, but follow a similar trend.

\subsection{Binary evolution with \sevn}\label{sect:sevn}

To investigate the influence of different mass loss on evolution of binary populations, we used the state-of-the-art population synthesis code Stellar Evolution for N-body \citep[\sevn,][]{Spera2015, Spera2017, speraMergingBlackHole2019a, iorioCompactObjectMergers2023}. \sevn\footnote{\url{https://gitlab.com/sevncodes/sevn}} is a highly flexible tool for population synthesis, utilizing on-the-fly interpolation of pre-computed stellar evolution tracks, making it both efficient and adaptable for different evolutionary scenarios.
To study the effect of the new stellar winds prescriptions on the distributions of the final binary compact objects, we used the \textsc{trackcruncher}\footnote{\url{https://gitlab.com/sevncodes/trackcruncher}} tool to adopt the new stellar evolutionary tracks with different winds to include in \sevn. For stars with masses \mzams\ $<$ 100 \msun, we use the stellar models from \citet{CostaShepherd2025} with standard stellar winds, while for stars with \mzams\ $\geq$ 100 \msun, we also use the new models presented in this paper. 
%We implemented the tracks produced with our new prescriptions for stellar winds in the state-of-the-art population synthesis code Stellar Evolution for N-body (\sevn) \cite{speraVeryMassiveStars2017, speraMergingBlackHole2019a, iorioCompactObjectMergers2023}. \sevn\ is a highly flexible tool for population synthesis, utilizing on-the-fly interpolation of pre-computed stellar evolution tracks, making it both efficient and adaptable for different evolutionary scenarios.\par
\sevn\ includes various prescriptions accounting for both the final fate of massive stars (e.g., core-collapse supernovae (CCSN), pulsational pair-instability supernovae (PPISN), and pair-instability supernovae (PISN)), and binary stellar evolution processes (e.g., Roche lobe overflows, common envelope phases, SN kicks). In this work, we use the standard \sevn\ code configuration, in its version 2, extensively described by \citet{Iorio2023}.

For the final fate of massive stars, we employ the delayed model for CCSN from \citet{fryerCOMPACTREMNANTMASS2012}, while for the PPISN and PISN regimes, and direct BH collapse (DBH), we use the formalism from \citet{Spera2017,Mapelli2020}. For the treatment of binary stellar evolution processes, we rely on the fiducial choices by \citep{Iorio2023}, such as common envelope (CE) efficiency $\alpha=1$, mass transfer efficiency $f_{MT}=0.5$, and the standard treatment for the stability of mass transfer events is determined as in \citep{Hurley2002}, with the addition that mass transfer is assumed to always be stable for donor stars with radiative envelopes.

To create our population of binary stars, we generate the initial conditions as follows. For each binary system, the mass of the primary star is extracted from a \citet{Kroupa2001} initial mass function (IMF). We specifically sample masses ranging from 20 to 350 solar masses (\msun) to focus on the effects of massive star winds on the spectrum of remnant masses. Thus, we obtain the mass of the primary star $M_{\textrm{ZAMS},1}$ as
\begin{equation}
    \textrm{pdf}(M_{\textrm{ZAMS},1})\propto M_{\textrm{ZAMS},1}^{-2.3}\quad\text{with}~M_{\textrm{ZAMS},1}\in[20, 350].
    \label{eq: Primary distribution}
\end{equation}

We extract the binary properties following the \citet{Sana2012} distributions. 
Therefore, the mass ratio $q$ is extracted as
\begin{align}
    &\textrm{pdf}(q)\propto q^{-0.1}\quad\text{with}\ q=\dfrac{M_{\textrm{ZAMS},2}}{M_{\textrm{ZAMS},1}}\in[q_{min}, 1.0]\\
    &q_{min}= \textrm{max} \Bigg( \dfrac{5.0}{M_{\textrm{ZAMS},1}},\, 0.1 \Bigg)
    \label{eq: Mass ratio distribution}
\end{align}
where $M_{\textrm{ZAMS},2}$ is the mass of the secondary star, and it is determined simply as
\begin{equation}
    M_{\textrm{ZAMS},2}=qM_{\textrm{ZAMS},1}.
    \label{eq: Secondary mass}
\end{equation}
Finally, the initial orbital periods $P$ and eccentricities $e$ are sampled as
\begin{equation}
    \textrm{pdf}(\mathcal{P}) \propto \mathcal{P}^{-0.55}, \quad \text{with}\  \mathcal{P} = \log(P/\text{day}) \in [0.15, 5.5]
    \label{eq: Period distribution}
\end{equation}
and:
\begin{equation}
    \textrm{pdf}(e) \propto e^{-0.42}, \quad \text{with}\ e\in [0, 0.9].
    \label{eq: Period distribution2}
\end{equation}

We generated an initial population of $10^7$ binaries at the ZAMS and used them as initial conditions for all of our simulations. Each population makes use of the same treatments for binary evolution processes and final fates described above.
Such populations are then evolved with the three different sets of stellar tracks computed with the various mass loss recipes presented in Sec. \ref{sect:winds}, to study the effect of different stellar winds prescriptions on the evolution of stellar binaries.

\section{Results}\label{sect:results}

\subsection{Evolution on the HRD}\label{sect:HRD}

\begin{figure}
    \centering
    \includegraphics[width=0.9\linewidth]{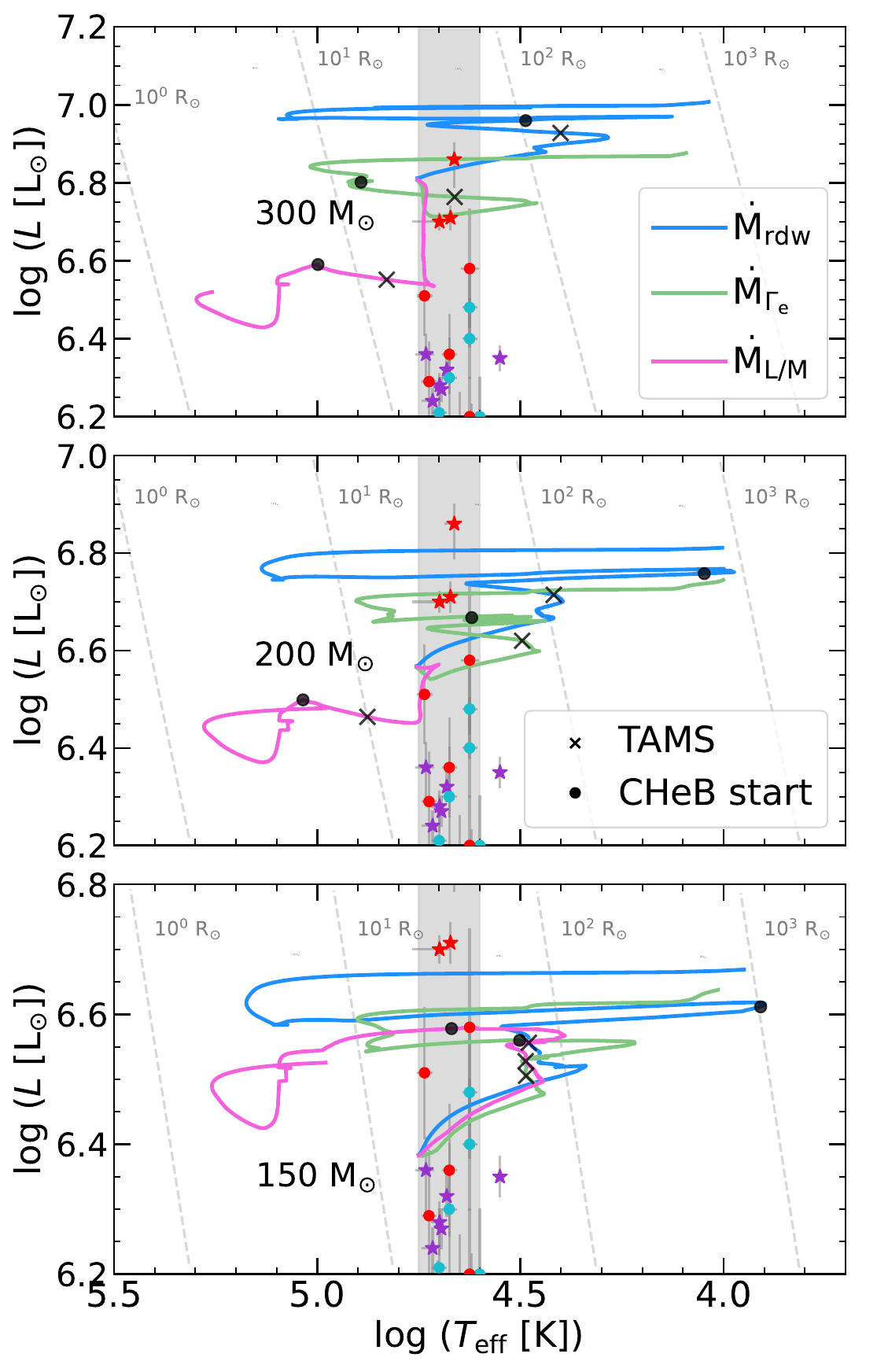}
    \caption{HRD showing the evolution of stellar tracks computed with the \mrdw, \mgam, and  \mlm\ winds, in blue, green, and pink, respectively. Models are shown with ZAMS masses of 300 \msun\ (top), 200 \msun\ (middle), and 150 \msun\ (bottom). The data symbols and shaded gray area are the same as in Fig. \ref{fig:HRD_LMC}. Black crosses and dots indicate the model's location at the TAMS and the start of CHeB.
    Diagonal dashed gray lines show constant stellar radii as labeled. 
    }
    \label{fig:HRD_Z0.006}
\end{figure}

Fig. \ref{fig:HRD_Z0.006} shows the HRD of selected tracks computed with the three different mass loss recipes at LMC metallicity. 
The evolution of the star with \mzams\ = 150 \msun\ shows a fairly similar evolution on the MS despite different mass loss prescriptions. We find a similar behavior for stars below 150 \msun. They all increase in luminosity and move to lower \teff, with the \mrdw\ track evolving at slightly higher luminosity. The similar evolution arises because these models do not enter the high-\gammae\ regime.
Therefore, they have mass loss rates similar to the standard rates of \mrdw\ (before reaching a bistability jump). The greatest difference occurs after the MS. The \mrdw\ ignites core-helium burning (CHeB) at cooler \teff\ than the other models, as a yellow supergiant (YSG). Meanwhile, the \mgam\ and \mlm\ tracks ignite CHeB as WN stars, near log \teff\ 4.5 - 4.6. 

At larger \mzams, more significant differences in the evolutionary paths begin to emerge. As can be seen in the comparison of 200  and 300 \msun\ stellar tracks, the \mgam\ and \mlm\ tracks encompass a narrower \teff\ range on the MS when compared to the \mrdw\ tracks. At the start of the MS, the \mgam\ models show a slight luminosity decrease and then evolve to cooler \teff. Then, they become WNhs shortly before reaching the bistability jump at log \teff\ $\simeq$ 4.4,  where the mass loss rate increases and the track moves to higher \teff. The jump is also experienced by the \mrdw\ models here. 
The evolution of the 200 and 300 \msun\ tracks is markedly different, computed with the \mlm\ rates. These tracks show a sharp decrease in luminosity early on the MS and continue to evolve nearly vertically until just before the end of the MS. These tracks become WNh stars earlier than the other two tracks, and spend up to 30 \% of the MS as WR stars.

We find that stellar tracks computed with the \mgam\ and \mlm\ winds spend more time in the observed \teff\ region and expand less compared to the \mrdw\ tracks. The tracks with enhanced winds can even evolve vertically down on the MS, depending on the ZAMS mass. Concerning the lifetimes, table \ref{table:lifetime} lists the cumulative time spent within the \teff\ range of observed VMS transitioning to the WNh sequence. The 200 and 300 \msun\ models with \mgam\ and \mlm\ stay within the observed \teff\ range for $\sim$ 1.5 - 1.7 Myr, while the \mrdw\ models spend less than 1 Myr within the observed \teff\ range. Models with \mrdw\ spend $<$ 1 Myr in the observed limits, \mgam\ models spend $\sim$ 1.6 Myr within the range, and \mlm\ models spend from 1.2 - 2.1 Myr within the limits.

\begin{table}[t]
    \centering
    \caption{Time spent in the observed temperature range $\log \teff$ between 4.75 and 4.6.}
    \renewcommand{\arraystretch}{1.2}
    \begin{tabular}{cccc}
        \hline
         \mzams & \mrdw & \mgam  & \mlm  \\
         
         [\msun] & [Myr] & [Myr]  & [Myr] \\
         \hline
         100 & 1.53 & 1.54 & 1.51 \\
         150 & 1.05 & 1.59 & 1.34 \\
         200 & 0.95  & 1.64 & 1.53 \\
         250 & 0.92 & 1.57 & 1.21 \\
         300 & 0.91 & 1.64 & 1.76 \\
         350 & 0.88 & 1.64 & 2.15 \\
         400 & 0.81 & 1.57 & 2.08 \\
         450 & 0.73 & 1.49 & 1.96 \\
         500 & 0.63 & 1.42 & 1.86 \\
         \hline

    \end{tabular}

    \label{table:lifetime}
\end{table}

Given that the R136 cluster is estimated to be between 1 - 2.5 Myr old \citep{Crowther2016, Bestenlehner2020, Brands2022} and the individual stars in R136a are estimated to have ages of approximately $\sim$ 1 - 1.7 Myr \citep{Crowther2010, Bestenlehner2020, Brands2022}, tracks with \mrdw\ would be too cool to match the data unless the stars were very young (< 1 Myr). In comparison, the \mgam\ and \mlm\ models spend most of their lives within the observed ranges. Thus, the \mgam\ and \mlm\ models can better reproduce the observed VMS properties in the R136 star cluster than the \mrdw\ models.

To further investigate this aspect, we perform a fitting procedure exploiting the \textsc{param} statistical tool (described in Sec \ref{subsec:param}) to fit the three most massive stars of the cluster (namely, R136a1, R136a2, and R136a3) with our stellar tracks with standard and new winds prescriptions. As input values, we adopt the current luminosity, effective temperature, and metallicity of stars provided by \citet{Brands2022}. As output of the fitting procedure, we obtain the stellar parameters listed in Table~\ref{table:r136}, i.e., the initial mass, the current mass, and the age of the three stars for each wind recipe.
The analysis shows that stellar evolutionary tracks calculated using the \mrdw\ and \mlm\ wind prescriptions predict ages for the three stars that are consistently too young. In contrast, the tracks computed with \mgam\ yield ages that align very well with the cluster age reported by other studies \citep{Crowther2010, Bestenlehner2020, Brands2022}. This analysis emphasizes that \mgam\ is the most effective wind prescription for modeling the VMS WNh stars in the R136 cluster.

\begin{table}[t]
    \begin{center}
    \caption{Best fit of the stellar tracks within the 68\% confidence interval for the massive stars R136a1, R136a2, and R136a3.}
    \renewcommand{\arraystretch}{1.2}
    \begin{tabular}{ m{3em}  m{1.5cm} m{1.5cm} m{1.5cm} }
    %\begin{tabular}{cccc}
        \hline
         $\dot{M}$ & \mzams  & $M_{\rm current}$ & Age \\
         
         & [\msun] & [\msun] & [Myr] \\
         \hline
         \multicolumn{4}{c}{R136a1} \\
         \hline
         \mrdw & 315 $^{+11}_{-13}$ & 295 $^{+14}_{-16}$ & 0.61 $^{+0.14}_{-0.12}$ \\
         \mgam\ &  389 $^{+ 1 }_{- 64 }$ &  202 $^{+ 12 }_{- 66 }$ &  1.52 $^{+ 0.65  }_{- 0.17 }$ \\

         \mlm\ &  359 $^{+ 30 }_{- 10 }$ &  254 $^{+ 22 }_{- 44 }$ &  0.52 $^{+ 0.44  }_{- 0.28 }$ \\
         
         \hline
         \multicolumn{4}{c}{R136a2} \\
         \hline
         
         \mrdw\ &  240 $ ^{ +6}_{ -6 }$ &  255 $ ^{+ 9 }_{ -8 }$ & 0.81 $ ^{+0.12  }_{-0.15 }$ \\
         \mgam\ &   296 $^{+ 12 }_{- 20 }$ &  169 $^{+ 13 }_{- 13 }$ &  1.58 $^{+ 0.13  }_{- 0.16 }$ \\
         \mlm\ & 260 $^{+130}_{- 1 }$& 201 $^{+32}_{- 80}$ & 0.92 $^{+0.47}_{- 0.99}$\\
        
         \hline
         \multicolumn{4}{c}{R136a3} \\
         \hline
         \mrdw &   241 $^{+ 5}_{- 5 }$ &  232 $^{+ 9 }_{- 9 }$ &  0.53 $^{+ 0.21  }_{- 0.27 }$ \\
         \mgam\ &  286 $^{+ 4 }_{- 25 }$ &  180 $^{+ 42 }_{- 14 }$ &  1.34 $^{+ 0.23  }_{- 0.77 }$ \\
         \mlm\ & 257 $^{+39}_{-4}$ & 220 $^{+17}_{-41}$ & 0.64 $^{+0.68}_{-0.33}$ \\
         
         \hline
    \end{tabular}
            
    \end{center}
    \footnotesize{Column 1: Mass loss recipe. Column 2: ZAMS mass of the model. Column 3: Current mass of the model. Column 4: Current age of the model.}
    \label{table:r136}
\end{table}

Our results establish a strict upper limit on the possible initial mass for R136a1 since no models with an initial mass below 300 \msun\ fit the data within the uncertainty. 
These results are in line with other works that estimate the mass of R136a1: \citet{Crowther2010} found the maximum initial mass to be 320 $^{+100}_{-40}$ \msun, while \citet{Brands2022} found M$_{\rm ini}$ = 273 $^{+25}_{-36}$ \msun. For R136a2, all of our predictions for the initial mass are in agreement with \citet{Crowther2010}, who suggested 240 $^{+45}_{-45}$ \msun, while our initial mass suggestions are slightly larger than \citet{Bestenlehner2020} and \citet{Brands2022}, who provided an initial mass of 211 $^{+31}_{-32}$ \msun\ and 221 $^{+16}_{-12}$ \msun, respectively. Finally, all of our derived initial masses for R136a3 are larger than suggested by other authors, who give 165 $^{+30}_{-30}$ \msun\ from \citet{Crowther2010} and  181 $^{+29}_{-31}$ \msun\ from \citet{Bestenlehner2020}, and 213 $^{+12}_{-11}$ \msun\ from \citet{Brands2022}. The derived initial and current masses for R136a2 and R136a3 are listed in Table \ref{table:R136_masses_all} in App. \ref{app:r136a2}. 
%This slight discrepancy arises from different treatments for the stellar winds.

Our derived stellar ages for R136a show the best agreement with other estimates from literature for the tracks computed with the \mgam\ wind prescription. Specifically, these derived ages are slightly older than those provided by \citet{Brands2022}, but remain within their reported uncertainty. Similarly, our stellar ages are slightly younger than the age estimates from \citet{Crowther2010}, which also fall within their uncertainty. In contrast, using the \mrdw\ and \mlm\ prescriptions results in much younger derived ages, less than 1 Myr, which is inconsistent with the ages reported in other studies.

\begin{figure}
    \centering
    \includegraphics[width=\linewidth]{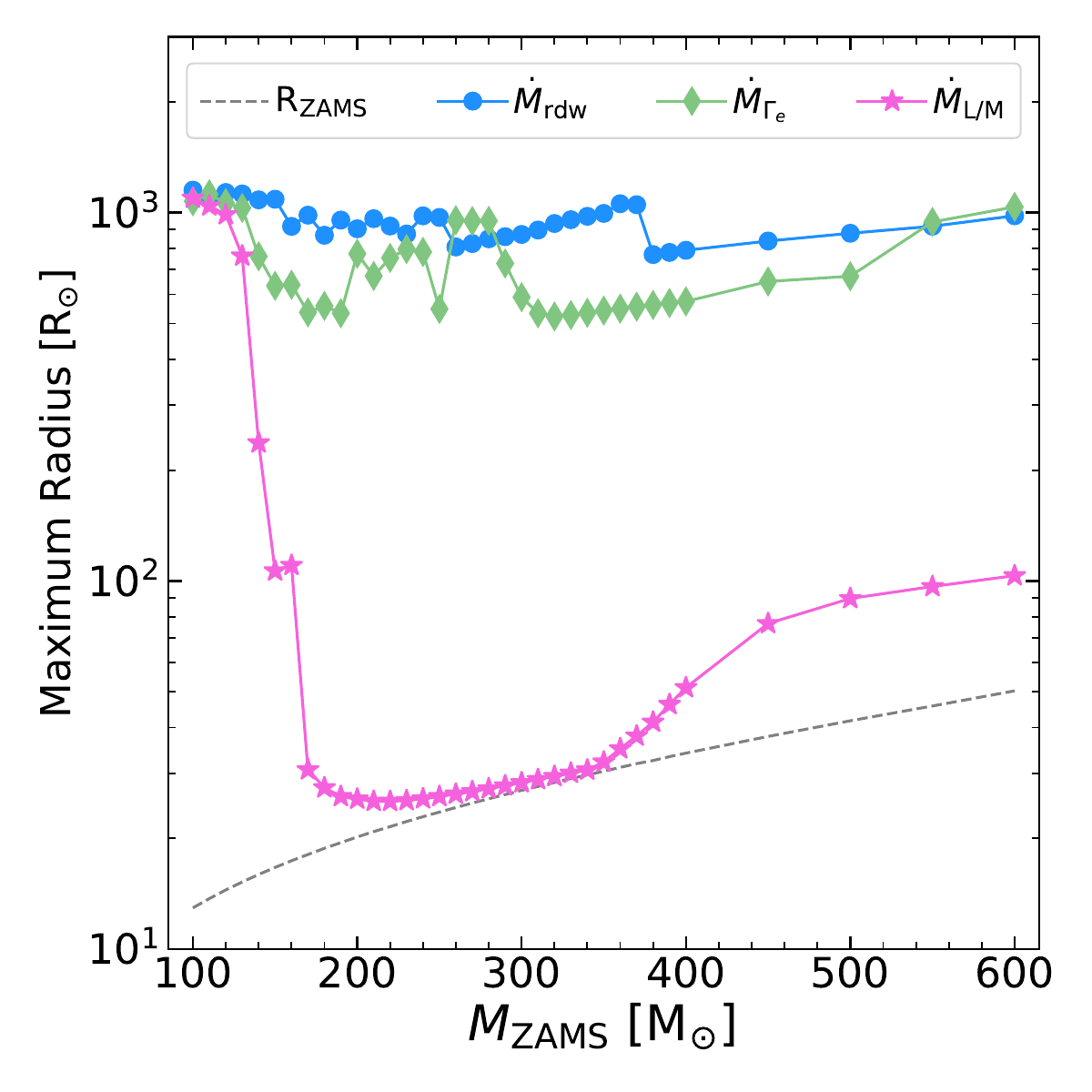}
    \caption{Maximum radius reached for each ZAMS mass for tracks computed with different mass loss rates, indicated by the color and marker. The blue circles, green diamonds, and pink stars show the maximum radius for tracks computed with \mrdw, \mgam, and \mlm\ winds, respectively. The dashed gray line shows the ZAMS radius. }
    \label{fig:Radii_0.006}
\end{figure}

Figure \ref{fig:Radii_0.006} shows the maximum radii reached for the different mass loss rates as a function of the initial mass. The stellar tracks with \mlm\ have the smallest radii, as they lose their outer envelope and large fractions of mass early in the evolution. Consequently, they typically do not expand during shell-burning phases, like other tracks do. Indeed, these tracks with \mlm\ and \mzams $\sim$ 160 - 350 \msun\ have maximum radii reached during the MS and are near their ZAMS radius, about tens to 100 \rsun. Meanwhile, the stellar tracks with \mrdw\ and \mgam\ reach their maximum radii in the post-MS phases. Tracks with \mrdw\ and ZAMS masses below $\sim$ 200 \msun\ generally reach their maximum radii of $\sim$ 500 - 1\,000 \rsun\ at the start of the CHeB phase. Similarly, tracks with \mgam\ and \mzams\ $\leq$ 130 \msun\ reach their maximum radii of $\sim$ 1\,000 of \rsun\ at the start of CHeB. For larger masses of both sets, the maximum radius is reached only during the final phase of evolution, when the star quickly crosses the HRD in a timescale as short as just a few years. The \mrdw\ tracks reach $\sim$ 700 - 1\,000 \rsun\ in this stage, while the \mgam\ tracks similarly reach $\sim$ 500 - 1\,000 \rsun. The time spent at larger radii directly influences the likelihood and type of binary interactions, as discussed in Sect. \ref{subsec:wind_accretion}.

\subsection{Structure and chemical evolution}\label{sect:chemi}

\begin{figure*}
    \centering
    \includegraphics[width=\linewidth]{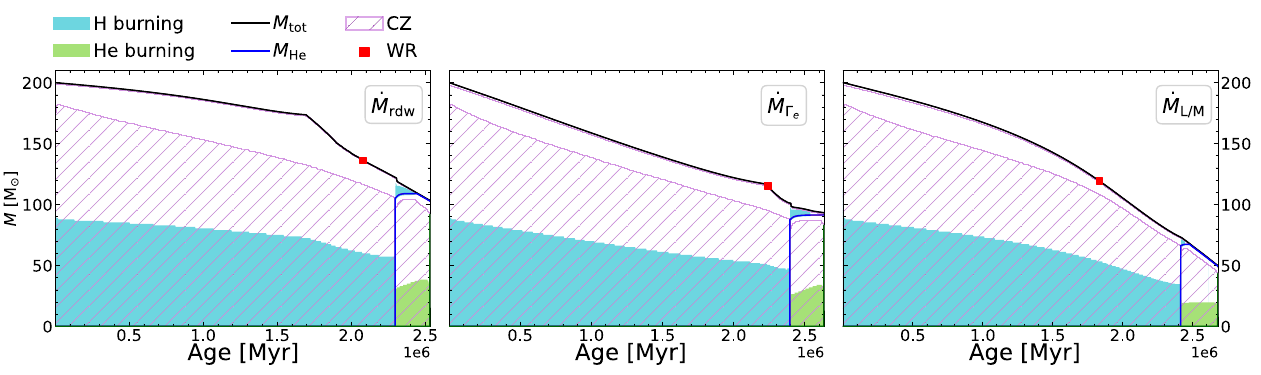}
    \caption{Kippenhahn diagrams of the stellar tracks computed with the three mass loss rates with \Mz\ = 200 \msun. The solid blue and green regions show the H and He burning zones, while the purple hatched region shows convective zones. Solid black and navy lines indicate the total mass and mass of the He core. The red square indicates when the star enters the WR phase (i.e., when \xs\ $<$ 0.3).}
    \label{fig:kipp_M200_Z0.006}
\end{figure*}

Figure \ref{fig:kipp_M200_Z0.006} shows the Kippenhan diagrams for three different mass loss rates of a \mzams\ = 200 \msun\ stellar track. The differences in the mass loss rates are most pronounced during the MS, where stars spend most of their life. As mentioned above, all of these tracks become WRs on the MS. The \mlm\ track is the first to become a WNh at $t$ = 1.83 Myr, as the stronger winds peel off the star's H envelope faster than the other two wind prescriptions. The enhanced winds of \mgam\ and \mlm\ stellar tracks result in a smaller H burning region by the end of the MS, producing smaller He cores compared to the case of standard winds.
By the end of the MS, the \mrdw\ star has lost 80 \msun\ of material, the \mgam\ one has lost 100 \msun, and the \mlm\ has lost 130 \msun\ of its material due to stellar winds.

In the post-MS phases, tracks with \mrdw\ tend to expand, becoming YSGs before igniting CHeB (as shown in Fig. \ref{fig:HRD_Z0.006}).
In contrast, the tracks computed with the new winds ignite CHeB at higher \teff, typically as WN stars. 
All 200 \msun\ models have a small envelope at the start of CHeB.% which is lost during this phase, leading them to expose their bare cores. 
The \mgam\ model is the only one that maintains a He-rich envelope throughout its evolution, while the other models lose it and expose their bare cores at the end of evolution. 
In general, \mgam\ and \mlm\ tracks stay hotter than the \mrdw\ one for the remainder of the evolution.

\begin{figure*}
    \centering
    \includegraphics[width=\linewidth]{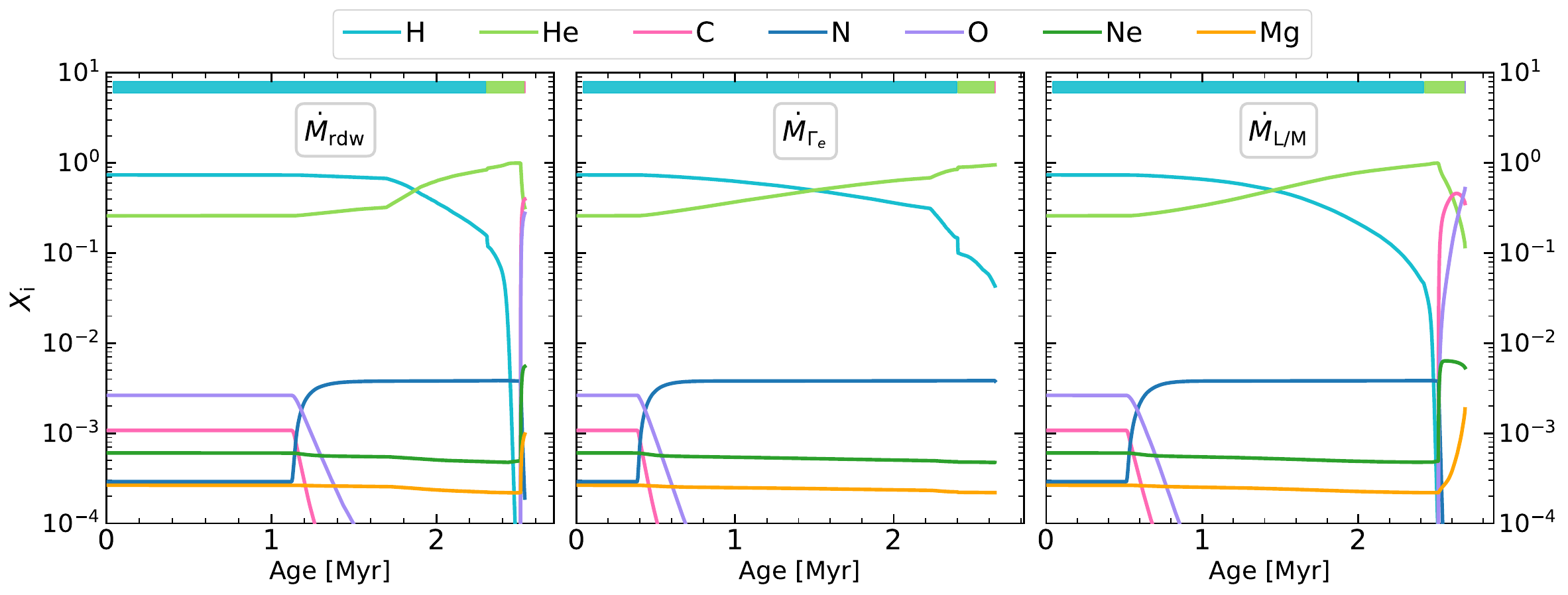}
    \caption{Surface abundances of a \Mz\ = 200 \msun\ star for three mass loss rates. The blue, green, and pink shaded bars indicate the time of H, He, and C burning, respectively.}
    \label{fig:abund_M200_Z0.006}
\end{figure*}

The stellar wind prescriptions used not only affect the structural evolution of stars described above but also influence the abundances and nucleosynthetic products. 
Figure \ref{fig:abund_M200_Z0.006} shows the evolution of the surface abundances for the different mass loss rates of the same tracks with \Mz\ = 200 \msun\ presented in Figure~\ref{fig:kipp_M200_Z0.006}. 
The \mrdw\ model becomes a WNh star at log \teff\ $\sim$ 4.4 and spends the last 10\% of its MS as a WNh.
This star retains more of its envelope during the MS and early CHeB phases. The He-shell burning produces C and O, which are exposed to the surface before the start of CCB. This star ends evolution as a PISN during CCB, with surface abundances composed of $X\mathrm{_{He}^{S}}=$ 0.31,  $X\mathrm{_{C}^{S}}=$ 0.39, and  $X\mathrm{_{O}^{S}}=$ 0.27, making it a WC star.

The retained envelope of the \mgam\ model sustains the He-dominated surface and limits the exposure of the inner He shell-burning products. It becomes a WNh star around log \teff\ $\simeq$ 4.5 and retains a small fraction of H on the surface during the rest of its evolution.
%This star ends evolution after igniting Neon in its core, and shortly after becoming unstable to PI. 
This star ends it's evolution with a He dominated surface, $X\mathrm{_{He}^{S}}=$ 0.95, some remnant H,  $X\mathrm{_{H}^{S}}~=$ 0.04, and trace amounts of C and O, $X\mathrm{_{C}^{S}} \sim~10^{-5}$ and $X\mathrm{_{O}^{S}}~\sim~ 10^{-5}$.

The \mlm\ track is the first to become a WNh at log \teff\ $\simeq$ 4.7 and spends the last 15\% of its MS as a WNh star. Later, during the CHeB phase, the star becomes a bare core, where He burning products are directly exposed on the stellar surface. This star loses the most mass with respect to the other two models. This significant reduction also affects the core mass, enabling the star to continue igniting subsequent burning phases until core oxygen burning, before becoming dynamically unstable to PI. The \mlm\ model ends as a WO star, with $X\mathrm{_{C}^{S}}=$ 0.35, and $X\mathrm{_{O}^{S}}=$ 0.51, and low He abundance, $X\mathrm{_{He}^{S}}=$ 0.11.

It is worth noting that in this analysis, we do not include rotation in our tracks, as it was not the paper's focus. Other authors, as \citet{Sabhahit2022}, find that moderately rotating models ($\Omega/\Omega_{\rm crit} = 0.2$) with \Mz\ $\gtrsim$ 200 \msun\ become chemically homogeneous during the MS, however, the evolution still remains dominated by the mass loss. To explore this aspect, in Appendix \ref{app:rotating}, we present and describe the evolution of selected VMS stellar tracks with new stellar winds and rotation. In agreement with \citet{Sabhahit2022}, we find that for all \Mz\ $\gtrsim$ 200 \msun, the evolution is completely dominated by the mass loss rather than rotation, even for very rapid rotational rates ($\Omega/\Omega_{\rm crit} = 0.8$).

\subsection{Pre-SN masses single BH masses }\label{sect:fates}

\begin{table}[t]
\renewcommand{\arraystretch}{1.2}
\centering
     \caption{Pre-SN masses in \msun\ for three different wind recipes.}
    \begin{tabularx}{\columnwidth}{@{\extracolsep{\fill}}cccc}
    \toprule 
    
    \multirow{3}{*}{M$_{\mathrm{ZAMS}}$ [\msun]} & \multicolumn{3}{c}{$M_{\mathrm{Pre-SN}}$ [\msun]} \\ \cmidrule(lr){2-4}
        
          & \mrdw & \mgam & \mlm \\
        
        \toprule

        100 & 48.75 &  52.67 & 42.91 \\
        150 & 75.10 & 74.99 & 54.46 \\
        200 & 103.13 & 93.21 & 49.33\\
        250 & 132.44 & 107.84 & 51.27 \\
        300 & 160.22 & 122.29 & 54.96 \\
        350 & 186.10 & 131.87 & 58.98 \\
        400 & 212.21 & 143.23 & 63.53\\
        500 & 264.74  & 164.22 & 71.95\\
        600 & 291.65 & 174.40 & 76.19 \\
         \hline

    \end{tabularx}

\label{table:masses}
\end{table}

\begin{figure}
    \centering
    \includegraphics[width=\linewidth]{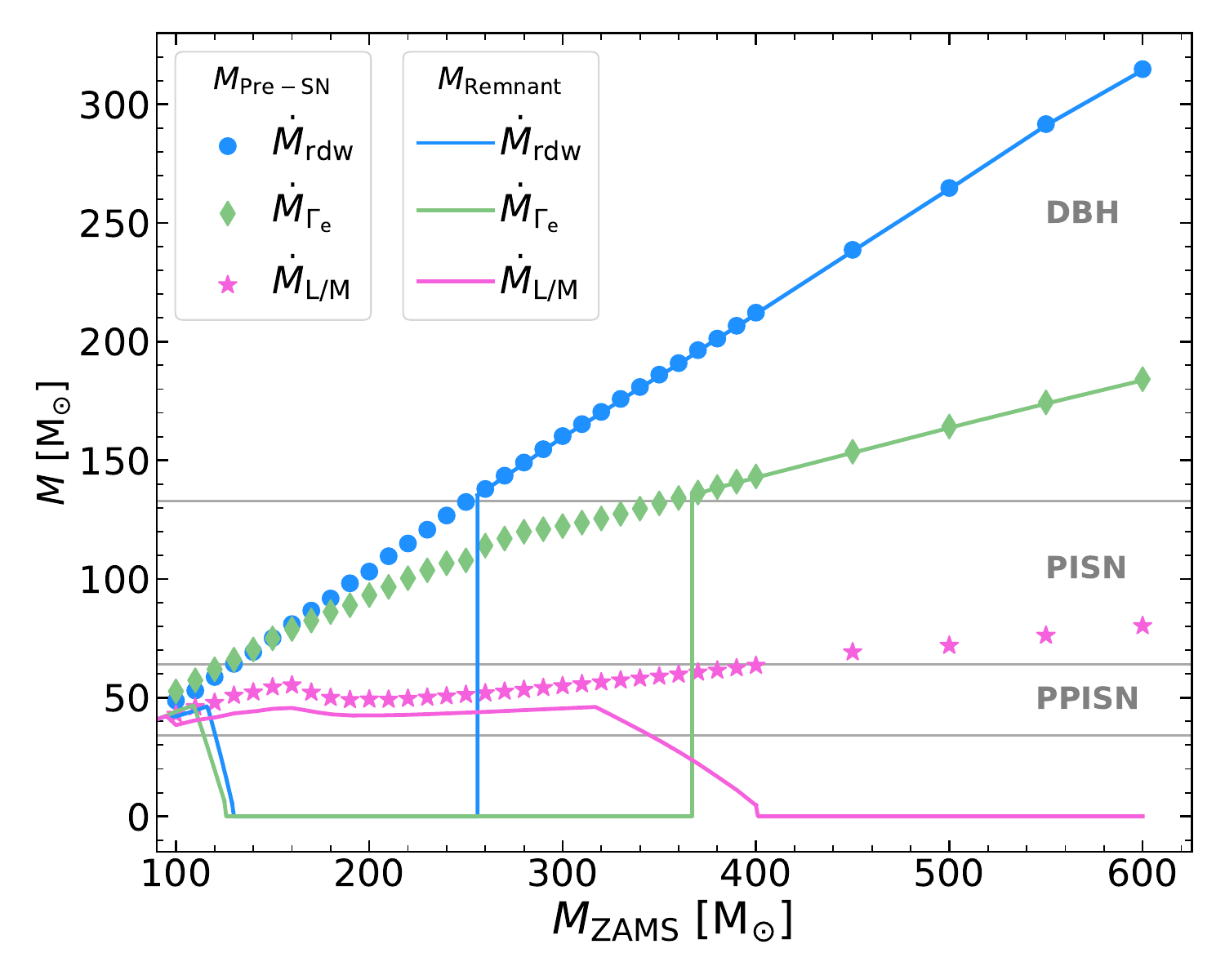}
    \caption{Pre-SN mass (markers) and remnant mass (lines) as a function of the ZAMS mass for the three mass loss rates, indicated by the marker style and color. 
    The gray horizontal lines indicate different fates (DBH, PISN, and PPISN) based on the final He core mass. }
    \label{fig:Mini_Mfin_0.006}
\end{figure}

Figure \ref{fig:Mini_Mfin_0.006} shows the pre-SN and remnant masses for all stellar tracks we computed. Pre-SN masses are shown for selected ZAMS masses in Table \ref{table:masses}.
All of these stars' pre-SN mass is nearly that of the He core. As mentioned previously, 'lower mass' models (M $\lesssim$ 100 \msun) evolve similarly for the different mass loss rates and result in comparable pre-SN masses. In contrast, more massive stars show greater differences with increasing \mzams. Stellar tracks computed with \mrdw\ end their evolution with masses roughly half that of their ZAMS masses, regardless of the initial mass. On the other hand, models computed with \mgam\ end with pre-SN masses $\sim$ 30 - 50 \% of their ZAMS mass, and \mlm\ models end with $\sim$ 13 - 40 \% of their initial masses. In both cases, the percentage decreases with increasing ZAMS mass. The largest pre-SN masses for the three wind recipes are 314 \msun, 184 \msun, and 80 \msun\ for the \mrdw, \mgam, and \mlm\ winds, respectively, all originating from the \mzams\ = 600 \msun\ models.

We use \sevn\ version 2.13 to compute the remnant masses based on the pre-SN properties of the stellar tracks.%, and find that the ZAMS mass ranges determining the fate are shifted for each wind recipe. %Models with ZAMS masses below $\sim$ 100 \msun\ show the same remnant mass distribution, and are not shown.
The models computed with the \mrdw\ rate produce a gap in the BH mass spectrum of 46 \msun\ - 135 \msun, (originating from stars having initial masses $\sim$ 130 $\leq$ \mzams/\msun\ $\leq$ 256) in general agreement with other works at LMC metallicity \citep{speraVeryMassiveStars2017, Woosley2017, Farag2022}. The BH mass gap is the same for models with \mgam\ (as we use the same prescriptions for the fate, see Sect. \ref{sect:sevn}), but the ZAMS masses defining the gap widens, to 125 $\leq$ \mzams/\msun\ $\leq$ 367. The \mgam\ BH masses predicted above the gap are between $\sim$ 0.3 - 0.4 times lighter than the BH masses of the \mrdw\ rate. The \mlm\ models do not have any upper limit to the mass gap in this mass range, as all models above 400 \msun\ die as a PISN and leave no remnant. The \mlm\ models are the only ones to produce BHs from stars with ZAMS masses of $\sim$ 140 - 250 \msun. The maximum BH mass achieved with the \mlm\ wind prescription is 46 \msun\ at LMC metallicity, which is significantly lower than predictions from other studies at the same metallicity \citep{Spera2017, Giacobbo2018, Farmer2019}. 
%The most massive BH from the \mlm\ tracks is much smaller than the other tracks, at just 46 \msun.

%\section{Results from Binary Evolution} \label{sect:binary}

\subsection{Wind accretion}
\label{subsec:wind_accretion}

To understand how binary evolution, particularly with enhanced winds, contributes to forming VMS like those observed in R136, we first examine the process of wind accretion. This mechanism is crucial, as the stellar winds' feedback onto the orbit will both produce more massive secondary stars and more loose binaries, as discussed in \citep{iorioCompactObjectMergers2023}, thus resulting in differences in the observed properties of the stars. 
Enhanced stellar winds can influence the evolution and interactions of binary systems by modifying the evolution of each star, affecting key parameters such as the maximum stellar radius ($R_{\mathrm{max}}$), the envelope mass ($M_{\mathrm{envelope}}$), and the total mass ($M_{\mathrm{total}}$).
Moreover, stellar winds can determine whether or not a star expands in the RSG stage, thereby setting the conditions for binary interaction through mass transfer episodes. Additionally, the presence or absence of a stellar envelope will impact the likelihood of CE occurrences. Winds influence all of these processes and ultimately leave their footprint on the final mass of the stars, as discussed in Sect. \ref{sect:fates}.

\begin{figure}[t]
    \centering
    \includegraphics[width=\linewidth]{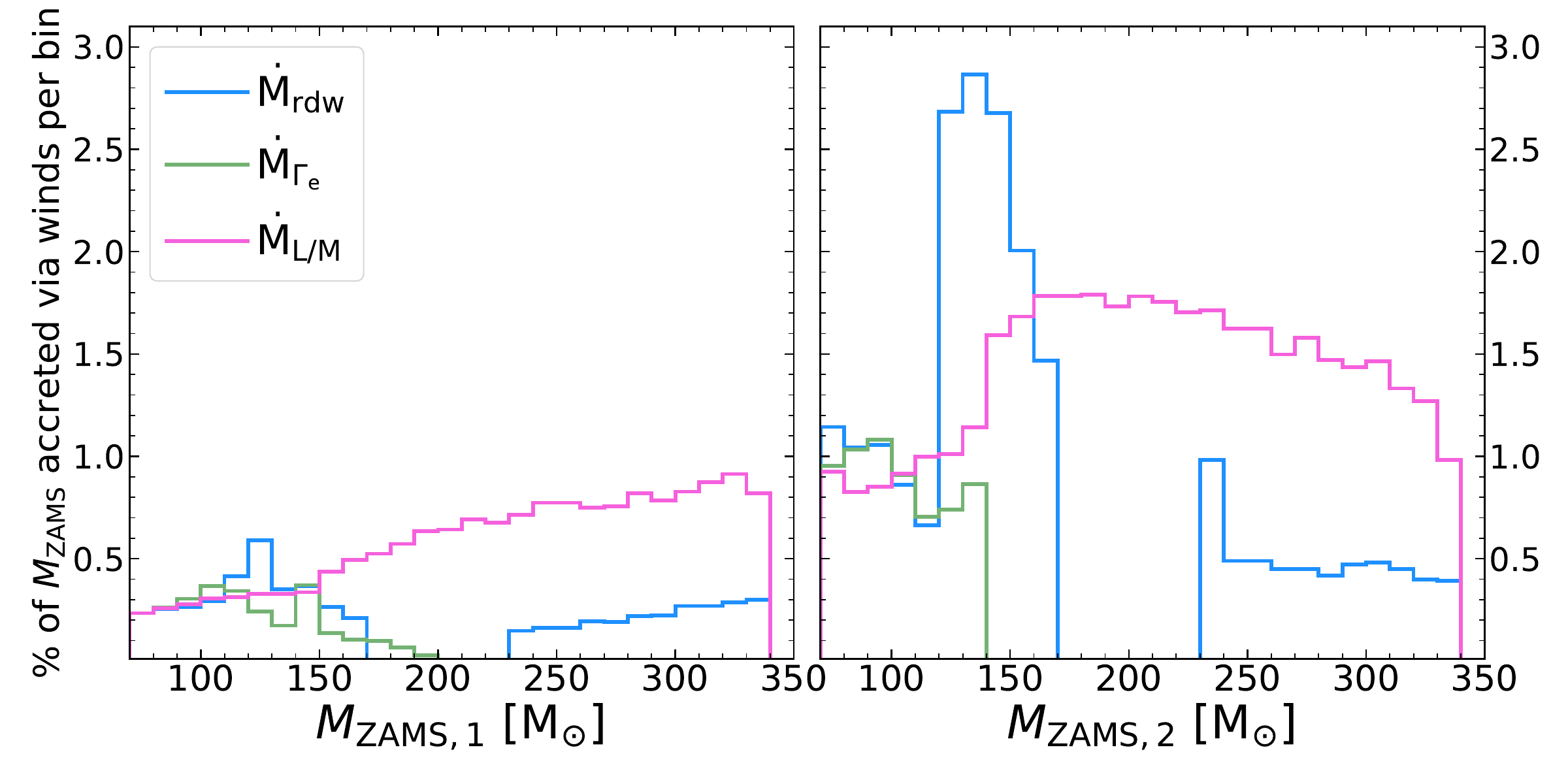}
    \caption{The averaged percentage of ZAMS mass accreted onto the primary (left) and secondary (right) per mass bin. The bins are 10 \msun\ each. The colors are the same as Fig. \ref{fig:Mini_Mfin_0.006}. }
    \label{fig:Mass_accr}
\end{figure}

Stellar models with enhanced winds experience different evolutionary processes in a binary system, for example, by widening the orbital separation and by transferring large amounts of mass to the companion via stellar winds. Figure \ref{fig:Mass_accr} shows the percentage of ZAMS mass accreted onto the primary and secondary star during the evolution averaged over a mass bin of 10 \msun, for all binaries that contain at least one BH component at the end of the simulation and do not merge during stellar evolution. The total accreted material goes to zero if a star leaves a massless remnant (i.e., a PISN or the ejection of the companion star), which causes the percentages drop near ZAMS masses $\sim$ 170 - 230 \msun\ for the \mrdw\ and \mgam\ cases.
The amount of mass accreted by the star, in the case of enhanced winds, can be up to $\sim$ 3\% of its ZAMS mass; in one case, a star of the \mlm\ simulation accretes 39 \msun\ of material from its companion from just stellar winds alone. In comparison, the greatest amount of accreted material for the \mgam\ and \mrdw\ simulations is 24 \msun\ and 16 \msun,  respectively. 

The secondary star accretes more mass in general, as the primary is more massive and, therefore, loses more material, which is then available to be accreted. The simulation using \mgam\ has smaller averaged accretion around $M_{\rm ZAMS, ~2} \simeq$ 120 to 160 \msun\ compared to the \mrdw\ one. This is because these binaries have primaries with ZAMS masses $M_{\rm ZAMS, ~1} \simeq$ 120 to 345 \msun, where most stars end evolution as PISN, thus leaving a massless remnant (see Fig. \ref{fig:Mini_Mfin_0.006}) where the total accretion goes to zero. The \mrdw\ simulation with secondary masses of $M_{\rm ZAMS, ~2} \simeq$ 120 to 160 \msun\ also have primaries with $M_{\rm ZAMS, ~1} \simeq$ 120 to 348 \msun, where primaries with masses $M_{\rm ZAMS, ~1} \gtrsim$ 250 \msun\ do not enter the PISN regime, and thus their accretion onto the secondary is accounted for. 
For the \mlm\ simulation, the absence of PISN below 350 \msun\, and the infrequency of stellar mergers allow it to have consistently high accretion rates for ZAMS masses above 150 \msun.

\subsection{Stellar mergers}\label{subsec:stellar mergers}
Stellar mergers represent a critical outcome of binary evolution, and have been proposed as the origin of VMS \citep{PortegiesZwart2002}. When two stars in a binary approach one another too closely, gravitational interactions can lead to a collision or a stellar merger. These mergers can result in a single, more massive star, exhibiting properties similar to an isolated star born with a greater ZAMS mass. The stellar radii, separation, and accretion are the main parameters determining whether the binary will merge to form a more massive star.
To investigate the viability of stellar mergers being the origin of the observed VMS in R136, this section presents the characteristics of stellar mergers predicted by our simulations using different wind prescription, specifically focusing on whether the merged stars match the observed properties of R136a.

We search our simulations for systems with \mprim\ $ \geq$ 50~\msun\ that undergo a stellar merger and continue their evolution as a more massive star before forming a BH. We select merged objects whose luminosities and effective temperatures lie within the observational uncertainties of R136a. 

\begin{figure}[t]
    \centering
    \includegraphics[width=\linewidth]{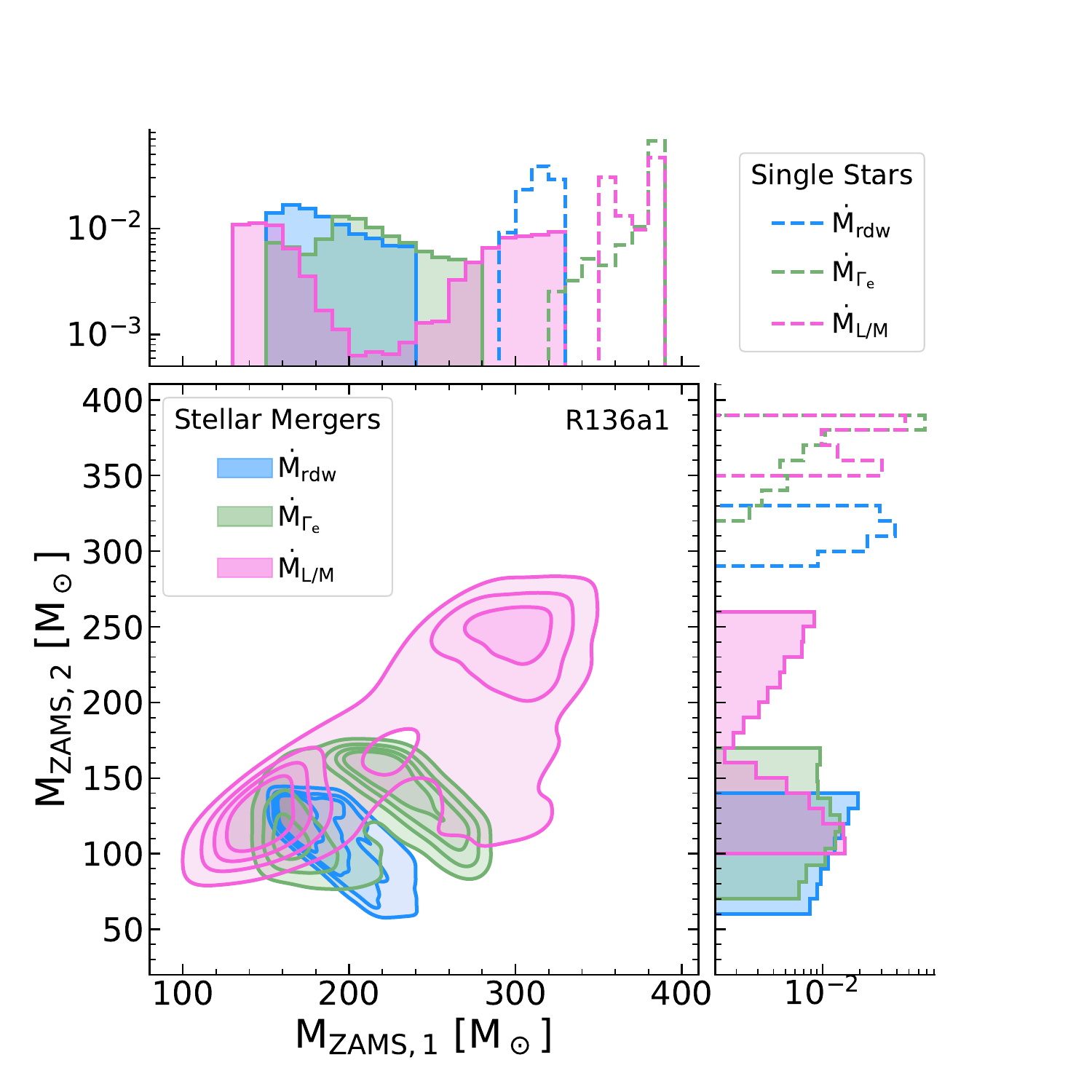}
    \caption{The possible initial masses of R136a1 from single star and stellar merger origins.
    The shaded areas show the primary and secondary ZAMS masses of stars that merge during evolution and overlap in the HRD with the observed location of R136a1. The blue, green, and pink show results from \mrdw\, \mgam, and \mlm, respectively. The dashed histograms show the initial mass from single stellar evolution. }
    \label{fig:R136a1_masses}
\end{figure}

Figure \ref{fig:R136a1_masses} shows the ZAMS mass distributions capable of reproducing the observed position of R136a1 on the HRD, from single stellar evolution and stellar mergers. Our binary simulations indicate that systems merging to form more massive stars matching R136a's HRD position originate from primary stars with initial masses that vary depending on the wind prescription used. For the \mrdw\ simulation, the most probable initial primary mass is 186 $^{+ 53 }_{ - 30 }$ \msun\ in the 68\% confidence interval (1$\sigma$). The \mgam\ simulations suggest a larger most probable initial primary mass of 207 $^{ + 69 }_{- 56 }$ \msun, and the \mlm\ simulations predict an even larger most probable initial mass and larger 1$\sigma$ interval, of 239 $^{+ 87 }_{- 108 }$ \msun.
In contrast, the single star models reproducing the observed HRD position of R136a1 suggest initial single star masses over 100 \msun\ larger than from binary evolution: the \mrdw\ and \mgam\ tracks give $M_{\rm ZAMS ~1}$ = 315 $^{ + 11 }_{- 13 }$ \msun\ and \Mz\ = 389 $^{ +1 }_{- 64 }$ \msun, respectively. The single star result from the \mlm\ tracks give 359 $^{+ 30 }_{- 10 }$ \msun\ as the most probable initial mass within 1$\sigma$. %The \mlm\ simulation shows a bimodal peak around $M_{\rm ZAMS ~1} \sim$ 150 \msun\ and $M_{\rm ZAMS ~1} \sim$ 300 \msun. This arises because primaries with initial masses around 150 \msun\ merge to form stars that subsequently evolve similarly to a single star born with a ZAMS mass of $\sim$ 300 \msun, which begins its MS very near to the R136a1 star's location on the HRD. Similarly, the peak around 300 \msun\ arises from mergers which resemble stars born with ZAMS masses of $\sim$ 600 \msun, which directly traverses the HRD location of R136a1.
%These results highlight a key implication for the IMF of dense star-forming regions like R136: if the observed VMS are formed predominantly through binary stellar mergers, a lower initial mass range, and possibly a less top-heavy IMF, would be sufficient to explain the VMS's existence compared to a single stellar evolution scenario. 
These results highlight a key implication for the initial mass required to form a very massive object like R136a1: if the observed VMS was formed from a binary stellar merger, less heavy initial masses, and possibly a lower upper limit on the IMF, are needed to explain the star's existence, compared to a single stellar evolution scenario. 

For stellar merger products of systems that match R136a1's HRD location, the post-merger masses within 1$\sigma$ are 278 $^{ +25 }_{- 18 }$ \msun\ for the \mrdw\ simulation,  200 $^{+ 10 }_{- 6 }$ \msun\ for the \mgam\ simulation, and 191 $^{ +28 }_{- 19 }$ \msun\ in the \mlm\ simulation. 
Single stellar evolution models predict comparable current mass for stars that occupy the same HRD location, listed in Table \ref{table:r136}.
This significant overlap in predicted current masses implies that the current mass of R136a1 is not a strong diagnostic for discriminating between an origin of a single star or a stellar merger product.
We provide and discuss the possible initial and current masses for R136a2 and R136a3 in App \ref{app:r136a2}.

Stellar mergers occur while the primary is on the MS 83\% of the time for the \mrdw\ simulations, and 82\% for both the \mgam\ and \mlm\ simulations. The \mrdw\ simulation is also more likely to merge during CHeB than the other simulations, as stars with \mrdw\ winds expand more during this phase. On the other hand, the simulations with enhanced winds have more stellar mergers during H and He-shell burning phases than the \mrdw\ simulation.

The most common stellar merger pathway for all three wind recipes is by beginning a stable RLOF that circularizes the orbit, and ends in the merger of the two stars. A RLOF occurs when the radius of a star becomes equivalent to or larger than the RL radius and transfers mass until it leads to a merger or a CE configuration. The RLOF pathway leading directly to a stellar merger occurs in 66\% of all stellar mergers, for all wind recipes, while the RLOF leading to a CE phase and then a merger occurs in 3.9\% of all stellar mergers.

A stellar merger via a direct collision between two stars is also possible. It occurs when the sum of the two radii is larger than the stellar separation at the periastron. For all three simulations, this happens in 22\% of all merging stellar systems.

\subsection{Remnant mass spectrum}\label{subsec:Remnants}

\begin{figure}[t]
    \centering
    \includegraphics[width=\linewidth]{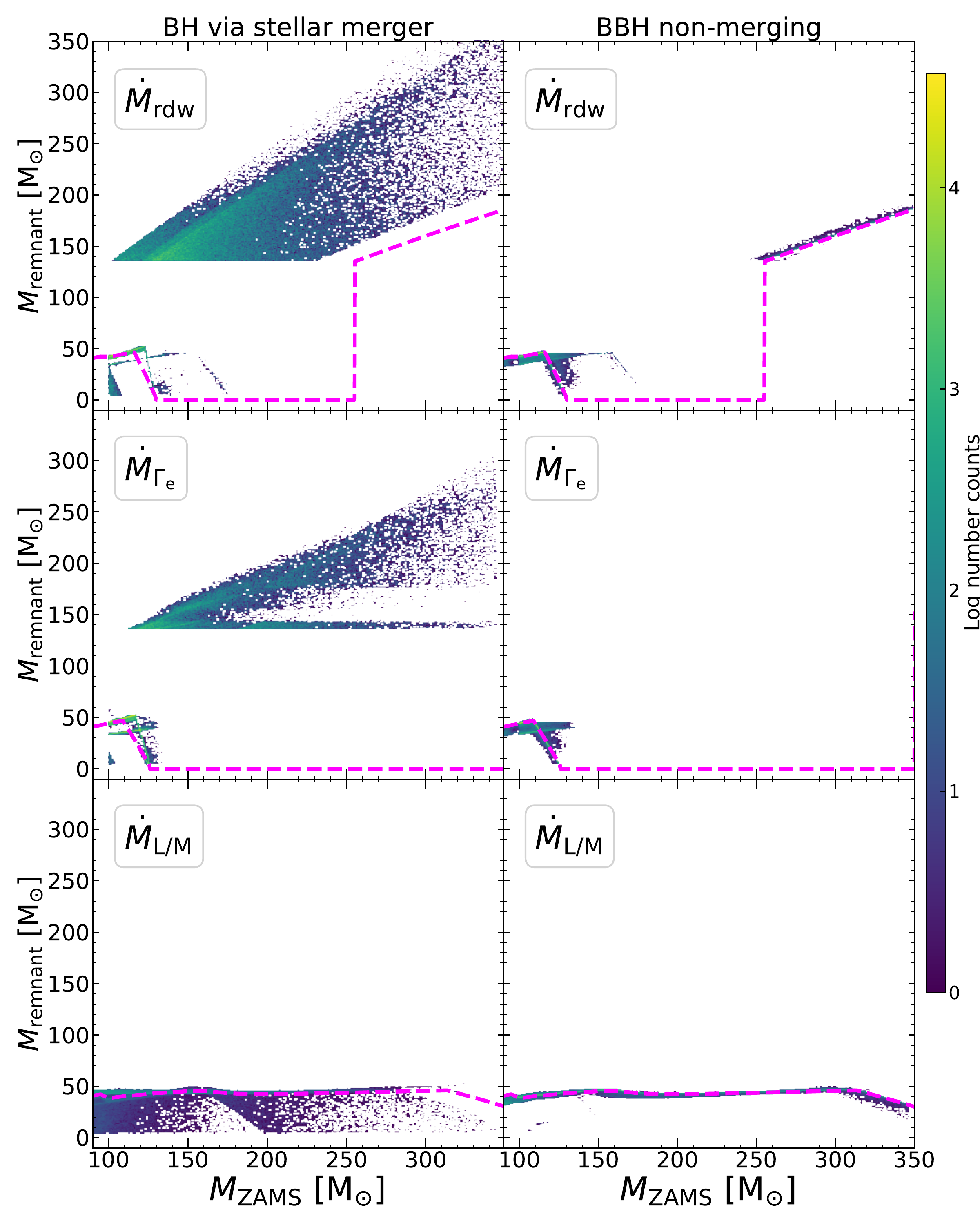}
    \caption{BH number counts for BHs formed from stellar mergers (left) and non-merging BBHs (right) for the simulations with \mrdw\ (top), \mgam\ (middle), and \mlm\ (bottom).
    The dashed pink lines indicate BH masses from single stellar evolution. 
}
    \label{fig:all_BHs_grid}
\end{figure}

Our investigation into the possible origin and evolution of the VMS observed in the Tarantula Nebula via single and binary evolution with new winds naturally leads to the final fate of such stars. The masses and formation channels of BHs are linked to the evolution of the progenitor star, therefore, the properties of predicted BH populations serve as an indirect test of the stellar and binary evolution models. This section presents the predicted BH populations and their evolutionary paths from our simulations, to highlight the consequences stellar winds have on binary evolution and the properties of BHs.

Two BH populations arising from different binary evolution pathways are shown in Figure \ref{fig:all_BHs_grid}. For stars with ZAMS masses below 100 \msun, we use the same stellar tracks using standard wind recipes for all three simulations, and therefore, the BH populations have similar distributions for lower ZAMS masses and are not shown in the figure for clarity.
From our simulations, we find that the distribution of single BHs formed via stellar mergers changes considerably depending on the stellar winds assumed. Simulations with enhanced winds form fewer massive BHs, particularly with remnant masses above 100 \msun. As mentioned in \ref{sect:HRD}, these stars expand less during the evolution due to the substantial mass loss, making them less likely to fill their Roche Lobe. The enhanced mass loss also widens the orbit, allowing the stars to evolve with fewer binary interactions or stellar mergers than the standard winds case. %As mentioned in \ref{sect:HRD}, these stars expand less during the evolution due to the substantial mass loss, weakening the gravitational pull and making it less likely to merge during the evolution, and also retain less mass at the end of evolution. 
The extreme winds of the \mlm\ simulation erode away the stars' mass so much that even the VMS that merge during evolution can not form a BH above $\sim$ 50 \msun. This is a stark contrast with the \mrdw\ and \mgam\ simulations, which generally form more massive BHs ($M_{\rm BH} > $ 100 \msun) after a stellar merger when compared to the resulting BH mass from single stellar evolution. 

The populations of binary BHs that do not merge within a Hubble time are shown in the right panel of Fig. \ref{fig:all_BHs_grid}. The \mrdw\ is the only one that can form non-merging BHs above the PI mass gap. These BHs arise from binaries that experience stable mass transfer before the first BH formation. %i.e. channel 1
There is a lack of non-merging binary BHs from the \mgam\ simulation with ZAMS masses $\gtrsim$ 140 \msun, due to the stars experiencing PISN. Finally, the \mlm\ simulation forms non-merging BHs with masses near the result for single stellar evolution, as they too arise from binaries experiencing episodes of stable mass transfer.
%\GC{Here, we should comment on the BBH nonmerging and BH from GW merger distributions. Interestingly, the BHs distribution partially fills the PI mass gap found by single stars! Moreover, there is a small peak in the BHm \mgam\ and \ml\ distributions that does not appear in \mrdw.}

\begin{figure}[ht]
    \centering
    \includegraphics[width=\linewidth]{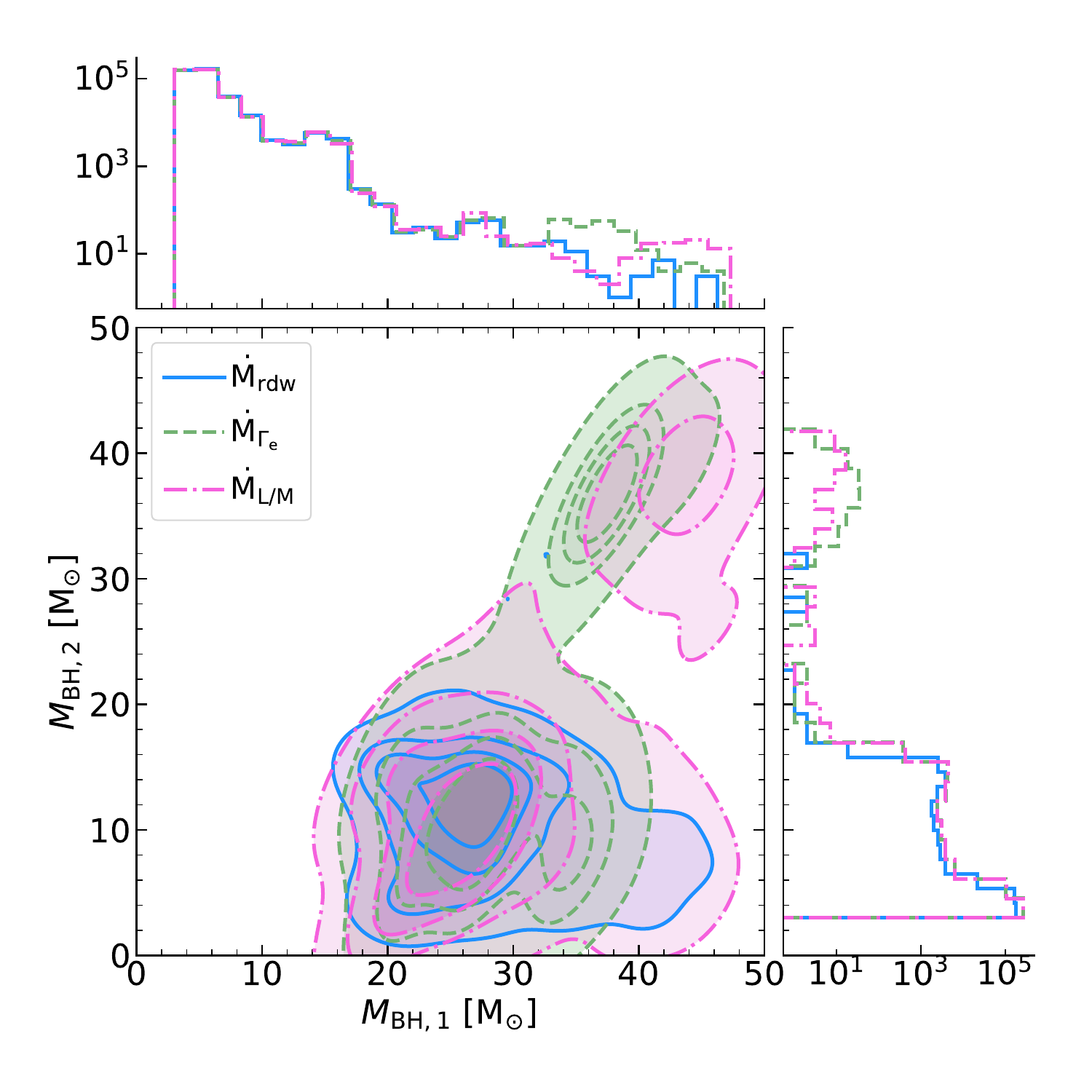}
    \caption{Primary versus secondary BH masses for BBHs that merge in a Hubble time. The contour plot shows the mergers for binaries with $M_{\rm BH,1} \geq$ 20 \msun, while the distributions show the full mass range 
    of primary and secondary BH masses on the top and right.
    The solid blue, dashed green, and dash-dotted pink lines refer to simulations with stellar tracks computed with the \mrdw, \mgam, and \mlm\ rates, respectively.}
    \label{fig:mergers}
\end{figure}

Finally, the BBH's that do merge within a Hubble time originate primarily from ‘lower-mass’ progenitors (less than $\sim$ 80 \msun), so the differences between wind prescriptions are subtle. 
To highlight the differences in the component masses of these mergers, Fig. \ref{fig:mergers} shows the primary ($M_{\rm BH, ~1}$) and secondary ($M_{\rm BH, ~2}$) BH masses of BBHs that merge within a Hubble time. 
The most common primary BH masses lie between 3 and 9 \msun\ for all mass loss rates. These BHs originate from the more numerous lower-mass stars with ZAMS masses in the range of 20 - 40 \msun, which are not the focus of this study and are not shown in the contour plot. More striking differences appear in the high-mass tail of the distributions: the \mgam\ and \mlm\ simulations yield many more massive primaries ($M_{\rm{BH, ~1}}~ \geq$ 30 \msun), with a peak around $\sim$ 35 \msun\ for \mgam\ and at $\sim$ 45 \msun\ for \mlm. These systems also produce a peak in the secondary masses around $M_{\rm{BH, ~2}} \simeq$ 40 \msun.
This is noteworthy because it suggests the emergence of such peaks in the high-mass side is not only due to prescriptions employed for the final fate and the binary stellar evolution processes 
\citep[e.g.,][]{vansonNoPeaksValleys2022a,brielUnderstandingHighmassBinary2023,farrahPreferentialGrowthChannel2023,golombPhysicalModelsAstrophysical2023,hendriksPulsationalPairinstabilitySupernovae2023,dorozsmaiImportanceStableMass2024b, Ugolini_bump_2025}; rather, they result purely from imposing different stellar wind prescriptions in the stellar tracks. We find that the simulations with enhanced winds tend to have primary and secondary components with similar masses, at variance with standard wind simulations that show a larger distribution of the component mass ratios.
Additionally, the LIGO-Virgo-Kagra (LVK) collaboration has detected gravitational wave emission from BBH mergers with secondary BH masses $\gtrsim$ 40 \msun\ \citep{Abbott2020a, Abbott2023, Abbott2024}, a mass range only reproduced with our \mgam\ and \mlm\ simulations. Therefore, to form such massive BHs maintaining this metallicity using the \mrdw\ winds, other formation scenarios are needed, such as dynamical interaction in a cluster and/or hierarchical BH mergers. Features of the merging BBHs, like the masses, formation channels, and binary properties, are discussed in further detail in App. \ref{app:merging_BBH}.

%these features suggest that the onset of peaks in the high-mass side of the primary BH mass distributions are not only due to prescriptions employed for the final fate and the binary stellar evolution processes 

%\GC{Here, I'll put the description of the secondary distribution as done for the primaries. Without mixing the description with other information and details of the system. After the description of Fig. 10 is complete, we can discuss the properties of the systems that create merging BHs and their characteristics, also referring to new figures in the appendices.}

\begin{figure*}[t]
    \centering
    \includegraphics[width=\textwidth]{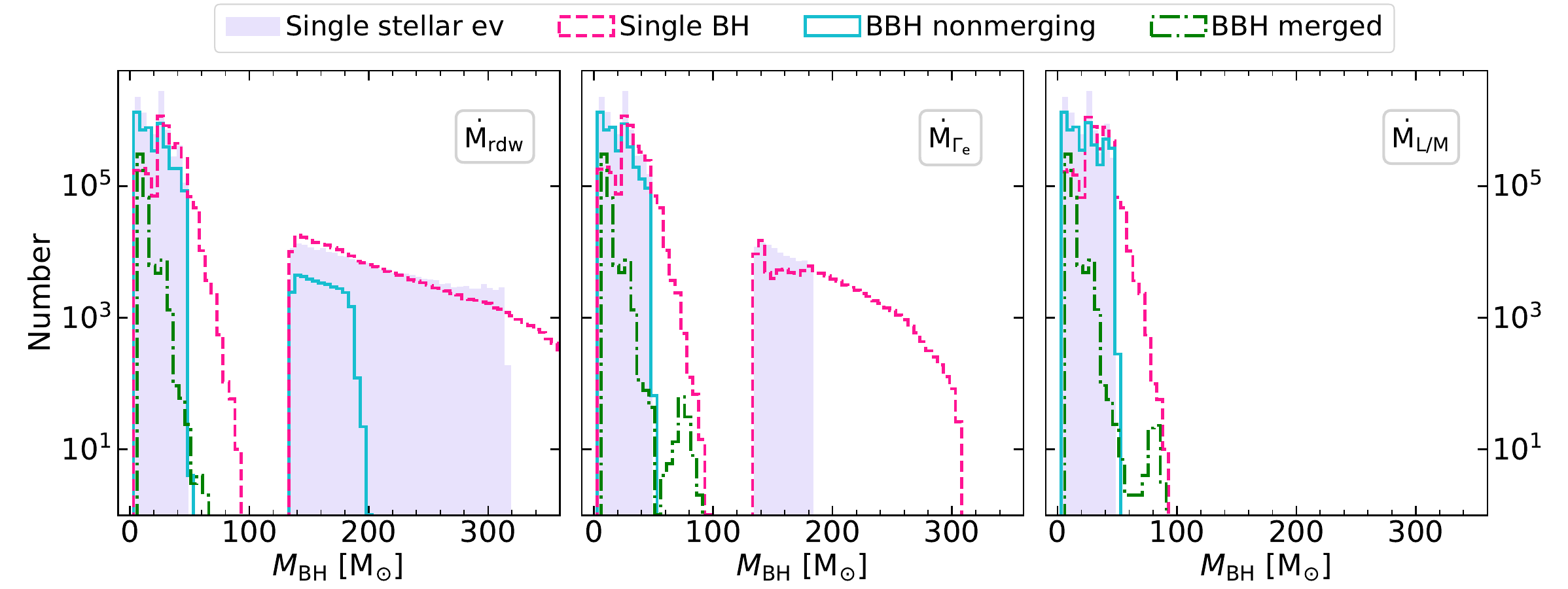}
    \caption{Number counts of BHs from the simulations with \mrdw\ (left), \mgam\ (center), and \mlm\ (right). The shaded purple shows the result from single stellar evolution. The dashed pink, solid blue, and dash-dotted green lines show results from binary evolution of single BHs, binary BHs that do not merge within a Hubble time, and the resulting BH mass of a BBH merger that merged within a Hubble time, respectively. }
    \label{fig:all_BHs_hist}
\end{figure*}

Fig. \ref{fig:all_BHs_hist} shows the BH mass histograms of different BH populations computed with tracks using different stellar winds. In all three simulations, the single BHs (formed from either ejecting or merging with the companion) can partially fill the BH mass gap from single stars, forming BHs with masses up to $\sim$ 90 \msun. These BHs account for 59\%, 58\%, and 57\% of all BHs formed from the \mrdw, \mgam, and \mlm\ simulations. The binary BHs that do not merge in a Hubble time have a maximum BH mass of 198 \msun\ for the \mrdw\ case, while for the \mgam\ and \mlm\ cases it is 50 \msun. The non-merging BBHs account for 35.7\% (\mrdw), 36.4\% (\mgam), and 37.8\% (\mlm) of BHs formed from the simulations. 
Interestingly, the \mgam\ and \mlm\ simulations are able to fill part of the single stars' BH mass gap with their merged BH masses, up to $\sim$ 90 \msun. This is in contrast with the \mrdw\ simulation, in which the most massive BH from the merger of a BBH within a Hubble time is 64 \msun. These BH mergers account for 5.7\%, 5.9\%, and 5.2\% of BHs formed in the \mrdw, \mgam, and \mlm\ simulations.

\section{Discussion}\label{sect:discussion}

While our simulations provide interesting insights into massive star evolution with updated stellar winds and their implications for VMS formation and remnant populations, it is necessary to acknowledge a few uncertainties and limitations in the modeling. Here, we outline a few key caveats that should be considered in this work.

A specific source of uncertainty in single stellar evolution lies in the validity of the new mass loss prescriptions in the post-MS phases. However, as stars spend most of their lives on the MS and indeed lose the most mass via stellar winds during this stage, the post-MS mass loss rates have a much smaller impact on the final masses of the star when compared to MS mass loss.

Another fundamental physical process that is intricately involved with stellar evolution and mass loss, and that remains challenging to model comprehensively in population synthesis studies, is rotation. 
Stellar rotation can increase the mixing of chemical elements, altering the nuclear burning lifetimes and abundance composition at the surface \citep{Maeder2009}. We ran models with rotation (see App. \ref{app:rotating}), and found that the evolution is more greatly affected by the mass loss from stellar winds than rotational mixing for moderate rotational rates ($\Omega/\Omega_{\rm crit} \lesssim$ 0.6). This is because VMSs have large convective cores during the MS, and moderate rotation has a negligible effect on VMS evolution \citep{Yusof2013}. In addition, \citet{Bestenlehner2014} found rotational mixing plays a relatively unimportant role in luminous massive stars with very strong mass loss, instead, the mass loss rate has a strong correlation with the surface He enrichment due to winds stripping the outer layers. 

Uncertainties in the final fate of VMS propagate into uncertainties in the remnant properties. Of these, the edges of the PI mass gap pose as a source of uncertainty in our predictions. The conditions that give rise to pair creation depend on the CO core mass formed and its composition (C/O), which are affected by evolutionary processes such as mass loss, convection, overshooting, and nuclear reactions. Many studies have shown how the edges of the PI mass gap change when altering major sources of uncertainty, such as variations in the $^{12}\mathrm{C(\alpha, \gamma)^{16}O}$ reaction rates \citep{Farmer2019, Farmer2020}, dredge-up episodes \citep{Costa2021}, pulsation-driven mass loss \citep{Volpato2023} with rotation \citep{Volpato2024}, and other uncertainties in massive star evolution \citep{Woosley2021, Vink2021}. Additionally, the prescriptions adopted to compute the final remnant mass are also subject to uncertainty.

Binary evolution modeling propagates the uncertainties found in single star models, and adds additional sources and challenges, the main ones are described in \citet{Iorio2023}. The CE phase is one of the most uncertain processes in binary stellar evolution, due to the hefty computational challenges \citep{Ivanova2013b, MacLeod2017, Fragos2019, ivanova2020}. As a result, hydrodynamical simulations face challenges in modeling the full CE evolution: simulations may omit key physical processes, including recombination energy, nuclear reactions in accreted material, neutrino cooling, and energy transport mechanisms \citep{roepke2023}. 
Moreover, the binding energy definition and the core-mass definition used can alter the resulting binding energy of the star, affecting the binary interactions and the merger rate density (Sgalletta et al. in prep.). The mass transfer stability conditions are also uncertain, and can heavily affect the merger rate densities \citep{Olejak2021, Dorozsmai2024}.
Finally, due to the scale of these simulations and the necessary computational efficiency, the precise tracking of surface abundances from stellar mergers is not accurately modeled. Therefore, we cannot provide constraints on the surface chemical compositions of the post-merger products discussed in section \ref{subsec:stellar mergers}. However hydrodynamical simulations following the evolution of a post-collision star suggest the stellar surface would show an enrichment of helium \citep{Costa2022}. Future works will be dedicated to hydrodynamic modeling of stellar mergers in order to trace the surface composition in detail.

%We ran models with rotation (see App. \ref{app:rotating}), and found that the evolution is more greatly affected by the mass loss from stellar winds than rotational mixing for moderate rotational rates ($\Omega/\Omega_{\rm crit} \lesssim$ 0.6). 

Finally, it is worth mentioning that we cannot make conclusive comparisons with the LVK data by using the newly implemented prescriptions for stellar winds due to our focus on VMSs in the LMC at a single metallicity of Z=0.006. Future studies will examine the effects of these new prescriptions on the evolution of single and binary stars across different metallicities.

\section{Conclusion}\label{sect:concl}
In this paper, we have presented new stellar evolution tracks computed with enhanced stellar winds for VMS. We ran single stellar models with 100 $\leq$ \mzams/\msun\ $\leq$ 600 at LMC metallicity (Z = 0.006), for the three mass loss rates discussed in this paper, based on theoretical and observational studies suggesting a heavier dependence on the Eddington ratio when VMSs approach the Eddington limit. 

We presented the possible evolution of a population of massive binaries by employing our new stellar models in the \sevn\ binary population synthesis code. We investigated stellar mergers as a possible origin of observed VMSs, like those in R136a, and compared the implications with the results from single stellar evolution. We characterized the different BH populations and formation channels that arise from altering the wind prescription. Below, we summarize our main results.

\begin{itemize}
    \item The stellar tracks computed with the new winds, \mgam\ and \mrdw, expand less on the MS, remaining hotter and more compact as a result of the enhanced mass loss, compared to models computed with the standard rates, \mrdw. The tracks with stronger winds spend more time within the observed narrow \teff\ range of observed VMS stars in R136a, and better reproduce their observed properties.
    
    \item Using the \textsc{param} statistical tool, we found that only stellar tracks computed with the \mgam\ wind prescription can fit the observed features of R136a with stellar age estimates of $\sim$ 1.3 - 1.6 Myr. This age determination is in agreement age estimates from other studies. 
    
    \item We find a notable variation in the amount of mass retained by stars at the pre-SN stage depending on the adopted wind prescription. With the \mrdw\ winds, stars typically retain 48\% to 52\% of their ZAMS mass. In contrast, the \mgam\ and \mlm\ models result in appreciably higher mass loss, and the percentage of ZAMS mass retained depends on the ZAMS mass. For \mgam, the retained mass ranges from 30\% to 52\%, decreasing as the ZAMS mass increases. The \mlm\ winds predict the most extreme mass loss, resulting in stars retaining only 13\% to 42\% of their ZAMS mass at the pre-SN stage, also decreasing with increasing initial mass.

    \item We investigated the possible origins of the VMS R136a1 from single and binary evolution. %Our results demonstrate that the inferred upper mass limit of the IMF is significantly dependent on the origin of the star. 
    If binary stellar mergers are the origin of R136a1, then our models suggest a primary initial upper mass limit of 239 \msun\ (for the \mrdw\ case), 276 \msun\ (for the \mgam\ case), and 325 \msun\ (for the \mlm\ case). Conversely, explaining the origin of this VMS through single stellar evolution necessitates a higher upper initial mass limit, from 325 \msun\ (for the \mrdw\ case) to 390 \msun\ (for the enhanced winds).  In either case, we establish a strict upper limit on the ZAMS mass for R136a1, which must be $\lesssim$ 400 \msun, regardless of wind prescription.

    \item From the simulation of binary populations with different winds, we find unique BH distributions for each wind recipe. Simulations with enhanced winds are less likely to merge during stellar evolution because the expansion is suppressed on the MS. As a consequence, these systems have fewer stellar mergers and produce fewer intermediate mass BHs (with masses above the PISN gap), when compared to standard stellar winds.
    
    \item For binary BHs that merge within a Hubble time, only the \mgam\ and \mlm\ simulations were able to have secondaries with masses $>$ 32 \msun, in line with observed BBH mergers. Moreover, these two simulations produced many more primaries with BH masses $>$ 30 \msun\ compared to the \mrdw\ simulation.

\end{itemize}

\begin{acknowledgements}
The authors thank Iorio, G. for the helpful discussions.
We acknowledge the Italian Ministerial grant PRIN2022, ``Radiative opacities for astrophysical applications'', no. 2022NEXMP8.
GC acknowledges partial financial support from European Union—Next Generation EU, Mission 4, Component 2, CUP: C93C24004920006, project ‘FIRES'.
MS, CU, and CS acknowledge their participation in the 2024 Unconventional Thinking Tank Conference, supported by INAF, during which part of this work was developed. MS acknowledges financial support from Large Grant INAF 2024 ``Envisioning Tomorrow: prospects and challenges for multi-messenger astronomy in the era of Rubin and Einstein Telescope'', from Fondazione ICSC, Spoke 3 Astrophysics and Cosmos Observations, National Recovery and Resilience Plan (Piano Nazionale di Ripresa e Resilienza, PNRR) Project ID CN\_00000013 ``Italian Research Center on High-Performance Computing, Big Data and Quantum Computing'' funded by MUR Missione 4 Componente 2 Investimento 1.4: Potenziamento strutture di ricerca e creazione di ``campioni nazionali di R\&S (M4C2-19 )'' - Next Generation EU (NGEU), and from the program ``Data Science methods for Multi-Messenger Astrophysics \& Multi-Survey Cosmology'' funded by the Italian Ministry of University and Research, Programmazione triennale 2021/2023 (DM n.2503 dd. 09/12/2019), Programma Congiunto Scuole.

\end{acknowledgements}

\bibliographystyle{aa} % style aa.bst
%\bibliography{biblio} 

\begin{thebibliography}{115}
\expandafter\ifx\csname natexlab\endcsname\relax\def\natexlab#1{#1}\fi

\bibitem[{Abbott {et~al.}(2016)Abbott, Abbott, Abbott, Abernathy, Acernese, Ackley, Adams, Adams, Addesso, Adhikari, Adya, Affeldt, Agathos, Agatsuma, Aggarwal, Aguiar, Aiello, Ain, Ajith, Allen, Allocca, Altin, Anderson, Anderson, Arai, Arain, Araya, Arceneaux, Areeda, Arnaud, Arun, Ascenzi, Ashton, Ast, Aston, Astone, Aufmuth, Aulbert, Yam, Yamamoto, Yancey, Yap, Yu, Yvert, Zadro\ifmmode~\dot{z}\else \.{z}\fi{}ny, Zangrando, Zanolin, Zendri, Zevin, Zhang, Zhang, Zhang, Zhang, Zhao, Zhou, Zhou, Zhu, Zucker, Zuraw, \& Zweizig}]{Abbot2016}
Abbott, B.~P., Abbott, R., Abbott, T.~D., {et~al.} 2016, Phys. Rev. Lett., 116, 061102

\bibitem[{{Abbott} {et~al.}(2020){Abbott}, {Abbott}, {Abraham}, {Acernese}, {Ackley}, {Adams}, {Adhikari}, {Adya}, {Affeldt}, {Agathos}, {Agatsuma}, {Aggarwal}, {Aguiar}, {Aich}, {Aiello}, {Ain}, {Ajith}, {Akcay}, {Allen}, {Allocca}, {Altin}, {Amato}, {Anand}, {Ananyeva}, {Anderson}, {Anderson}, {Angelova}, {Ansoldi}, {Antier}, {Appert}, {Arai}, {Araya}, {Areeda}, {Ar{\`e}ne}, {Arnaud}, {Chao}, {Charlton}, {Chase}, {Chassande-Mottin}, {Chatterjee}, {Chaturvedi}, {Chatziioannou}, {Chen}, \& {Chen}}]{Abbott2020a}
{Abbott}, R., {Abbott}, T.~D., {Abraham}, S., {et~al.} 2020, \prl, 125, 101102

\bibitem[{{Abbott} {et~al.}(2023){Abbott}, {Abbott}, {Acernese}, {Ackley}, {Adams}, {Adhikari}, {Adhikari}, {Adya}, {Affeldt}, {Agarwal}, {Agathos}, {Agatsuma}, {Aggarwal}, {Aguiar}, {Aiello}, {Ain}, {Ajith}, {Akcay}, {Akutsu}, {Albanesi}, {Allocca}, {Altin}, {Amato}, {Brandt}, {Brau}, {Breschi}, {Briant}, {Briggs}, {Brillet}, {Brinkmann}, {Brockill}, {Brooks}, {Brooks}, {Brown}, {Brunett}, {Bruno}, {Bruntz}, {Bryant}, {Bulik}, {Bulten}, {Buonanno}, {Buscicchio}, {Buskulic}, {Buy}, {Byer}, {Davies}, {Cadonati}, {Cagnoli}, {Cahillane}, {Bustillo}, {Callaghan}, {Callister}, {Calloni}, {Cameron}, {Camp}, {Canepa}, {Canevarolo}, {Cannavacciuolo}, {Cannon}, {Cao}, {Cao}, {Capocasa}, {Capote}, {Carapella}, \& {Carbognani}}]{Abbott2023}
{Abbott}, R., {Abbott}, T.~D., {Acernese}, F., {et~al.} 2023, Physical Review X, 13, 041039

\bibitem[{{Abbott} {et~al.}(2024){Abbott}, {Abbott}, {Acernese}, {Ackley}, {Adams}, {Adhikari}, {Adhikari}, {Adya}, {Affeldt}, {Agarwal}, {Agathos}, {Agatsuma}, {Aggarwal}, {Aguiar}, {Aiello}, {Ain}, {Ajith}, {Albanesi}, {Allocca}, {Altin}, {Amato}, {Anand}, {Anand}, {Ananyeva}, {Anderson}, {Anderson}, {Andrade}, {Andres}, {Cannon}, {Cao}, {Capote}, {Carapella}, {Carbognani}, {Carlin}, {Carney}, {Carpinelli}, {Carrillo}, {Carullo}, {Carver}, {Diaz}, {Casentini}, {Castaldi}, {Caudill}, {Cavagli{\`a}}, {Cavalier}, {Cavalieri}, {Ceasar}, {Cella}, {Cerd{\'a}-Dur{\'a}n}, {Cesarini}, \& {Chaibi}}]{Abbott2024}
{Abbott}, R., {Abbott}, T.~D., {Acernese}, F., {et~al.} 2024, \prd, 109, 022001

\bibitem[{{Bestenlehner}(2020)}]{Bestenlehner2020a}
{Bestenlehner}, J.~M. 2020, \mnras, 493, 3938

\bibitem[{{Bestenlehner} {et~al.}(2020){Bestenlehner}, {Crowther}, {Caballero-Nieves}, {Schneider}, {Sim{\'o}n-D{\'\i}az}, {Brands}, {de Koter}, {Gr{\"a}fener}, {Herrero}, {Langer}, {Lennon}, {Ma{\'\i}z Apell{\'a}niz}, {Puls}, \& {Vink}}]{Bestenlehner2020}
{Bestenlehner}, J.~M., {Crowther}, P.~A., {Caballero-Nieves}, S.~M., {et~al.} 2020, \mnras, 499, 1918

\bibitem[{{Bestenlehner} {et~al.}(2014){Bestenlehner}, {Gr{\"a}fener}, {Vink}, {Najarro}, {de Koter}, {Sana}, {Evans}, {Crowther}, {H{\'e}nault-Brunet}, {Herrero}, {Langer}, {Schneider}, {Sim{\'o}n-D{\'\i}az}, {Taylor}, \& {Walborn}}]{Bestenlehner2014}
{Bestenlehner}, J.~M., {Gr{\"a}fener}, G., {Vink}, J.~S., {et~al.} 2014, \aap, 570, A38

\bibitem[{{Bj{\"o}rklund} {et~al.}(2023){Bj{\"o}rklund}, {Sundqvist}, {Singh}, {Puls}, \& {Najarro}}]{Bjorklund2023}
{Bj{\"o}rklund}, R., {Sundqvist}, J.~O., {Singh}, S.~M., {Puls}, J., \& {Najarro}, F. 2023, \aap, 676, A109

\bibitem[{{B{\"{o}}hm-Vitense}(1958)}]{Bohm-Vitense1958}
{B{\"{o}}hm-Vitense}, E. 1958, \zap, 46, 108

\bibitem[{{Brands} {et~al.}(2022){Brands}, {de Koter}, {Bestenlehner}, {Crowther}, {Sundqvist}, {Puls}, {Caballero-Nieves}, {Abdul-Masih}, {Driessen}, {Garc{\'\i}a}, {Geen}, {Gr{\"a}fener}, {Hawcroft}, {Kaper}, {Keszthelyi}, {Langer}, {Sana}, {Schneider}, {Shenar}, \& {Vink}}]{Brands2022}
{Brands}, S.~A., {de Koter}, A., {Bestenlehner}, J.~M., {et~al.} 2022, \aap, 663, A36

\bibitem[{{Bressan} {et~al.}(2012){Bressan}, {Marigo}, {Girardi}, {Salasnich}, {Dal Cero}, {Rubele}, \& {Nanni}}]{Bressan2012}
{Bressan}, A., {Marigo}, P., {Girardi}, L., {et~al.} 2012, \mnras, 427, 127

\bibitem[{{Bressan} {et~al.}(1981){Bressan}, {Chiosi}, \& {Bertelli}}]{Bressan1981}
{Bressan}, A.~G., {Chiosi}, C., \& {Bertelli}, G. 1981, \aap, 102, 25

\bibitem[{Briel {et~al.}(2023)Briel, Stevance, \& Eldridge}]{brielUnderstandingHighmassBinary2023}
Briel, M.~M., Stevance, H.~F., \& Eldridge, J.~J. 2023, Monthly Notices of the Royal Astronomical Society, 520, 5724

\bibitem[{{Broekgaarden} {et~al.}(2022){Broekgaarden}, {Stevenson}, \& {Thrane}}]{Broekgaarden2022}
{Broekgaarden}, F.~S., {Stevenson}, S., \& {Thrane}, E. 2022, \apj, 938, 45

\bibitem[{{Brott} {et~al.}(2011){Brott}, {de Mink}, {Cantiello}, {Langer}, {de Koter}, {Evans}, {Hunter}, {Trundle}, \& {Vink}}]{Brott2011}
{Brott}, I., {de Mink}, S.~E., {Cantiello}, M., {et~al.} 2011, \aap, 530, A115

\bibitem[{{Caffau} {et~al.}(2011){Caffau}, {Ludwig}, {Steffen}, {Freytag}, \& {Bonifacio}}]{Caffau2011}
{Caffau}, E., {Ludwig}, H.~G., {Steffen}, M., {Freytag}, B., \& {Bonifacio}, P. 2011, \solphys, 268, 255

\bibitem[{{Cassinelli} {et~al.}(1981){Cassinelli}, {Mathis}, \& {Savage}}]{Cassinelli1981}
{Cassinelli}, J.~P., {Mathis}, J.~S., \& {Savage}, B.~D. 1981, Science, 212, 1497

\bibitem[{{Castor} {et~al.}(1975){Castor}, {Abbott}, \& {Klein}}]{CAK1975}
{Castor}, J.~I., {Abbott}, D.~C., \& {Klein}, R.~I. 1975, \apj, 195, 157

\bibitem[{{Chen} {et~al.}(2015){Chen}, {Bressan}, {Girardi}, {Marigo}, {Kong}, \& {Lanza}}]{Chen2015}
{Chen}, Y., {Bressan}, A., {Girardi}, L., {et~al.} 2015, \mnras, 452, 1068

\bibitem[{{Choudhury} {et~al.}(2021){Choudhury}, {de Grijs}, {Bekki}, {Cioni}, {Ivanov}, {van Loon}, {Miller}, {Niederhofer}, {Oliveira}, {Ripepi}, {Sun}, \& {Subramanian}}]{Choudhury2021}
{Choudhury}, S., {de Grijs}, R., {Bekki}, K., {et~al.} 2021, \mnras, 507, 4752

\bibitem[{{Choudhury} {et~al.}(2016){Choudhury}, {Subramaniam}, \& {Cole}}]{Choudhury2016}
{Choudhury}, S., {Subramaniam}, A., \& {Cole}, A.~A. 2016, \mnras, 455, 1855

\bibitem[{{Costa} {et~al.}(2022){Costa}, {Ballone}, {Mapelli}, \& {Bressan}}]{Costa2022}
{Costa}, G., {Ballone}, A., {Mapelli}, M., \& {Bressan}, A. 2022, \mnras, 516, 1072

\bibitem[{{Costa} {et~al.}(2021){Costa}, {Bressan}, {Mapelli}, {Marigo}, {Iorio}, \& {Spera}}]{Costa2021}
{Costa}, G., {Bressan}, A., {Mapelli}, M., {et~al.} 2021, \mnras, 501, 4514

\bibitem[{{Costa} {et~al.}(2019{\natexlab{a}}){Costa}, {Girardi}, {Bressan}, {Chen}, {Goudfrooij}, {Marigo}, {Rodrigues}, \& {Lanza}}]{Costa2019a}
{Costa}, G., {Girardi}, L., {Bressan}, A., {et~al.} 2019{\natexlab{a}}, \aap, 631, A128

\bibitem[{{Costa} {et~al.}(2019{\natexlab{b}}){Costa}, {Girardi}, {Bressan}, {Marigo}, {Rodrigues}, {Chen}, {Lanza}, \& {Goudfrooij}}]{Costa2019b}
{Costa}, G., {Girardi}, L., {Bressan}, A., {et~al.} 2019{\natexlab{b}}, \mnras, 485, 4641

\bibitem[{{Costa} {et~al.}(2025){Costa}, {Shepherd}, {Bressan}, {Addari}, {Chen}, {Fu}, {Volpato}, {Nguyen}, {Girardi}, {Marigo}, {Mazzi}, {Pastorelli}, {Trabucchi}, {Bossini}, \& {Zaggia}}]{CostaShepherd2025}
{Costa}, G., {Shepherd}, K.~G., {Bressan}, A., {et~al.} 2025, arXiv e-prints, arXiv:2501.12917

\bibitem[{{Crowther} {et~al.}(2016){Crowther}, {Caballero-Nieves}, {Bostroem}, {Ma{\'\i}z Apell{\'a}niz}, {Schneider}, {Walborn}, {Angus}, {Brott}, {Bonanos}, {de Koter}, {de Mink}, {Evans}, {Gr{\"a}fener}, {Herrero}, {Howarth}, {Langer}, {Lennon}, {Puls}, {Sana}, \& {Vink}}]{Crowther2016}
{Crowther}, P.~A., {Caballero-Nieves}, S.~M., {Bostroem}, K.~A., {et~al.} 2016, \mnras, 458, 624

\bibitem[{{Crowther} {et~al.}(2010){Crowther}, {Schnurr}, {Hirschi}, {Yusof}, {Parker}, {Goodwin}, \& {Kassim}}]{Crowther2010}
{Crowther}, P.~A., {Schnurr}, O., {Hirschi}, R., {et~al.} 2010, \mnras, 408, 731

\bibitem[{{da Silva} {et~al.}(2006){da Silva}, {Girardi}, {Pasquini}, {Setiawan}, {von der L{\"u}he}, {de Medeiros}, {Hatzes}, {D{\"o}llinger}, \& {Weiss}}]{daSilva2006}
{da Silva}, L., {Girardi}, L., {Pasquini}, L., {et~al.} 2006, \aap, 458, 609

\bibitem[{{de Jager} {et~al.}(1988){de Jager}, {Nieuwenhuijden}, \& {van der Hucht}}]{deJager1988}
{de Jager}, C., {Nieuwenhuijden}, H., \& {van der Hucht}, K.~A. 1988, Bulletin d'Information du Centre de Donnees Stellaires, 35, 141

\bibitem[{{de Koter} {et~al.}(1998){de Koter}, {Heap}, \& {Hubeny}}]{dekoter1998}
{de Koter}, A., {Heap}, S.~R., \& {Hubeny}, I. 1998, \apj, 509, 879

\bibitem[{Dorozsmai \& Toonen(2024)}]{dorozsmaiImportanceStableMass2024b}
Dorozsmai, A. \& Toonen, S. 2024, Monthly Notices of the Royal Astronomical Society, 530, 3706

\bibitem[{{Dorozsmai} \& {Toonen}(2024)}]{Dorozsmai2024}
{Dorozsmai}, A. \& {Toonen}, S. 2024, \mnras, 530, 3706

\bibitem[{{Evans} {et~al.}(2011){Evans}, {Taylor}, {H{\'e}nault-Brunet}, {Sana}, {de Koter}, {Sim{\'o}n-D{\'\i}az}, {Carraro}, {Bagnoli}, {Bastian}, {Bestenlehner}, {Bonanos}, {Bressert}, {Brott}, {Campbell}, {Cantiello}, {Clark}, {Costa}, {Crowther}, {de Mink}, {Doran}, {Dufton}, {Dunstall}, {Friedrich}, {Garcia}, {Gieles}, {Gr{\"a}fener}, {Herrero}, {Howarth}, {Izzard}, {Langer}, {Lennon}, {Ma{\'\i}z Apell{\'a}niz}, {Markova}, {Najarro}, {Puls}, {Ramirez}, {Sab{\'\i}n-Sanjuli{\'a}n}, {Smartt}, {Stroud}, {van Loon}, {Vink}, \& {Walborn}}]{Evans2011}
{Evans}, C.~J., {Taylor}, W.~D., {H{\'e}nault-Brunet}, V., {et~al.} 2011, \aap, 530, A108

\bibitem[{{Farag} {et~al.}(2022){Farag}, {Renzo}, {Farmer}, {Chidester}, \& {Timmes}}]{Farag2022}
{Farag}, E., {Renzo}, M., {Farmer}, R., {Chidester}, M.~T., \& {Timmes}, F.~X. 2022, \apj, 937, 112

\bibitem[{{Farmer} {et~al.}(2020){Farmer}, {Renzo}, {de Mink}, {Fishbach}, \& {Justham}}]{Farmer2020}
{Farmer}, R., {Renzo}, M., {de Mink}, S.~E., {Fishbach}, M., \& {Justham}, S. 2020, \apjl, 902, L36

\bibitem[{{Farmer} {et~al.}(2019){Farmer}, {Renzo}, {de Mink}, {Marchant}, \& {Justham}}]{Farmer2019}
{Farmer}, R., {Renzo}, M., {de Mink}, S.~E., {Marchant}, P., \& {Justham}, S. 2019, \apj, 887, 53

\bibitem[{Farrah {et~al.}(2023)Farrah, Petty, Croker, Tarl{\'e}, Zevin, Hatziminaoglou, Shankar, Wang, Clements, Efstathiou, Lacy, Nishimura, Afonso, Pearson, \& Pitchford}]{farrahPreferentialGrowthChannel2023}
Farrah, D., Petty, S., Croker, K.~S., {et~al.} 2023, ApJ, 943, 133

\bibitem[{{Fragos} {et~al.}(2019){Fragos}, {Andrews}, {Ramirez-Ruiz}, {Meynet}, {Kalogera}, {Taam}, \& {Zezas}}]{Fragos2019}
{Fragos}, T., {Andrews}, J.~J., {Ramirez-Ruiz}, E., {et~al.} 2019, \apjl, 883, L45

\bibitem[{{Freitag} {et~al.}(2006){Freitag}, {G{\"u}rkan}, \& {Rasio}}]{Freitag2006}
{Freitag}, M., {G{\"u}rkan}, M.~A., \& {Rasio}, F.~A. 2006, \mnras, 368, 141

\bibitem[{Fryer {et~al.}(2012)Fryer, Belczynski, Wiktorowicz, Dominik, Kalogera, \& Holz}]{fryerCOMPACTREMNANTMASS2012}
Fryer, C.~L., Belczynski, K., Wiktorowicz, G., {et~al.} 2012, The Astrophysical Journal

\bibitem[{{Fu} {et~al.}(2018){Fu}, {Bressan}, {Marigo}, {Girardi}, {Montalb{\'a}n}, {Chen}, \& {Nanni}}]{Fu2018}
{Fu}, X., {Bressan}, A., {Marigo}, P., {et~al.} 2018, \mnras, 476, 496

\bibitem[{{Gallagher}(1989)}]{Gallagher}
{Gallagher}, J.~S. 1989, in Astrophysics and Space Science Library, Vol. 157, IAU Colloq. 113: Physics of Luminous Blue Variables, ed. K.~{Davidson}, A.~F.~J. {Moffat}, \& H.~J.~G.~L.~M. {Lamers}, 185

\bibitem[{{Georgy} {et~al.}(2013){Georgy}, {Ekstr{\"o}m}, {Eggenberger}, {Meynet}, {Haemmerl{\'e}}, {Maeder}, {Granada}, {Groh}, {Hirschi}, {Mowlavi}, {Yusof}, {Charbonnel}, {Decressin}, \& {Barblan}}]{Georgy2013}
{Georgy}, C., {Ekstr{\"o}m}, S., {Eggenberger}, P., {et~al.} 2013, \aap, 558, A103

\bibitem[{{Giacobbo} \& {Mapelli}(2018)}]{Giacobbo2018}
{Giacobbo}, N. \& {Mapelli}, M. 2018, \mnras, 480, 2011

\bibitem[{Golomb {et~al.}(2023)Golomb, Isi, \& Farr}]{golombPhysicalModelsAstrophysical2023}
Golomb, J., Isi, M., \& Farr, W. 2023, Physical {{Models}} for the {{Astrophysical Population}} of {{Black Holes}}: {{Application}} to the {{Bump}} in the {{Mass Distribution}} of {{Gravitational Wave Sources}}

\bibitem[{{Gormaz-Matamala} {et~al.}(2024){Gormaz-Matamala}, {Cuadra}, {Ekstr{\"o}m}, {Meynet}, {Cur{\'e}}, \& {Belczynski}}]{Gormaz2024}
{Gormaz-Matamala}, A.~C., {Cuadra}, J., {Ekstr{\"o}m}, S., {et~al.} 2024, \aap, 687, A290

\bibitem[{{Gr{\"a}fener} \& {Hamann}(2008)}]{Grafener2008}
{Gr{\"a}fener}, G. \& {Hamann}, W.~R. 2008, \aap, 482, 945

\bibitem[{{Gr{\"a}fener} {et~al.}(2011){Gr{\"a}fener}, {Vink}, {de Koter}, \& {Langer}}]{Grafener2011}
{Gr{\"a}fener}, G., {Vink}, J.~S., {de Koter}, A., \& {Langer}, N. 2011, \aap, 535, A56

\bibitem[{{Heger} {et~al.}(2000){Heger}, {Langer}, \& {Woosley}}]{Heger2000}
{Heger}, A., {Langer}, N., \& {Woosley}, S.~E. 2000, \apj, 528, 368

\bibitem[{Hendriks {et~al.}(2023)Hendriks, {van Son}, Renzo, Izzard, \& Farmer}]{hendriksPulsationalPairinstabilitySupernovae2023}
Hendriks, D.~D., {van Son}, L. A.~C., Renzo, M., Izzard, R.~G., \& Farmer, R. 2023, Pulsational Pair-Instability Supernovae in Gravitational-Wave and Electromagnetic Transients

\bibitem[{{Humphreys} \& {Davidson}(1979)}]{Humphreys1979}
{Humphreys}, R.~M. \& {Davidson}, K. 1979, \apj, 232, 409

\bibitem[{{Hurley} {et~al.}(2002){Hurley}, {Tout}, \& {Pols}}]{Hurley2002}
{Hurley}, J.~R., {Tout}, C.~A., \& {Pols}, O.~R. 2002, Monthly Notices of the Royal Astronomical Society, 329, 897

\bibitem[{Iorio {et~al.}(2023)Iorio, Mapelli, Costa, Spera, Escobar, Sgalletta, Trani, Korb, Santoliquido, Dall'Amico, Gaspari, \& Bressan}]{iorioCompactObjectMergers2023}
Iorio, G., Mapelli, M., Costa, G., {et~al.} 2023, Monthly Notices of the Royal Astronomical Society, 524, 426

\bibitem[{{Iorio} {et~al.}(2023){Iorio}, {Mapelli}, {Costa}, {Spera}, {Escobar}, {Sgalletta}, {Trani}, {Korb}, {Santoliquido}, {Dall'Amico}, {Gaspari}, \& {Bressan}}]{Iorio2023}
{Iorio}, G., {Mapelli}, M., {Costa}, G., {et~al.} 2023, \mnras, 524, 426

\bibitem[{{Ivanova} {et~al.}(2013){Ivanova}, {Justham}, {Chen}, {De Marco}, {Fryer}, {Gaburov}, {Ge}, {Glebbeek}, {Han}, {Li}, {Lu}, {Marsh}, {Podsiadlowski}, {Potter}, {Soker}, {Taam}, {Tauris}, {van den Heuvel}, \& {Webbink}}]{Ivanova2013b}
{Ivanova}, N., {Justham}, S., {Chen}, X., {et~al.} 2013, \aapr, 21, 59

\bibitem[{Ivanova {et~al.}(2020)Ivanova, Justham, \& Ricker}]{ivanova2020}
Ivanova, N., Justham, S., \& Ricker, P. 2020, Common Envelope Evolution, 2514-3433 (IOP Publishing)

\bibitem[{{Josiek} {et~al.}(2024){Josiek}, {Ekstr{\"o}m}, \& {Sander}}]{Josiek2024}
{Josiek}, J., {Ekstr{\"o}m}, S., \& {Sander}, A.~A.~C. 2024, \aap, 688, A71

\bibitem[{{Justham} {et~al.}(2014){Justham}, {Podsiadlowski}, \& {Vink}}]{Justham2014}
{Justham}, S., {Podsiadlowski}, P., \& {Vink}, J.~S. 2014, \apj, 796, 121

\bibitem[{{Kobulnicky} \& {Fryer}(2007)}]{Kobulnicky2007}
{Kobulnicky}, H.~A. \& {Fryer}, C.~L. 2007, \apj, 670, 747

\bibitem[{{K{\"o}hler} {et~al.}(2015){K{\"o}hler}, {Langer}, {de Koter}, {de Mink}, {Crowther}, {Evans}, {Gr{\"a}fener}, {Sana}, {Sanyal}, {Schneider}, \& {Vink}}]{Kohler2015}
{K{\"o}hler}, K., {Langer}, N., {de Koter}, A., {et~al.} 2015, \aap, 573, A71

\bibitem[{{Kroupa}(2001)}]{Kroupa2001}
{Kroupa}, P. 2001, \mnras, 322, 231

\bibitem[{{Lattanzi} {et~al.}(1994){Lattanzi}, {Hershey}, {Burg}, {Taff}, {Holfeltz}, {Bucciarelli}, {Evans}, {Gilmozzi}, {Pringle}, \& {Walborn}}]{Lattanzi1994}
{Lattanzi}, M.~G., {Hershey}, J.~L., {Burg}, R., {et~al.} 1994, \apjl, 427, L21

\bibitem[{{MacLeod} {et~al.}(2017){MacLeod}, {Antoni}, {Murguia-Berthier}, {Macias}, \& {Ramirez-Ruiz}}]{MacLeod2017}
{MacLeod}, M., {Antoni}, A., {Murguia-Berthier}, A., {Macias}, P., \& {Ramirez-Ruiz}, E. 2017, \apj, 838, 56

\bibitem[{{Maeder}(2009)}]{Maeder2009}
{Maeder}, A. 2009, {Physics, Formation and Evolution of Rotating Stars}

\bibitem[{{Mapelli} {et~al.}(2020){Mapelli}, {Spera}, {Montanari}, {Limongi}, {Chieffi}, {Giacobbo}, {Bressan}, \& {Bouffanais}}]{Mapelli2020}
{Mapelli}, M., {Spera}, M., {Montanari}, E., {et~al.} 2020, \apj, 888, 76

\bibitem[{{Martinet} {et~al.}(2023){Martinet}, {Meynet}, {Ekstr{\"o}m}, {Georgy}, \& {Hirschi}}]{Martinet2023}
{Martinet}, S., {Meynet}, G., {Ekstr{\"o}m}, S., {Georgy}, C., \& {Hirschi}, R. 2023, \aap, 679, A137

\bibitem[{{Martins} {et~al.}(2008){Martins}, {Hillier}, {Paumard}, {Eisenhauer}, {Ott}, \& {Genzel}}]{Martins2008}
{Martins}, F., {Hillier}, D.~J., {Paumard}, T., {et~al.} 2008, \aap, 478, 219

\bibitem[{{Mason} {et~al.}(2009){Mason}, {Hartkopf}, {Gies}, {Henry}, \& {Helsel}}]{Mason2009}
{Mason}, B.~D., {Hartkopf}, W.~I., {Gies}, D.~R., {Henry}, T.~J., \& {Helsel}, J.~W. 2009, \aj, 137, 3358

\bibitem[{{Massey} \& {Hunter}(1998)}]{Massey1998}
{Massey}, P. \& {Hunter}, D.~A. 1998, \apj, 493, 180

\bibitem[{{McEvoy} {et~al.}(2015){McEvoy}, {Dufton}, {Evans}, {Kalari}, {Markova}, {Sim{\'o}n-D{\'\i}az}, {Vink}, {Walborn}, {Crowther}, {de Koter}, {de Mink}, {Dunstall}, {H{\'e}nault-Brunet}, {Herrero}, {Langer}, {Lennon}, {Ma{\'\i}z Apell{\'a}niz}, {Najarro}, {Puls}, {Sana}, {Schneider}, \& {Taylor}}]{McEvoy2015}
{McEvoy}, C.~M., {Dufton}, P.~L., {Evans}, C.~J., {et~al.} 2015, \aap, 575, A70

\bibitem[{{Menon} \& {Heger}(2017)}]{Menon2017}
{Menon}, A. \& {Heger}, A. 2017, \mnras, 469, 4649

\bibitem[{{Nguyen} {et~al.}(2022){Nguyen}, {Costa}, {Girardi}, {Volpato}, {Bressan}, {Chen}, {Marigo}, {Fu}, \& {Goudfrooij}}]{Nguyen2022}
{Nguyen}, C.~T., {Costa}, G., {Girardi}, L., {et~al.} 2022, \aap, 665, A126

\bibitem[{{Nugis} \& {Lamers}(2000)}]{Nugis2000}
{Nugis}, T. \& {Lamers}, H.~J.~G.~L.~M. 2000, \aap, 360, 227

\bibitem[{{Olejak} {et~al.}(2021){Olejak}, {Belczynski}, \& {Ivanova}}]{Olejak2021}
{Olejak}, A., {Belczynski}, K., \& {Ivanova}, N. 2021, \aap, 651, A100

\bibitem[{{Petrovic} {et~al.}(2005){Petrovic}, {Langer}, \& {van der Hucht}}]{Petrovic2005}
{Petrovic}, J., {Langer}, N., \& {van der Hucht}, K.~A. 2005, \aap, 435, 1013

\bibitem[{{Podsiadlowski} {et~al.}(1990){Podsiadlowski}, {Joss}, \& {Rappaport}}]{Podsiadlowski1990}
{Podsiadlowski}, P., {Joss}, P.~C., \& {Rappaport}, S. 1990, \aap, 227, L9

\bibitem[{{Portegies Zwart} \& {McMillan}(2002)}]{PortegiesZwart2002}
{Portegies Zwart}, S.~F. \& {McMillan}, S. L.~W. 2002, \apj, 576, 899

\bibitem[{{Ram{\'\i}rez-Agudelo} {et~al.}(2017){Ram{\'\i}rez-Agudelo}, {Sana}, {de Koter}, {Tramper}, {Grin}, {Schneider}, {Langer}, {Puls}, {Markova}, {Bestenlehner}, {Castro}, {Crowther}, {Evans}, {Garc{\'\i}a}, {Gr{\"a}fener}, {Herrero}, {van Kempen}, {Lennon}, {Ma{\'\i}z Apell{\'a}niz}, {Najarro}, {Sab{\'\i}n-Sanjuli{\'a}n}, {Sim{\'o}n-D{\'\i}az}, {Taylor}, \& {Vink}}]{Ramirez2017}
{Ram{\'\i}rez-Agudelo}, O.~H., {Sana}, H., {de Koter}, A., {et~al.} 2017, \aap, 600, A81

\bibitem[{{Rodrigues} {et~al.}(2017){Rodrigues}, {Bossini}, {Miglio}, {Girardi}, {Montalb{\'a}n}, {Noels}, {Trabucchi}, {Coelho}, \& {Marigo}}]{Rodrigues2017}
{Rodrigues}, T.~S., {Bossini}, D., {Miglio}, A., {et~al.} 2017, \mnras, 467, 1433

\bibitem[{{Rodrigues} {et~al.}(2014){Rodrigues}, {Girardi}, {Miglio}, {Bossini}, {Bovy}, {Epstein}, {Pinsonneault}, {Stello}, {Zasowski}, {Allende Prieto}, {Chaplin}, {Hekker}, {Johnson}, {M{\'e}sz{\'a}ros}, {Mosser}, {Anders}, {Basu}, {Beers}, {Chiappini}, {da Costa}, {Elsworth}, {Garc{\'\i}a}, {Garc{\'\i}a P{\'e}rez}, {Hearty}, {Maia}, {Majewski}, {Mathur}, {Montalb{\'a}n}, {Nidever}, {Santiago}, {Schultheis}, {Serenelli}, \& {Shetrone}}]{Rodrigues2014}
{Rodrigues}, T.~S., {Girardi}, L., {Miglio}, A., {et~al.} 2014, \mnras, 445, 2758

\bibitem[{{R{\"o}pke} \& {De Marco}(2023)}]{roepke2023}
{R{\"o}pke}, F.~K. \& {De Marco}, O. 2023, Living Reviews in Computational Astrophysics, 9, 2

\bibitem[{{Sabhahit} {et~al.}(2022){Sabhahit}, {Vink}, {Higgins}, \& {Sander}}]{Sabhahit2022}
{Sabhahit}, G.~N., {Vink}, J.~S., {Higgins}, E.~R., \& {Sander}, A. A.~C. 2022, \mnras, 514, 3736

\bibitem[{{Sab{\'\i}n-Sanjuli{\'a}n} {et~al.}(2017){Sab{\'\i}n-Sanjuli{\'a}n}, {Sim{\'o}n-D{\'\i}az}, {Herrero}, {Puls}, {Schneider}, {Evans}, {Garcia}, {Najarro}, {Brott}, {Castro}, {Crowther}, {de Koter}, {de Mink}, {Gr{\"a}fener}, {Grin}, {Holgado}, {Langer}, {Lennon}, {Ma{\'\i}z Apell{\'a}niz}, {Ram{\'\i}rez-Agudelo}, {Sana}, {Taylor}, {Vink}, \& {Walborn}}]{SabinSanjulian2017}
{Sab{\'\i}n-Sanjuli{\'a}n}, C., {Sim{\'o}n-D{\'\i}az}, S., {Herrero}, A., {et~al.} 2017, \aap, 601, A79

\bibitem[{{Sab{\'\i}n-Sanjuli{\'a}n} {et~al.}(2014){Sab{\'\i}n-Sanjuli{\'a}n}, {Sim{\'o}n-D{\'\i}az}, {Herrero}, {Walborn}, {Puls}, {Ma{\'\i}z Apell{\'a}niz}, {Evans}, {Brott}, {de Koter}, {Garcia}, {Markova}, {Najarro}, {Ram{\'\i}rez-Agudelo}, {Sana}, {Taylor}, \& {Vink}}]{SabinSanjulian2014}
{Sab{\'\i}n-Sanjuli{\'a}n}, C., {Sim{\'o}n-D{\'\i}az}, S., {Herrero}, A., {et~al.} 2014, \aap, 564, A39

\bibitem[{{Sana} {et~al.}(2012){Sana}, {de Mink}, {de Koter}, {Langer}, {Evans}, {Gieles}, {Gosset}, {Izzard}, {Le Bouquin}, \& {Schneider}}]{Sana2012}
{Sana}, H., {de Mink}, S.~E., {de Koter}, A., {et~al.} 2012, Science, 337, 444

\bibitem[{{Sana} {et~al.}(2013){Sana}, {de Mink}, {de Koter}, {Langer}, {Evans}, {Gieles}, {Gosset}, {Izzard}, {Le Bouquin}, \& {Schneider}}]{Sana2013}
{Sana}, H., {de Mink}, S.~E., {de Koter}, A., {et~al.} 2013, in Astronomical Society of the Pacific Conference Series, Vol. 470, 370 Years of Astronomy in Utrecht, ed. G.~{Pugliese}, A.~{de Koter}, \& M.~{Wijburg}, 141

\bibitem[{{Sander} {et~al.}(2019){Sander}, {Hainich}, {Hamann}, {Ignace}, {Oskinova}, {Shenar}, {Todt}, \& {Vink}}]{Sander2019}
{Sander}, A., {Hainich}, R., {Hamann}, W.-R., {et~al.} 2019, {Anchoring mass-loss and metallicity for the WR population in M31 - A prototype study}, HST Proposal. Cycle 27, ID. \#15822

\bibitem[{{Savage} {et~al.}(1983){Savage}, {Fitzpatrick}, {Cassinelli}, \& {Ebbets}}]{Savage1983}
{Savage}, B.~D., {Fitzpatrick}, E.~L., {Cassinelli}, J.~P., \& {Ebbets}, D.~C. 1983, \apj, 273, 597

\bibitem[{{Schneider} {et~al.}(2018){Schneider}, {Sana}, {Evans}, {Bestenlehner}, {Castro}, {Fossati}, {Gr{\"a}fener}, {Langer}, {Ram{\'\i}rez-Agudelo}, {Sab{\'\i}n-Sanjuli{\'a}n}, {Sim{\'o}n-D{\'\i}az}, {Tramper}, {Crowther}, {de Koter}, {de Mink}, {Dufton}, {Garcia}, {Gieles}, {H{\'e}nault-Brunet}, {Herrero}, {Izzard}, {Kalari}, {Lennon}, {Ma{\'\i}z Apell{\'a}niz}, {Markova}, {Najarro}, {Podsiadlowski}, {Puls}, {Taylor}, {van Loon}, {Vink}, \& {Norman}}]{Schneider2018}
{Schneider}, F.~R.~N., {Sana}, H., {Evans}, C.~J., {et~al.} 2018, Science, 359, 69

\bibitem[{{Schwarzschild}(1958)}]{Schwarzschild1958}
{Schwarzschild}, M. 1958, {Structure and evolution of the stars.}

\bibitem[{Shenar {et~al.}(2023)Shenar, Sana, Crowther, Bostroem, Mahy, Najarro, Oskinova, \& Sander}]{shenar2023}
Shenar, T., Sana, H., Crowther, P.~A., {et~al.} 2023, Constraints on the multiplicity of the most massive stars known: R136 a1, a2, a3, and c

\bibitem[{{Smith}(2014)}]{Smith2014}
{Smith}, N. 2014, \araa, 52, 487

\bibitem[{{Smith} \& {Conti}(2008)}]{Smith2008}
{Smith}, N. \& {Conti}, P.~S. 2008, \apj, 679, 1467

\bibitem[{{Smith} \& {Tombleson}(2015)}]{Smith2015}
{Smith}, N. \& {Tombleson}, R. 2015, \mnras, 447, 598

\bibitem[{{Spera} \& {Mapelli}(2017)}]{Spera2017}
{Spera}, M. \& {Mapelli}, M. 2017, \mnras, 470, 4739

\bibitem[{Spera \& Mapelli(2017)}]{speraVeryMassiveStars2017}
Spera, M. \& Mapelli, M. 2017, Monthly Notices of the Royal Astronomical Society

\bibitem[{{Spera} {et~al.}(2015){Spera}, {Mapelli}, \& {Bressan}}]{Spera2015}
{Spera}, M., {Mapelli}, M., \& {Bressan}, A. 2015, \mnras, 451, 4086

\bibitem[{Spera {et~al.}(2019)Spera, Mapelli, Giacobbo, Trani, Bressan, \& Costa}]{speraMergingBlackHole2019a}
Spera, M., Mapelli, M., Giacobbo, N., {et~al.} 2019, Monthly Notices of the Royal Astronomical Society, 485, 889

\bibitem[{{Spera} {et~al.}(2022){Spera}, {Trani}, \& {Mencagli}}]{spera2022}
{Spera}, M., {Trani}, A.~A., \& {Mencagli}, M. 2022, Galaxies, 10, 76

\bibitem[{Ugolini {et~al.}(in prep)Ugolini, Sgalletta, Gabrielli, Limongi, Arca~Sedda, Paiella, Mestichelli, Di~Carlo, \& Spera}]{Ugolini_bump_2025}
Ugolini, C., Sgalletta, C., Gabrielli, F., {et~al.} in prep, Astronomy \& Astrophysics

\bibitem[{{van Son} {et~al.}(2022){van Son}, {de Mink}, Renzo, Justham, Zapartas, Breivik, Callister, Farr, \& Conroy}]{vansonNoPeaksValleys2022a}
{van Son}, L. A.~C., {de Mink}, S.~E., Renzo, M., {et~al.} 2022, The Astrophysical Journal, 940, 184

\bibitem[{{Vanbeveren} \& {Conti}(1980)}]{Vanbeveren1980}
{Vanbeveren}, D. \& {Conti}, P.~S. 1980, \aap, 88, 230

\bibitem[{{Vink}(2015)}]{Vink2015}
{Vink}, J.~S. 2015, in Wolf-Rayet Stars, ed. W.-R. {Hamann}, A.~{Sander}, \& H.~{Todt}, 133--138

\bibitem[{{Vink} {et~al.}(2000){Vink}, {de Koter}, \& {Lamers}}]{Vink2000}
{Vink}, J.~S., {de Koter}, A., \& {Lamers}, H.~J.~G.~L.~M. 2000, \aap, 362, 295

\bibitem[{{Vink} {et~al.}(2001){Vink}, {de Koter}, \& {Lamers}}]{Vink2001}
{Vink}, J.~S., {de Koter}, A., \& {Lamers}, H.~J.~G.~L.~M. 2001, \aap, 369, 574

\bibitem[{{Vink} \& {Gr{\"a}fener}(2012)}]{Vink2012}
{Vink}, J.~S. \& {Gr{\"a}fener}, G. 2012, \apjl, 751, L34

\bibitem[{{Vink} {et~al.}(2021){Vink}, {Higgins}, {Sander}, \& {Sabhahit}}]{Vink2021}
{Vink}, J.~S., {Higgins}, E.~R., {Sander}, A. A.~C., \& {Sabhahit}, G.~N. 2021, \mnras, 504, 146

\bibitem[{{Vink} {et~al.}(2011){Vink}, {Muijres}, {Anthonisse}, {de Koter}, {Gr{\"a}fener}, \& {Langer}}]{Vink2011}
{Vink}, J.~S., {Muijres}, L.~E., {Anthonisse}, B., {et~al.} 2011, \aap, 531, A132

\bibitem[{{Volpato} {et~al.}(2023){Volpato}, {Marigo}, {Costa}, {Bressan}, {Trabucchi}, \& {Girardi}}]{Volpato2023}
{Volpato}, G., {Marigo}, P., {Costa}, G., {et~al.} 2023, \apj, 944, 40

\bibitem[{{Volpato} {et~al.}(2024){Volpato}, {Marigo}, {Costa}, {Bressan}, {Trabucchi}, {Girardi}, \& {Addari}}]{Volpato2024}
{Volpato}, G., {Marigo}, P., {Costa}, G., {et~al.} 2024, \apj, 961, 89

\bibitem[{{Weigelt} \& {Baier}(1985)}]{Weigelt1985}
{Weigelt}, G. \& {Baier}, G. 1985, \aap, 150, L18

\bibitem[{{Woosley}(2017)}]{Woosley2017}
{Woosley}, S.~E. 2017, \apj, 836, 244

\bibitem[{{Woosley} \& {Heger}(2021)}]{Woosley2021}
{Woosley}, S.~E. \& {Heger}, A. 2021, \apjl, 912, L31

\bibitem[{{Yusof} {et~al.}(2013){Yusof}, {Hirschi}, {Meynet}, {Crowther}, {Ekstr{\"o}m}, {Frischknecht}, {Georgy}, {Abu Kassim}, \& {Schnurr}}]{Yusof2013}
{Yusof}, N., {Hirschi}, R., {Meynet}, G., {et~al.} 2013, \mnras, 433, 1114

\end{thebibliography}

%---------------------------------------------
%---------------------------------------------
%---------------------------------------------

\appendix
\section{Full mass loss recipes}\label{app:massloss}
The full equation for the mass loss prescription from \citet{Vink2001} for the hot side of the bistability jump is given as:
\begin{equation}
\begin{split}
    \text{log} ~ \dot{M}_{\mathrm{Vink2001}} = -6.697 + 2.194  \text{  log}  (L/\lsun / 10^5) \\
    -1.313  \text{  log}  (M/\msun / 30) \\
     -1.226 \text{  log}  ((v_{\infty}/v_{\mathrm{esc}})/2) \\
     + 0.933 \text{ log} (\teff /40000 ) \\
     -10.92 {\text{ log}(\teff / 40000)}^2 \\
     + 0.85 \text{  log}  (Z/\zsun)
\end{split} 
\label{eq:vink_01}
\end{equation}
Where $L, M, \teff$, and $Z$ are the star's current luminosity, mass, effective temperature, and metallicity. On the hot side of the bistability jump, the ratio of the terminal velocity to the escape velocity is given as $\nu_{\infty}$/$\nu_{\mathrm{esc}}$ = 2.6.
%Note the terminal velocity $\nu_{\infty}$ is a power law function of Z, $\nu_{\infty} \propto Z^{0.13}$ \citep{Leitherer1992}. 

%It is mostly metals that are responsible for line driving, and as in early-type stars, the main mechanism driving the wind is the radiation pressure on spectral lines; 
Metallicity is one of the most important factors driving stellar winds of VMS. Iron recombination (Fe IV $\rightarrow$ Fe III) is the origin of a so-called \textit{bi-stability jump} where the mass loss rate increases dramatically at \teff $\sim$ 25\,000 K. The exact \teff\ the jump occurs depends on Z; however, the mass loss rate is expected to increase about fivefold at galactic metallicities \citep{Vink2001}. %\cite{Vink2001} provides a complete mass loss recipe, depending on the location of the bistability jump equation. 
Equation \ref{eq:vink_01} provides the mass loss rates for the hot side of the bistability jump. The \teff\ of the jump can be calculated from the metallicity. 

%Iron is a complex structure and allows for millions of line transitions.
%Two other \textit{bistability jumps} have been found at \teff $\sim$ 15,000 and \teff $\sim$ 35,000 K. The first one is attributed to the recombination of Iron from Fe III $\rightarrow$ Fe II. However, the jump at \teff $\sim$ 35,000 K is due to the recombination of Carbon IV $\rightarrow$ Carbon III, and only occurs at low metallicity environments \cite{Vink2000}. 

Monte Carlo calculations of stars with initial masses up to 300 \msun\ have shown a kink in the mass loss rates for the largest masses considered \citep{Vink2011}. Below the kink, they found hydrodynamically consistent mass loss rates that agree with the $\beta$-velocity law from \cite{Vink2001}. However, the models of \cite{Vink2011} find that when \gammae\ exceeds a critical value (which is model dependent), there is a transition in the mass loss type from O-type stars to a more extreme WR type. Above the kink, in the optically thick regime, the mass loss rate has been found to have a steep dependence on \gammae.
Moreover, they find the spectral morphology of the He II line at 4686 \AA\  to change for differing \gammae. For increasing \gammae, models show stronger and broader He II 4686 \AA\  emission lines, a line which is characteristic of stars transitioning from Of to WN stars.
The mass loss rates follow different dependencies for high and low \gammae\ regimes \citep{Vink2011}:
\begin{equation}
    \Dot{M} \propto M^{0.68} \gammae^{2.2} \quad \text{for} \quad 0.4 < \gammae < 0.7 
\end{equation}

\begin{equation}
    \Dot{M} \propto M^{0.78} \gammae^{4.77} \quad \text{for} \quad 0.7 < \gammae < 0.95
\end{equation}

However, \cite{Vink2011} does not provide the full dependencies for the mass loss rates. We take the maximum between the \cite{Vink2001} and \cite{Vink2011}.

The full mass loss prescriptions derived in \cite{Sabhahit2022} are written as:

\begin{equation}
\begin{split}
     \text{  log } \Dot{M}_{\text{high, } \Gamma_e} = -9.552 + 4.77 \text{  log} (1 + X_s) \\
     + 4.77 \text{  log}  (L/\lsun / 10^5) \\
     -3.99 \text{  log} (M / \msun /30) \\
     -1.226 \text{  log}  ((v_{\infty}/v_{\mathrm{esc}})/2)\\
     + 0.5 \text{  log}  (Z/\zsun)
     \label{eq:gammae_sab_rate}
\end{split}
\end{equation}

\begin{equation}
\begin{split}
     \text{  log }  \Dot{M}_{\text{high, L/M}} = -8.445  \\
     + 4.77 \text{  log}  (L/\lsun / 10^5) \\
     -3.99 \text{  log}  (M / \msun /30) \\
     -1.226 \text{  log} ((v_{\infty}/v_{\mathrm{esc}})/2)\\
     + 0.761 \text{  log} (Z/\zsun)
     \label{eq:LM_sab_rate}
\end{split}
\end{equation}

The models from \cite{Sabhahit2022} take the maximum between \cite{Vink2001} and their new prescriptions. % To faithfully reproduce their results, we also only consider the maximum between  \citet{Vink2001} and the mass loss rate from considering equations  \ref{gammae_sab_rate} and \ref{eq:LM_sab_rate}.

In the models we present here, when \teff\ $>$ 10\,000 K, we choose the maximum between the low and high-\gammae\ mass loss regimes, such that the models use the 'low-\gammae' mass loss recipe of Eq. \ref{eq:vink_01} below the transition point, and use the 'high-\gammae' recipes of Eqs. \ref{eq:gammae_sab_rate} or \ref{eq:LM_sab_rate} above the transition point. When \teff\ $<$ 10\,000, we take the maximum between \cite{deJager1988} and Eqs. \ref{eq:gammae_sab_rate} or \ref{eq:LM_sab_rate}. We will refer to these models as \mgam\ and \mlm, shown in Eqs \ref{eq:mgam} and \ref{eq:mlm}.

\section{Mass estimates for R136a2 and R136a3}\label{app:r136a2}
\begin{figure}[t]
    \centering
    \includegraphics[width=\linewidth]{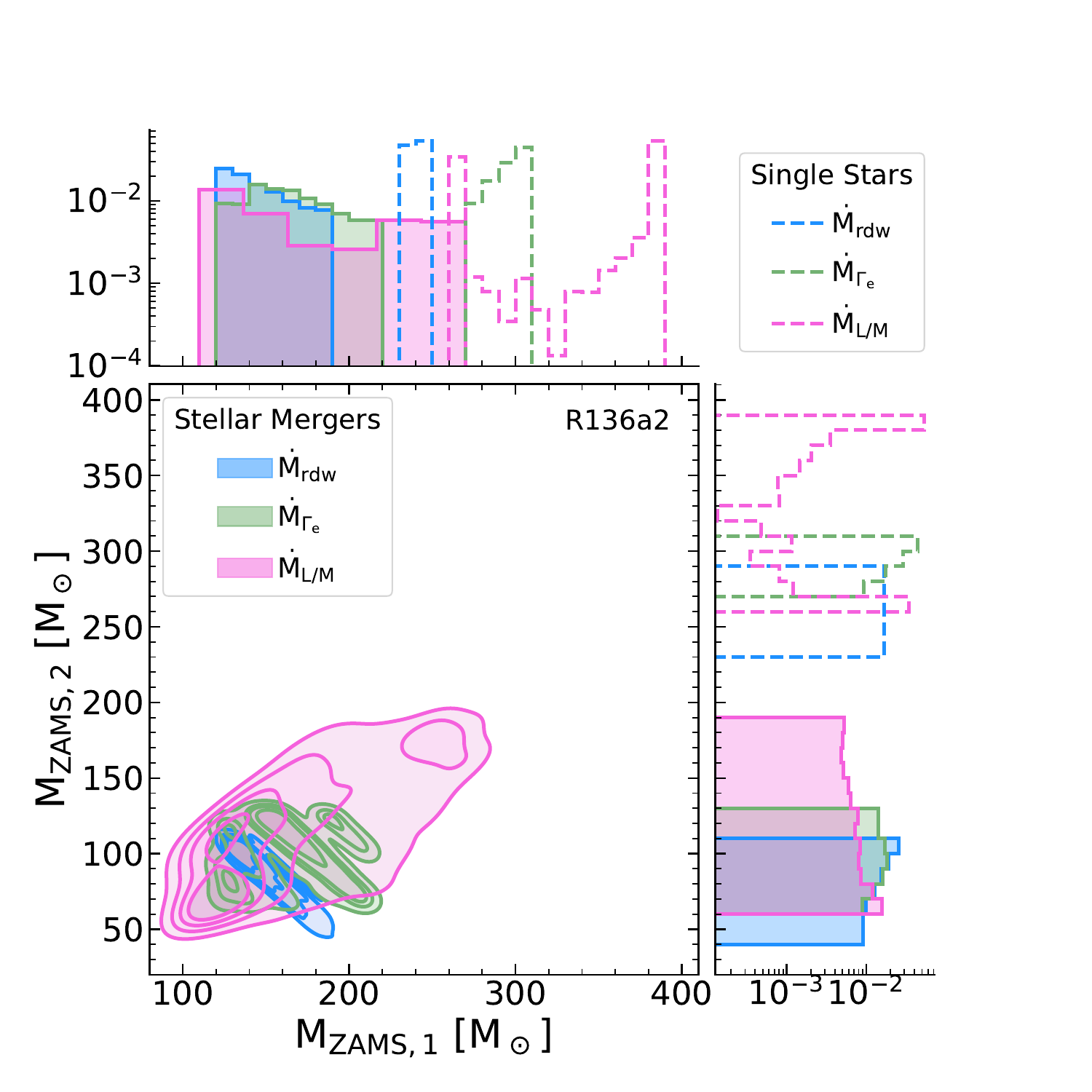}
    \caption{The possible initial masses of R136a2 from single star and stellar merger origins.
    The shaded areas show the primary and secondary ZAMS masses of stars that merge during evolution and overlap in the HRD with the observed location of R136a2. The blue, green, and pink show results from \mrdw\, \mgam, and \mlm, respectively. The dashed histograms show the initial mass from single stellar evolution. }
    \label{fig:R136a2_masses}
\end{figure}

Here we briefly discuss the possible origins of the VMSs R136a2 and R136a3. The initial and current mass derived from single and binary mergers that match the observed properties of the R136 stars are listed in Table \ref{table:R136_masses_all}, including mass estimates from literature.
Figure \ref{fig:R136a2_masses} shows the initial mass distributions for R136a2 from single and binary stellar merger origins. Similarly to our results of R136a1, a single star origin necessitates a ZAMS mass $\sim$ 100 \msun\ larger than the ZAMS masses from stellar merger origin. The estimates from a single star origin both from literature and from all of our wind recipes generally suggest ZAMS masses $>$ 200 \msun. Even so, the current mass derived is in general agreement between single and binary origin, for models using the same wind prescription.
%In particular, single star models predict R123a2's initial mass to be 240 $^{+ 6  }_{ - 6 }$ \msun\ (for the \mrdw\ tracks), 296 $^{+ 12  }_{ - 20 }$ \msun\ (for \mgam), and 260 $^{+ 130  }_{ - 1 }$ (for \mlm). Instead, if the origin is from a binary stellar merger, we find the primary initial mass to be 146 $^{+ 42  }_{ - 21 }$ \msun\ for the \mrdw\ simulation, 160 $^{+ 60  }_{ - 40 }$ \msun\ for the \mgam\ simulation, and 153 $^{+ 118  }_{ - 43 }$ \msun\ for the \mlm\ simulation.

%From single stellar evolution, the current mass of R136a2 is estimated to be 225 $^{+ 9  }_{ - 8 }$ \msun\ (\mrdw), 169 $^{+ 13 }_{ - 13 }$  \msun\ (\mgam), and 201$^{+ 32  }_{ - 80 }$ \msun\ (\mlm). The current mass estimate given by stellar mergers are much lower; the \mrdw\ simulation gives primary masses of 222 $^{+ 7  }_{ - 7 }$ \msun, the \mgam\ gives 164 $^{+ 24  }_{ - 4 }$ \msun, and the \mlm\ suggest current masses from 140 $^{+ 5  }_{ - 12 }$ \msun. 

\begin{table*}[t]
\renewcommand{\arraystretch}{1.2}
\centering
\caption{Mass estimates of R136a from single and binary stellar mergers for the winds described in this work.} 
    
    \begin{tabularx}{\textwidth}{@{\extracolsep{\fill}}lcccccc}
    \toprule
    \multirow{3}{*}{Method and Binarity} & \multicolumn{2}{c}{R136a1} & \multicolumn{2}{c}{R136a2} & \multicolumn{2}{c}{R136a3} \\
    \cmidrule(lr){2-3} \cmidrule(lr){4-5} \cmidrule(lr){6-7}
    & $M_{\rm ZAMS}$ & $M_{\rm current}$ & $M_{\rm ZAMS}$ & $M_{\rm current}$ & $M_{\rm ZAMS}$ & $M_{\rm current}$\\
    \midrule
     \mrdw\ (Single) & 315 $^{+11}_{-13}$ & 295 $^{+14}_{-16}$ & 240 $ ^{ +6}_{ -6 }$ & 255 $ ^{+ 9 }_{ -8 }$ & 241 $^{ +5 }_{- 5 }$ & 232 $^{+ 9 }_{- 9 }$ \\
    \mrdw\ (Stellar merger)  & 186 $^{+ 53 }_{ - 30 }$ & 278 $^{ +25 }_{- 18 }$ & 146 $^{+ 42 }_{ - 21 }$ & 222 $^{+ 7 }_{ - 7 }$ & 145 $^{+ 41 }_{- 21 }$ & 221 $^{+ 9 }_{- 9 }$\\
    \mgam\ (Single) & 389 $^{+ 1 }_{- 64 }$ & 202 $^{+ 12 }_{- 66 }$ & 296 $^{+ 12 }_{- 20 }$ & 169 $^{+ 13 }_{- 13 }$ & 286 $^{+ 4 }_{- 25 }$ &  180 $^{+ 42 }_{- 14 }$ \\
    \mgam\ (Stellar merger)  & 207 $^{ + 69 }_{- 56 }$ & 200 $^{+ 10 }_{- 6 }$ & 160 $^{+ 60 }_{ - 40 }$ & 164 $^{+ 24 }_{ - 4 }$  & 162 $^{+ 52 }_{- 36 }$ & 171 $^{ + 12 }_{- 11 }$ \\
    \mlm\ (Single)& 359 $^{+ 30 }_{- 10 }$ & 254 $^{+ 22 }_{- 44 }$ & 260 $^{+130}_{- 1 }$ & 201 $^{+32}_{- 80}$  & 257 $^{+39}_{-4}$  & 220 $^{+17}_{-41}$\\
    \mlm\ (Stellar merger)  & 239 $^{+ 87 }_{- 108 }$ & 191 $^{ +28 }_{- 19 }$ & 153 $^{+ 118 }_{ - 43 }$ & 140 $^{+ 5 }_{ - 12 }$  & 181 $^{ +90 }_{- 67 }$ &  139 $^{+ 9 }_{- 8 }$ \\
    \midrule
    \citet{Brands2022} & 273 $^{ +25}_{- 36 }$ & 222 $^{ +28 }_{- 29 }$ & 221 $^{ +16 }_{- 12 }$ & 186 $^{ +17 }_{- 15 }$  & 213$^{ +12}_{- 11}$  &  179 $^{ +16}_{- 11 }$ \\
    \citet{Bestenlehner2020} & 251 $^{ +48}_{- 35 }$ & 215$^{ +45}_{- 31 }$ & 211 $^{ +31}_{- 32 }$  & 187$^{ +23}_{- 33 }$ & 181 $^{ +29}_{- 31 }$ & 154$^{ +28}_{- 23 }$ \\
    \citet{Crowther2010} & 320 $^{ +100}_{- 40 }$ & 265 $^{ +80}_{- 35 }$ & 240$^{ +45}_{- 45 }$  & 195$^{ +35}_{- 35 }$  & 165 $^{ +30}_{- 30 }$ & 135$^{ +25}_{- 20 }$  \\
    \bottomrule
    \end{tabularx}
    \footnotesize{Column 1: The mass loss recipe used and the origin of the star. Column 2 and 3: ZAMS mass and current mass estimate of R136a1. Column 4 and 5: ZAMS mass and current mass estimate of R136a2. Column 5 and 6: ZAMS mass and current mass estimate of R136a3. In the case of a stellar merger, the \Mz\ shows the mass of the primary star only. The last three rows show comparisons from the literature, and all assume a single star origin. \citet{Crowther2010} results use the Geneva stellar evolution tracks, and the results from \citet{Bestenlehner2020} and \citet{Brands2022} both used the stellar models from  \citet{Brott2011} and \citet{Kohler2015}.  
    }
\label{table:R136_masses_all} 
\end{table*}

\begin{figure}[t]
    \centering
    \includegraphics[width=\linewidth]{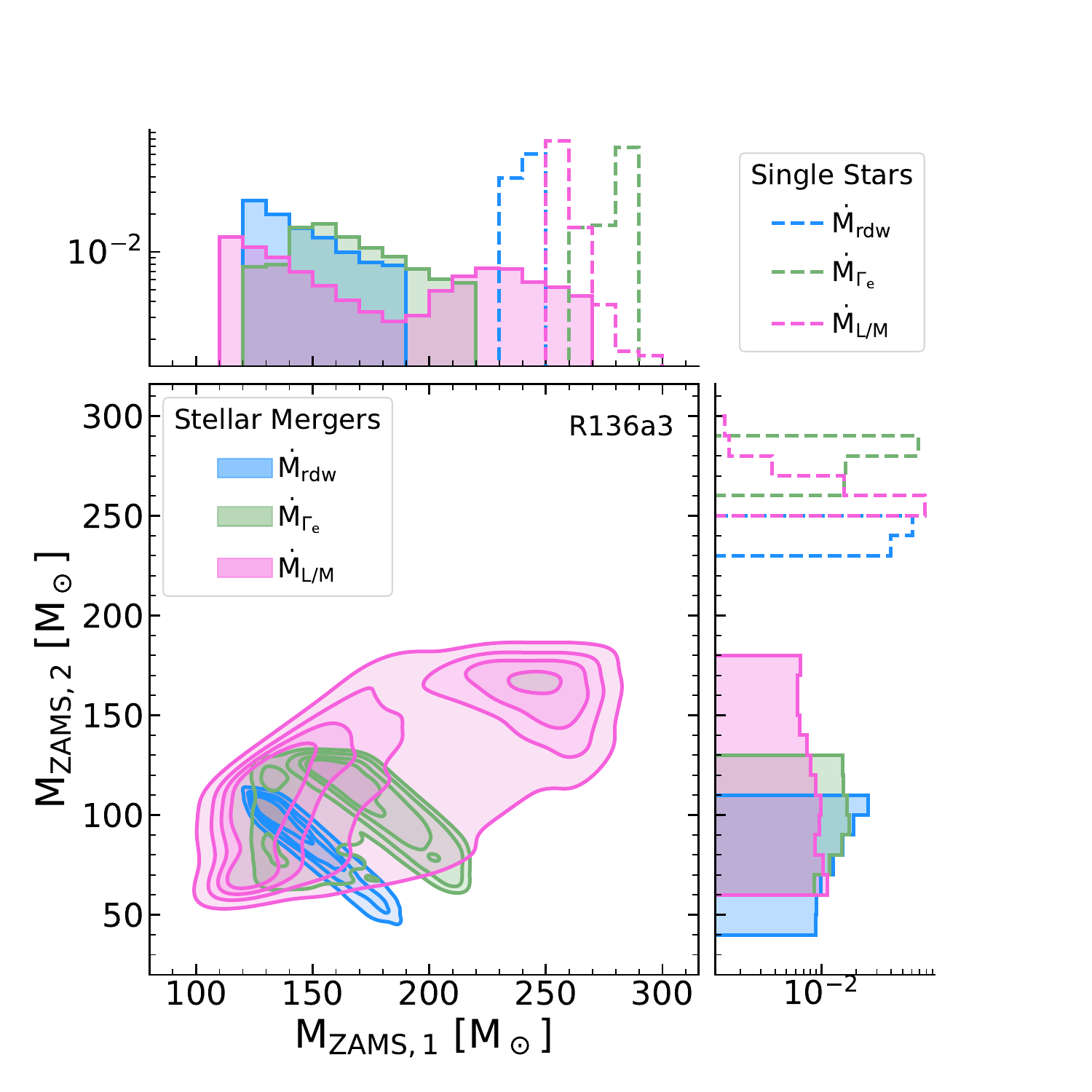}
    \caption{The same as Fig. \ref{fig:R136a2_masses} but for R136a3. }
    \label{fig:R136a3_masses}
\end{figure}

Finally, the 1$\sigma$ mass estimates from single and binary origin are shown in Figure \ref{fig:R136a3_masses} for the best fit to the R136a3 star. As expected, the stellar merger origin necessitates a primary ZAMS mass $\sim$ 100 \msun\ less than needed for a single star origin. Interestingly, both stellar mergers and single star current mass estimates from \mgam\ and \mrdw\ agree within the uncertainty of other mass estimates from literature, while the ones using \mrdw\ are much larger than other estimates. This is because the \mrdw\ tracks fit the stars better earlier in the evolution (see Table \ref{table:r136}), when they have not lost significant amounts of mass yet. The \mrdw\ binaries are more likely to merge earlier in evolution compared to the enhanced wind simulations, and therefore also retain more of their initial mass than the enhanced winds simulations.

\section{Rotating tracks }\label{app:rotating}
%\KS{they have higher core overshooting values -> more likely to have CHE/bigger core/etc. + check G1's 2019 paper }

Stellar rotation is a crucial aspect influencing the evolution of massive stars; for example, it may enhance mass loss through centrifugal effects. Rotation changes the stellar geometry by causing the star to become more oblate and affecting the surface temperature \citep[e.g., see][]{Maeder2009}. Rotation also increases the mixing of chemical elements, altering the nuclear burning lifetimes and abundance composition at the surface \citep{Maeder2009}. As such, rotation can also play a critical role in determining the final fate of the star.

We compute full stellar evolution models for stars with masses 100 $\leq$ \mzams/ \msun\ $\leq$ 350. The rotation processes are described in \citet{Costa2019a, Costa2019b, Nguyen2022}. Models are computed with initial rotation rates $\omega$ = 0.4, $\omega$ =0.6, and 100 $\leq$ \mzams/ \msun\ $\leq$ 200 for $\omega$ = 0.8. $\omega$ is the fraction of the initial angular velocity to the critical angular velocity, i.e.  $\omega$ = $\Omega_{\mathrm{ini}}/\Omega_{\mathrm{crit}}$. Our models take into account the mass loss enhancement due to rotation following the prescriptions from \cite{Heger2000}, which include a dependency on the ratio between the surface velocity and the break-up velocity.
We take into account mechanical mass loss when the stars reach the critical rotation \cite{Georgy2013}.

\begin{figure*}[t]
    \centering
    \includegraphics[width=\linewidth]{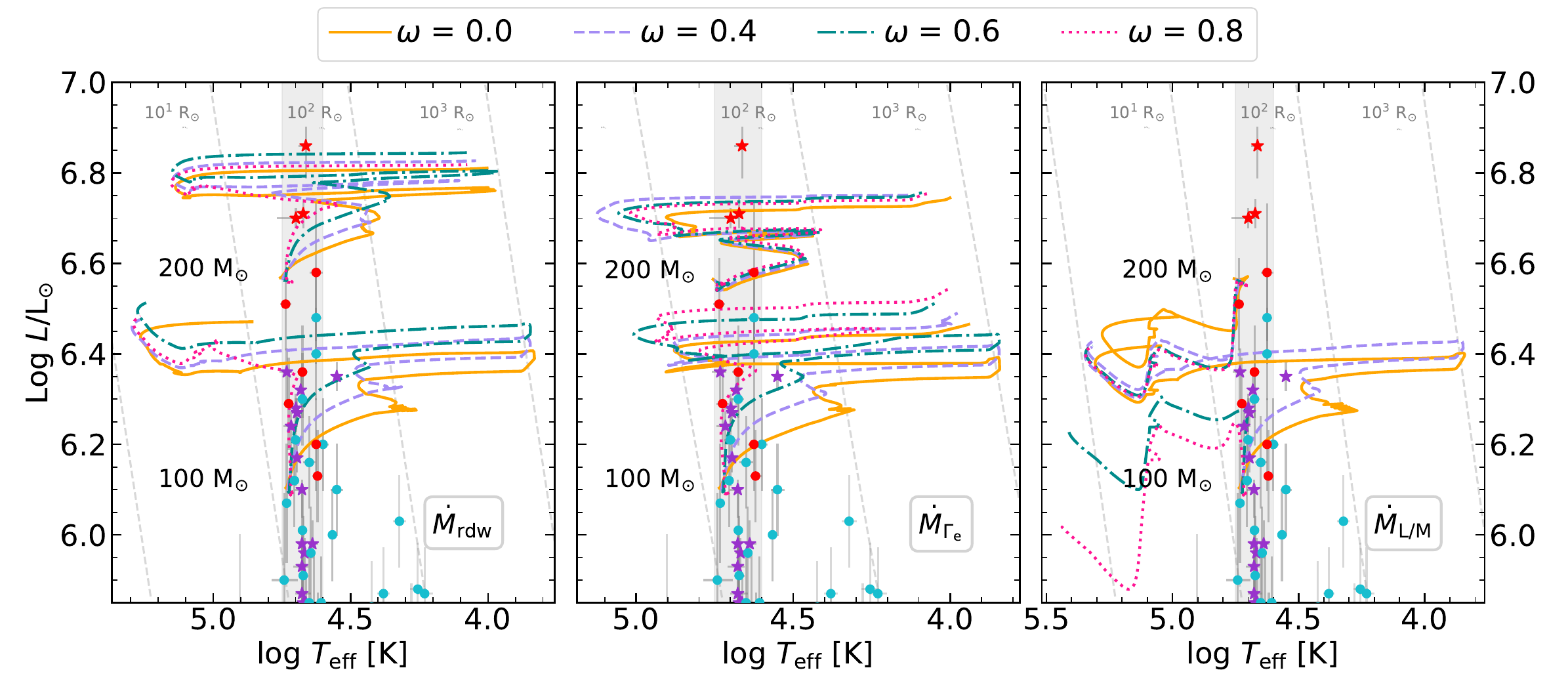}
    \caption{HRD for the different rotation rates as indicated by line color and style. The left, center, and right panel refer to the \mrdw, \mgam, and \mlm\ rates, respectively. The gray-shaded region indicates the observed temperature range in the Tarantula Nebula. Data of stars in 30 Doradus are shown in colored symbols, following the same notation as Fig. \ref{fig:HRD_LMC}.}
    \label{fig:HRD_rotating}
\end{figure*}

Figure \ref{fig:HRD_rotating} shows the HRD of selected tracks computed with four rotation rates for each mass loss recipe. For all the wind prescriptions, models with larger rotation rates evolve at higher luminosities and remain hotter when on the MS than nonrotating models. A major consequence of including rotation in the models is that they spend more of their evolution in the range of observed temperatures of VMS. %The final masses for the three rotation rates and different mass loss rates are displayed in Table \ref{table:masses}.

% Diffs in mrdw
For the \mrdw\ rate, the MS evolution occurs at higher luminosity than nonrotating models. The increase is moderate, except for the 100 \msun\ star with $\omega$ = 0.8, which evolves almost vertically up on the MS. This strong increase in luminosity is due to the surface enrichment of helium, %\GV{see 2nd paragraph of Sec 3.1 in Volpato+24}, 
which causes the increase in luminosity and \teff. The enhanced He can be attributed to the rotating models having increased rotational mixing, which brings fresh material from the envelope to the core. The convective cores thus make up a larger fraction of the total star's mass, compared to non-rotating models. %Regions above the convective core experience increased rotational mixing, which occurs even at lower metallicity \citep{Maeder2009}.
By the end of the MS, the surface He abundance can reach 93 $\%$ of the surface mass fraction for $\omega$ = 0.8, while for $\omega~\leq$ 0.6, the surface He mass fraction only reaches 68 - 85 $\%$, depending on the \mzams.
The rotating models of this rate show enhanced surface C, N, O, and Ne at the surface compared to the nonrotating models. We find no evidence of chemically homogeneous evolution for any masses or rotation rates considered in this work.
The 200 \msun\ models show less pronounced differences between rotating and nonrotating models, the greatest distinction is the increase in luminosity and decreased \teff\ range experienced by the stars on the MS.
%All 100 \msun\ models of this set are able to burn though central Ne before ending evolution.
%Rotation is also able to increase the surface N content. 

As rotation is expected to increase mass loss rates, it is interesting to note the unexpected trend that the final masses of rotating models with \mrdw\ rates tend to have larger final masses than non-rotating models (see Table \ref{table:masses_rotating}). This feature arises because the non-rotating models evolve to cooler \teff, where they encounter the bistability jump and have enhanced mass loss rates. If any rotating model does cross the bi-stability jump to experience stronger winds, it is for less time than the non-rotating models, thus resulting in less total mass lost due to the bistability jump.
Overall, the evolution of tracks with \mrdw\ rates are more strongly affected by stellar winds than the affects of rotation, with the exception of \mzams\ $\leq$ 100 \msun\ with $\omega$ = 0.8. % expand ...

% diffs in \mgam
The models computed with \mgam\ rates are shown in the central panel of Figure \ref{fig:HRD_rotating}. The 100 \msun\ models show the same differences in evolution as the \mrdw\ models, including the consequence of encountering the bistability jump has on the final mass.
%As expected, the rotating models show a boost in luminosity and hotter \teff\ on the MS. This is due to the enhanced He abundance, as in the \mrdw\ case. 
Compared to the \mrdw\ case, the \mgam\ 200 \msun\ models show even less variation in the HRD evolution for rotating models. Indeed, these models evolve through the same burning phases and end evolution with nearly identical final masses and surface compositions, composed of  $X\mathrm{_{He}^{S}}=$ 0.95 - 0.98,  $X\mathrm{_{H}^{S}}\sim~ 10^{-2}$, and  $X\mathrm{_{N}^{S}} \sim 10^{-3}$ in mass fraction. For these stars, the evolution is entirely dominated by the stellar winds and rotation plays a subdominant role in the evolution.
 %These models also show the same feature of the \mrdw\ models, of models resulting in slightly lighter final masses of rotating stars, due to 

% diffs in mlm
For the \mlm\ models with \mzams\ = 100 \msun, rotation begins to appreciably change the evolution on the HRD.  Like the \mrdw\ and \mgam\ rates, rotating models show a steep increase in luminosity at nearly constant \teff\ at the start of the MS, but the evolution is nearly vertical for $\omega \geq$ 0.6. The $\omega$ = 0.6 and $\omega$ = 0.8 models have their surface composition dominated by He earlier in the MS compared to the $\omega$ = 0.0 and $\omega$ = 0.4 models, causing the steeper increase in luminosity.
The increased rotation strengthens the mass loss, and models with $\omega \geq$ 0.6 become WR stars midway through the MS. By the end of the MS, these models have already lost 10 - 15 \msun\ more material than the $\omega \leq$ 0.4 models. By the end of evolution, the rapidly rotating model ends with only 1/2 of the mass of the nonrotating model. %highlighting the grand effect of the interplay between winds and rotation. 
On the other hand, the 200 \msun\ model is overall less influenced by rotation than it is by stellar winds, similar to the \mgam\ models. The final masses of rotating and nonrotating models are comparable. 

The initial rotation rate at which rotation dominates the evolution depends on the wind prescription, initial mass, and rotation rate. Overall, we found that rotation becomes the dominant factor affecting the evolution on the MS only when $\omega \geq$ 0.6 for \mrdw\ and \mlm\ for stars born with \mzams\ $\leq$ 150 \msun. For these stars, the variability of when rotational effects become dominant underscores the need for careful selection of wind models and highlights the complex interplay between the stellar winds, rotation rates, and ZAMS masses
For more massive stars, with \mzams $\geq$ 200 \msun\ and all wind prescriptions, it is mass loss and not rotation that dominates the evolution. This result is in agreement with other works \citep[e.g.][]{Sabhahit2022}, as rotation has a negligible effect on the evolution of VMS \citep{Yusof2013}. 

%when predicting evolutionary outcomes, particularly in the context of massive stars and their final fates.
%The disparity of rotational affects and ZAMS masses on the evolution and final mass of the star highlights the interplay between these two physical processes. 

%Rotation takes a subdominant role in determining the final masses for these stars. 
%his feature does not occur for models that remain hot throughout evolution, such as the \mzams\ = 200 \msun\ model computed with \mlm\ winds. 

%However, the surface He of the $\omega \geq$ 0.6 models is also stripped away earlier, and is no longer dominant essentially at the start of the CHeB phase, while the $\omega \leq$ 0.4 retain surface He for longer during CHeB. 
%Instead the stars become hotter and evolve at constant luminosity. It's at this point that the stars with $\omega~\geq$ 0.6 become WR stars. Shortly after, the surfaces are primarily composed of C and O. 

\begin{table}[t]
\renewcommand{\arraystretch}{1.2}
\centering
    \caption{Pre-SN masses in \msun\ for three different rotation rates.}
    \begin{tabularx}{\columnwidth}{@{\extracolsep{\fill}}ccccc}
    
    \toprule 
    
    \multirow{3}{*}{M$_{ZAMS}$} & \multicolumn{4}{c}{$\omega$} \\ %\cmidrule(lr){2-4}
        %\toprule
        & 0.0  & 0.4 & 0.6 & 0.8\\
        \cmidrule(r){1-1}  \cmidrule(lr){2-5}
        %\toprule
        \multicolumn{5}{c}{\mrdw} \\ \toprule
        100 & 48.75 & 50.36 & 54.69 & 51.40 \\
        150 & 75.10 & 78.66 & 82.50 & 77.27 \\
        200 & 103.13 & 106.77 & 111.15 & 104.71\\
        250 & 132.44 & 133.78 & 140.28 & 132.57 \\
        300 & 160.22 & 160.98 & 165.32 & - \\
        350 & 186.10 & 187.60 & 190.34 & - \\
         \hline
         \multicolumn{5}{c}{\mgam} \\ \toprule
         100 & 52.67 & 54.89 & 56.19  &  61.12 \\
         150 & 74.99 & 76.58 & 77.12 & 79.14 \\
         200 & 93.21 & 91.78 & 94.29 & 94.75 \\
         250 & 107.84 & 107.14 & 109.63 & - \\
        300 & 122.29 & 122.25 & 121.88 & - \\
        350 & 131.87 & 131.59  & 131.50 & - \\
        \hline
         \multicolumn{5}{c}{\mlm} \\ \toprule
         100 & 42.91 & 43.83 & 30.87  &  21.18 \\
         150 & 54.46 & 42.87 & 40.75 & 39.27 \\
         200 & 49.33 & 45.17 & 44.18 & 43.98\\
         250 & 51.27& 48.49 & 47.63& - \\
        300 & 54.96 & 52.26 & 51.34 & - \\
        350 & 58.98 & 56.36 & 55.62 & - \\

        \bottomrule
    \end{tabularx}

\label{table:masses_rotating}
\end{table}

\section{Merging binary BH formation channels}\label{app:merging_BBH}
Having characterized the overall population and formation scenarios of the BHs produced in our simulations in Sections \ref{subsec:stellar mergers} and \ref{subsec:Remnants}, we now focus on a particularly interesting subset: the BBHs that experience a gravitational wave-emitting merger within a Hubble time. Their formation and properties are a direct consequence of the binary evolution explored in the previous sections, which are also relevant to the formation of VMS in dense environments, like that of the Tarantula Nebula.

Binary BHs that coalesce in a Hubble time are the main source of gravitational waves and can provide valuable insights into binary evolution processes. Understanding how these systems form and evolve requires tracing their history through possible formation channels (e.g., \citealt{spera2022}). This section details the properties and formational channels of these BHs predicted by our simulations.

In all simulations, the most common formation channel is the so called 'channel I', following the definitions given by \citep{Broekgaarden2022, Iorio2023}: the progenitors undergo stable mass transfer before the first BH is formed, and after, evolve through at least one common envelope phase. This occurs in 91 \% of BBH mergers, regardless of the wind prescription. The second most common channel, occurring in 6\% of all BBH mergers, is channel IV, in which the progenitors undergo at least one CE phase before the first BH is formed, and the companion star has been stripped of its envelope before the BH formation. 

%5.1 p2
The frequency of different formation channels is generally very similar for all three simulations, as the majority of merging BBHs originate from progenitors with smaller ZAMS masses, whose evolution is not affected by the enhanced mass loss treatment presented in this work. However, the high-mass end of the BBH mergers (with $M_{\rm BH, 1} \geq$ 30 \msun) can arise from VMS progenitors, as described in Sect \ref{subsec:Remnants}, which influences the evolutionary pathway to BBH coalescence. For BBH mergers with a primary mass $M_{\rm BH, 1} \geq$ 30 \msun, a common formation channel is channel II. In this channel, the binary system evolves only through stable mass transfer episodes, and occurs in 82\%, 47\%, and 70\% of BBH mergers with \mrdw, \mgam, and \mlm, simulations, respectively. The most common pathway for \mgam\ is channel IV, with exactly one CE episode. The \mgam\ goes through channel IV 48\% of the time, while \mrdw\ and \mlm\ experience it just 1.5\% and 15\% of the time. Finally, there is the possibility that the binaries do not experience any mass transfer episodes, deemed channel 0. %this does not include accretion from winds
This channel arises in 13\%, 2\%, and 12\% of the massive BBH mergers for the \mrdw, \mgam, and \mlm\ cases.

Very massive primaries ($M_{\mathrm{ZAMS, 1}}$ > 140 \msun) in merging binaries computed with \mrdw\ tend to form with moderately massive secondaries (27 $\lesssim$ M$_{\rm ZAMS, 2}$/\msun $\lesssim$ 47), minor eccentric orbits ($e \in$ [0.01, 0.26]), and in close systems, $a \in$ [5.8 \rsun, 35.9 \rsun]. %However, this simulation produces only a handful of mergers with a secondary BH mass above 16 \msun, and none above 32 \msun.
Meanwhile, for the simulations with \mgam\ have %very massive primaries are able to merge as BHs with very massive secondaries (41 $\lesssim$ M$_{ZAMS, 2}$/\msun $\lesssim$ 147), at 
slightly larger semi-major axes $a \in $ [7.6 \rsun, 63.0 \rsun], and more circularized orbits ($e \in $ [0.01, 0.13]). This simulation resulted in a larger population of doubly massive merging systems than the \mrdw\ simulations. For the doubly massive systems computed with \mlm\ have %secondaries can be even more massive (38 $\leq$ M$_{ZAMS, 2}$/\msun $\leq$ 152), having 
close systems $a \in$ [14.3 \rsun, 60.3 \rsun], and a wider variety of eccentricities, $e \in $ [0.006, 0.38]. These simulations are able to produce many doubly massive merging systems, due in part to the remnant masses (see Fig. \ref{fig:Mini_Mfin_0.006}). 

%4.3 p5
The most massive primary BHs in the BBH merger population across all three simulations is $M_{\rm BH,1}$ $\sim$ 46 $\pm^1_1$ \msun, originating from progenitor stars with ZAMS masses of 115 \msun, 108 \msun, and 210 \msun\ for the \mrdw, \mgam, and \mlm\ winds, respectively. The \mgam\ and \mlm\ simulations produce up to an order of magnitude more BH primaries with $M_{\mathrm{BH, 1}} >$ 35 \msun\ than the \mrdw\ simulation (see Fig. \ref{fig:mergers}). They also yield significantly more massive secondaries, with $M_{\rm BH,2}$ $\sim$ 41.9 \msun\ and 41.7 \msun, originating from ZAMS masses of 147 \msun\ and 145 \msun. In contrast, that \mrdw\ simulation produces only a few mergers with secondary BHs above 16 \msun, and none above 32 \msun. These differences arise because the stronger winds in the \mgam\ and \mlm\ prescriptions increase mass loss and reduce stellar radii, lowering the likelihood of binary interactions and stellar mergers (see section \ref{subsec:stellar mergers}), thereby increasing the number of BBH systems that can later coalesce.

\end{document}